\documentclass[11pt,a4paper]{article}
\usepackage{amsmath}
\usepackage{enumerate}
\usepackage{amssymb}
\usepackage{graphicx}
\usepackage{multirow}
\usepackage{xcolor}
\usepackage{tablefootnote}
\usepackage{threeparttable}
\usepackage{bm}
\usepackage[normalem]{ulem}
\usepackage{soul}
\usepackage{jcappub}

\begin{document}

\title{The Age of the Universe with Globular Clusters III: Gaia distances and hierarchical modeling.}
\author[a]{David Valcin,}
\author[b,c]{Raul Jimenez,}
\author[a,d]{Uro\v{s} Seljak,}
\author[b,c]{Licia Verde}

\affiliation[a]{Berkeley Center for Cosmological Physics and Department of Physics, University of California, Berkeley, CA 94720, USA}
\affiliation[b]{ICC, University of Barcelona, Mart\' i i Franqu\` es, 1, E08028
Barcelona, Spain}
\affiliation[c]{ICREA, Pg. Lluis Companys 23, Barcelona, 08010, Spain.} 
\affiliation[d]{Lawrence Berkeley National Lab, 1 Cyclotron Road, Berkeley, CA 94720, USA}

\emailAdd{dvalcin@berkeley.edu}
\emailAdd{raul.jimenez@icc.ub.edu}
\emailAdd{useljak@bekeley.edu}
\emailAdd{liciaverde@icc.ub.edu}

\abstract{This is the third article in a series  aimed at computing accurate and precise ages of galactic globular clusters  from their full color-magnitude diagram in order to estimate the age of the Universe and in turn constrain the cosmological model. We update previous constraints using additional data and an improved methodology which allows us to vary the helium abundance and the reddening law in addition to the usual parameters (age, metallicity, alpha enhancement, distance and absorption) in the analysis. Even with the additional degrees of freedom, using the full color-magnitude diagram, now described as a Gaussian mixture bayesian hierarchical model, a tight constraint on the age(s) of the globular clusters and on the other parameters can be obtained and the statistical errors are fully subdominant to the systematic errors. We find that the age of the oldest globular clusters is $t_{\rm GC} = 13.39 \pm 0.10 ({\rm stat.}) \pm 0.23 ({\rm sys.})$ Gyr, resulting in an age of the Universe $t_{\rm U}=13.57^{+ 0.16}_{-0.14} ({\rm stat})\pm 0.23 ({\rm sys.})$ and a robust 95\% confidence upper limit of $t_U^{\leq}=13.92_{-0.1}^{+0.13}({\rm stat}) \pm{0.23} ({\rm sys})$. This is fully compatible with our previous estimates   
and with the model-dependent, Cosmic Microwave Background-derived age for the Universe of $t_{\rm U} =13.8 \pm 0.02$ Gyr for a $\Lambda$CDM model.}

\maketitle

\section{Introduction}
\label{sec:intro}

In the late 80s and early 90s of the past century, the so-called ``age tension" led to one of the first indications that new physics 
was needed to describe the Universe (see e.g., \cite{GC,Spinrad97}). The oldest objects in the Universe, galactic globular clusters and high redshift galaxies,   seemed to be older than the age  of the Universe itself, as predicted by the favored cosmological model at the time.  Indeed, the supernova results in the late 90s confirmed  the need to include the cosmological constant,  $\Lambda$ term,  or dark energy,  in Einstein's equations \cite{Perlmutter_sne, Riess_sne}, making the $\Lambda$CDM the  standard cosmological model. 

Today, the ``Hubble tension" (see e.g., recent reviews in~\cite{VTR2019,VerdeARAA,WendyJCAP,RiessJCAP} and references therein) has been hailed by some as a possible indication of a crack in the current cosmological paradigm. As with the discovery of dark energy, an accepted scientific  consensus can be  reached only after several independent lines of evidence and observational probes consistently confirm the need for new physics.   
In the case of the Hubble tension, the process of constructing different observational probes to provide several independent lines of evidence is now taking place.

Accurate estimates of the age of the Universe, inferred from the ages of the oldest stars, can help shed light on the  Hubble tension: such age determination is  independent of assumptions on the cosmological model and thus can be used to constrain it. It is a late-time probe, independent on the early-Universe physics. Further, it is completely independent of cosmological distances and relies only on  stellar nuclear reactions as an atomic clock to map the expansion history of the Universe.

It is now timely to revisit  these age estimates
given the recent advances in the modelling of stars, especially low mass ($M\lesssim M_{\odot}$) stars,  
the superb photometry of globular clusters obtained by the Hubble Space Telescope and, in the near future,  by the JWST telescope, the  accurate chemical abundances obtained via spectroscopy and exquisitely precise distances provided by the GAIA satellite. 

It is well known (e.g., \cite{JimenezPadoanLF,PadoanJimenezLF,JimenezPadoanGC})  that  richer information can be extracted from the full color magnitude diagram of GCs than simply using the main sequence turn-off luminosity.
In previous works~\cite{Valcin2020,Valcin2021} we have developed a  Bayesian framework to exploit the {\em full} color-magnitude diagram (CMD) of the resolved stellar population of a globular cluster (hereafter GC) to determine, simultaneously, age, metallicity [Fe/H],  alpha enhancement [$\alpha$/Fe], distance and reddening.  A precise and robust determination of the age of a GC requires a joint fit to all these quantities, and we developed a methodology to do so.

Our main results in~ \cite{Valcin2020, Valcin2021} were that it is possible to determine all these parameters jointly from the CMD, and that the age of the oldest GCs is $t_{\rm GC} = 13.32 \pm 0.1({\rm stat.}) \pm 0.23({\rm sys.})$ which translates into an age of the Universe of $t_{\rm U} = 13.5 \pm 0.15({\rm stat.}) \pm 0.33 ({\rm sys.})$ at 68\% confidence level, in agreement with the model-dependent value derived, within the $\Lambda$CDM model,  from the Cosmic Microwave Background observations of the Planck satellite ~\cite{planck}. In a subsequent paper, we studied the implications of these findings for the Hubble tension~\cite{BernalTriangles21}.
In the same vein, ~\cite{ChaboyerM92} obtained independently  age and other stellar parameters for a single GC (M92) in excellent statistical agreement (within one sigma) with our own previous estimate.

Missing from our previous studies was quantifying the effect of the (narrow) helium abundance prior adopted.
In this paper the helium abundance is  promoted to a free parameter, to be constrained by the data, and its impact on the age determination is explored quantitatively. We also include 
a more sophisticated  treatment of attenuation by dust.
Furthermore, we improve the methodology of the Bayesian inference framework by incorporating machine learning techniques that are capable of selecting better the GC star members and make results robust to the possible presence of multiple populations. Finally, we incorporate the latest GAIA distances to determine the adopted  distance prior.

This paper is organized as follows. In section \ref{sec:data} we describe the catalog and observations for the GCs used in this study, and review the key software components. We describe the improvements on the methodology compared to our previous work in section~\ref{sec:methods}. Parameter inference is described in section~\ref{sec:inference} as well as our treatment of systematic uncertainties. We present our main results in section~\ref{sec:results} and conclude with a summary in section~\ref{sec:summary}.

\section{Data and Stellar Models}
\label{sec:data}
 We use the same data,  stellar models and isochrone software packages as in \cite{Valcin2020, Valcin2021}, which we briefly summarize in this section. 
\subsection{Globular cluster sample and data quality cuts}

The  sample of \cite{Valcin2020, Valcin2021} (hereafter V20 and V21 respectively) consists of  68 globular clusters, selected from the  65 HST-ACS catalog ~\cite{Sarajedini2007} and additional 6 from~\cite{Dotter2010}, with additional quality cuts. Here we reintroduce a cluster that was ruled out in the previous studies for signs of multimodality in the previous selection, making the numbers of clusters in our sample 69. We refer to V20 for details on sample selection and data quality cuts, here we only summarize  them briefly. Only stars with high quality and high signal-to-noise photometry are included (see ~\cite{BayesianGC}). 
In addition, for each cluster a  ``functional'' magnitude interval is considered,  which ranges from a magnitude cut arbitrarily defined at $m_{F606W} = 27$ to the lowest apparent magnitude of the brightest stars \footnote{A further cut at low magnitudes is also introduced  to speed up the analysis without removing significant signal.}. Readers interested in the number and percentage of stars retained, can find  details in the appendix of V20.

\subsection{Software and stellar models}

A modified version of the software package \texttt{isochrones}\footnote{\url{https://github.com/timothydmorton/isochrones}, version 1.1-dev.}~\citep{isochrones} was developed  by 
V20, V21 and is used here.
The software reads synthetic photometry files provided by stellar models and  interpolates magnitudes along the stellar evolutionary tracks for fixed ages
correcting for absorption, given the input parameters.
While two stellar models  are implemented (\texttt{MIST}~\cite{MIST0,MIST1} and  \texttt{DSED}~\cite{dsed}), only \texttt{DSED} is used as it explores different abundances of $\alpha$ enhanced elements, parameterized by (non-zero values of) [$\alpha$/Fe].  These  elements, like O, Ne, Mg, Si, S, Ca and Ti,  are created via $\alpha-$particle  (helium nucleus) capture. Since the abundance of [$\alpha$/Fe] is partially degenerate with GC's age and metallicity, [$\alpha$/Fe] is treated as an independent  free parameter. 
 The ranges in parameter space covered by the \texttt{DSED} model photometry files in \texttt{isochrones} are specified in  Tab.~\ref{tab:Table1}.

The modifications to the original \texttt{isochrones} code  implemented by V20 and V21 
include the use of  equivalent isochrone evolutionary points (EEP)  to improve  the interpolation process\footnote{This method is also implemented in the newer releases of the \texttt{isochrones} package.}, additional interpolation in the [$\alpha$/Fe] parameter, and implementation of several corrections for extinction (see e.g.~\cite{O'Donnell, Fitzpatrick, Cardelli}) in the selected HST filters (here $F_{606W}$ and $F_{814W}$). The mixing length treatment is the same as in V21 as well as the systematic error estimate and propagation.

 \begin{table}
\centering
\begin{threeparttable}
\begin{tabular}{|c|c|}
\hline
Stellar model & DSED \\ \hline\hline
initial rotation rate $v/v_{crit}$ &  0.0 \\ \hline
Age range & 0.250-15 Gyr \\ \hline
Age sampling &  0.5 Gyr \\ \hline
number of EEPs per isochrone & $\simeq$ 270 \\ \hline
Metallicity range {[}Fe/H{]} & -2.5 to 0.5 dex\\ \hline
\begin{tabular}[c]{@{}c@{}}Helium fraction configuration\\ \end{tabular} &  $Y_{\rm init}$ = 0.245\tnote{\dag}, 0.33, 0.40 \tnote{\ddag} \\ \hline
 {[}$\alpha$/Fe{]} & -0.2 to 0.8\tnote{+} \\ \hline
\end{tabular}
\footnotesize
\begin{tablenotes}
\item[\dag] The varying Helium fraction configurations, $Y$, are defined in photometry files as $Y = Y_{\rm init} + 1.5 Z$ where $Z$ is the metal mass fraction and $Y_{\rm init}$ is the starting value.
\item[\ddag] Fixed Helium fraction configurations $Y=0.33 $ and 0.40 are only available for [Fe/H] $\leq$ 0.
\item[+] For the fixed Helium fraction configurations, only two options [$\alpha$/Fe]$=0$ or $+0.4$ are available.
\end{tablenotes}
\end{threeparttable}
\caption{Properties of the  \texttt{DSED} stellar models available in the \texttt{isochrones} package. We refer the reader to  the original Ref.~\cite{dsed} for more details.
}
\label{tab:Table1}
\end{table}
 
The fitted parameters for each GC are: age, distance, metallicity, [$\alpha$/Fe], helium fraction $Y$,  and two parameters for  absorption.

\section{Methodology improvements}
\label{sec:methods}
In this work, we implement the following  improvements over Refs.~\cite{Valcin2020, Valcin2021} (V20 and V21): better external priors on  distances,  more flexible extinction law, independent variation of the Helium fraction (i.e. not tied to the metallicity),  adoption of a Gaussian mixture model better suited to handle multiple populations in the CMD, better masking of spurious features in the CMD and  more sophisticated parameter inference. These are presented in this section, except for the latter, which is described in Sec. 4.

\subsection{Distances}
\label{sec:dist}
Distances evaluated through the distance modulus play an important role in the correction of magnitudes. In the literature, estimates of distances are usually based on Ref.~\cite{Harris} (hereafter Harris). 
However,  if these estimates were not to be perfect, assuming them as the distance prior could potentially result in biased  inference of the other cluster parameters. 
To mitigate  this, in our previous analysis, we use a wide prior around the Harris distance values.
While the distances recovered by our analysis were not dominated by the width of the adopted prior, potential biases introduced by the choice of the central value were not quantified.  

\begin{figure}[ht]
    \centering  \includegraphics[width=0.8\textwidth]{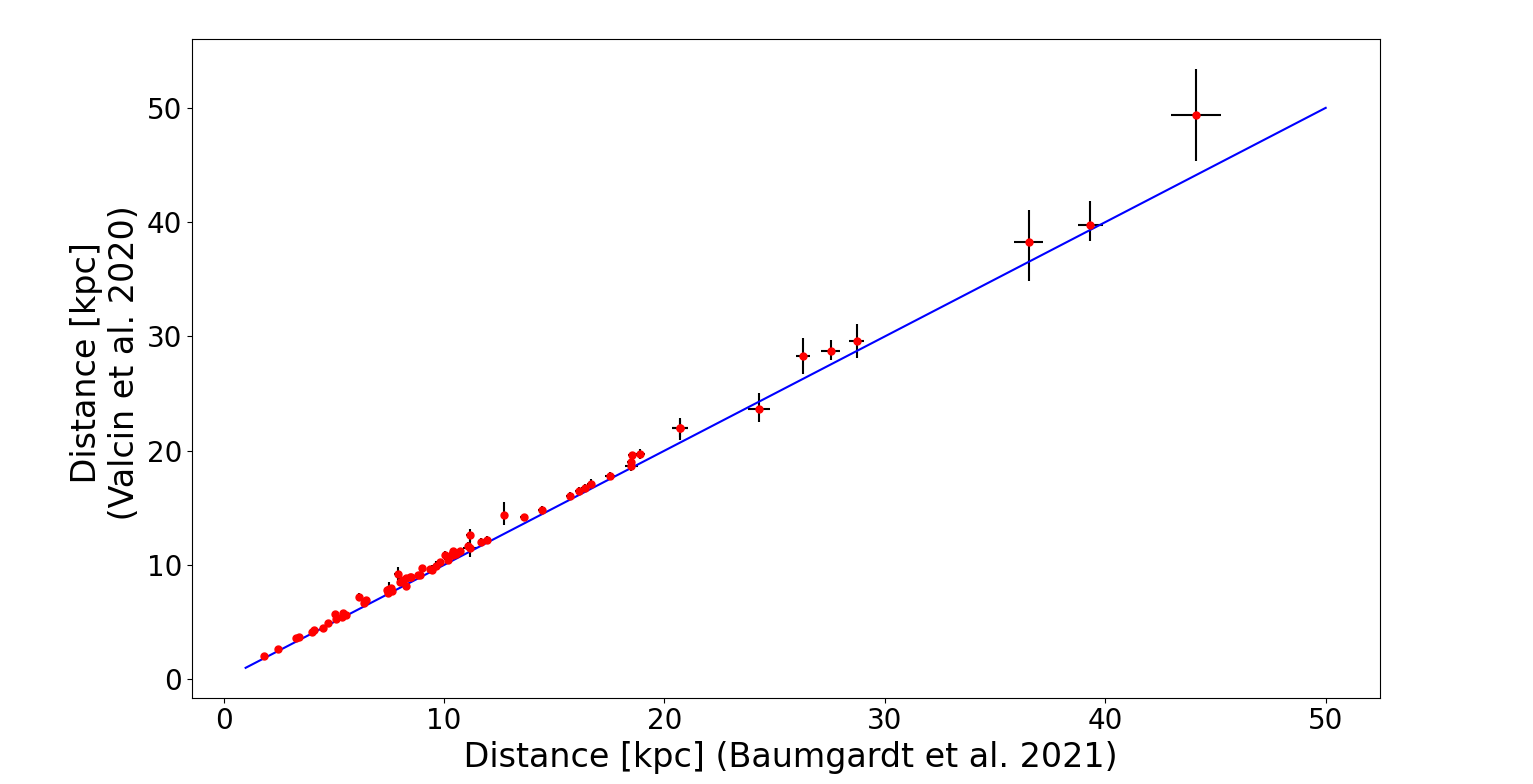}
\includegraphics[width=0.8\textwidth]{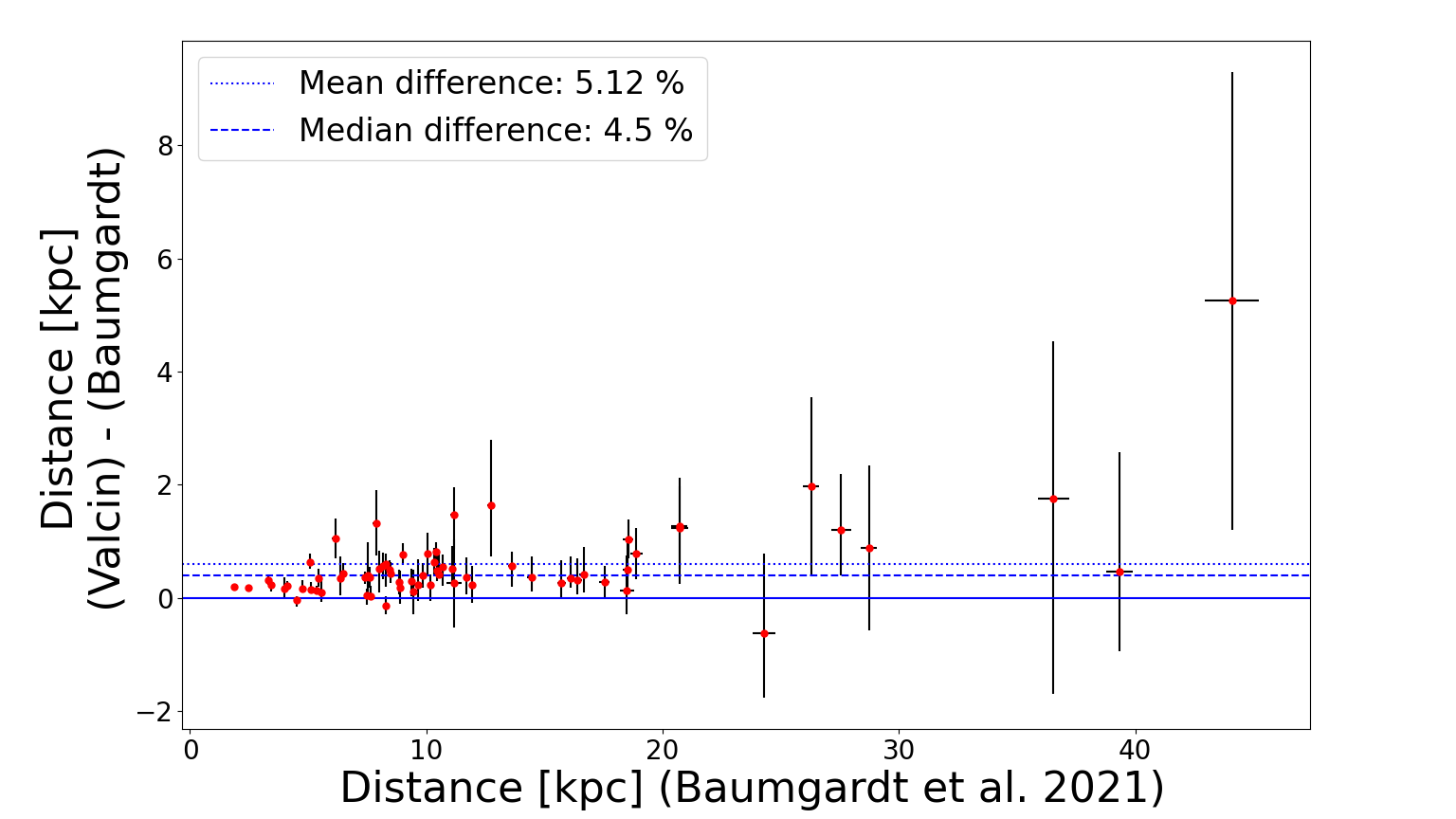}
    \caption{Comparison of distances between V20 (\cite{Valcin2020}) and Ref.~\cite{Baumgardt} for the full sample of GCs. The dashed (dotted) blue lines shows the median(mean) 5.12\% (4.5\%) offset. Here, we chose \cite{Baumgardt} estimates for our distance priors.}
    \label{fig:comp_gaia}
\end{figure}

 In Ref.~\cite{Baumgardt}, combining data from the Gaia Early Data Release 3 (EDR3) with distances based on Hubble Space Telescope (HST) data and literature-based distances,  derived distances to Galactic globular clusters and report their estimated errors, both statistical and potential systematics. In particular, \cite{Baumgardt} discuss that their distance estimates may be 2\% underestimated. 
In Fig.~\ref{fig:comp_gaia} we plot a comparison of their distance estimates with those obtained in V20. While the scatter broadly agrees with the errors,  V20 distances are $\sim$ 5\% larger than \cite{Baumgardt}.  The origin of this offset, larger than the 2\% possible systematic estimated by \cite{Baumgardt} is unclear. Here, departing from V20, we adopt the values from Ref.~\cite{Baumgardt} as our initial guess and prior on the distances. We will return to the impact of this choice, different from that in V20, in sec.~\ref{sec:V20V25}.

\subsection{Absorption and RV correction: an extra free parameter}
\label{sec:abs}
\begin{figure}[ht]
    \centering
    \includegraphics[width=0.9\textwidth]{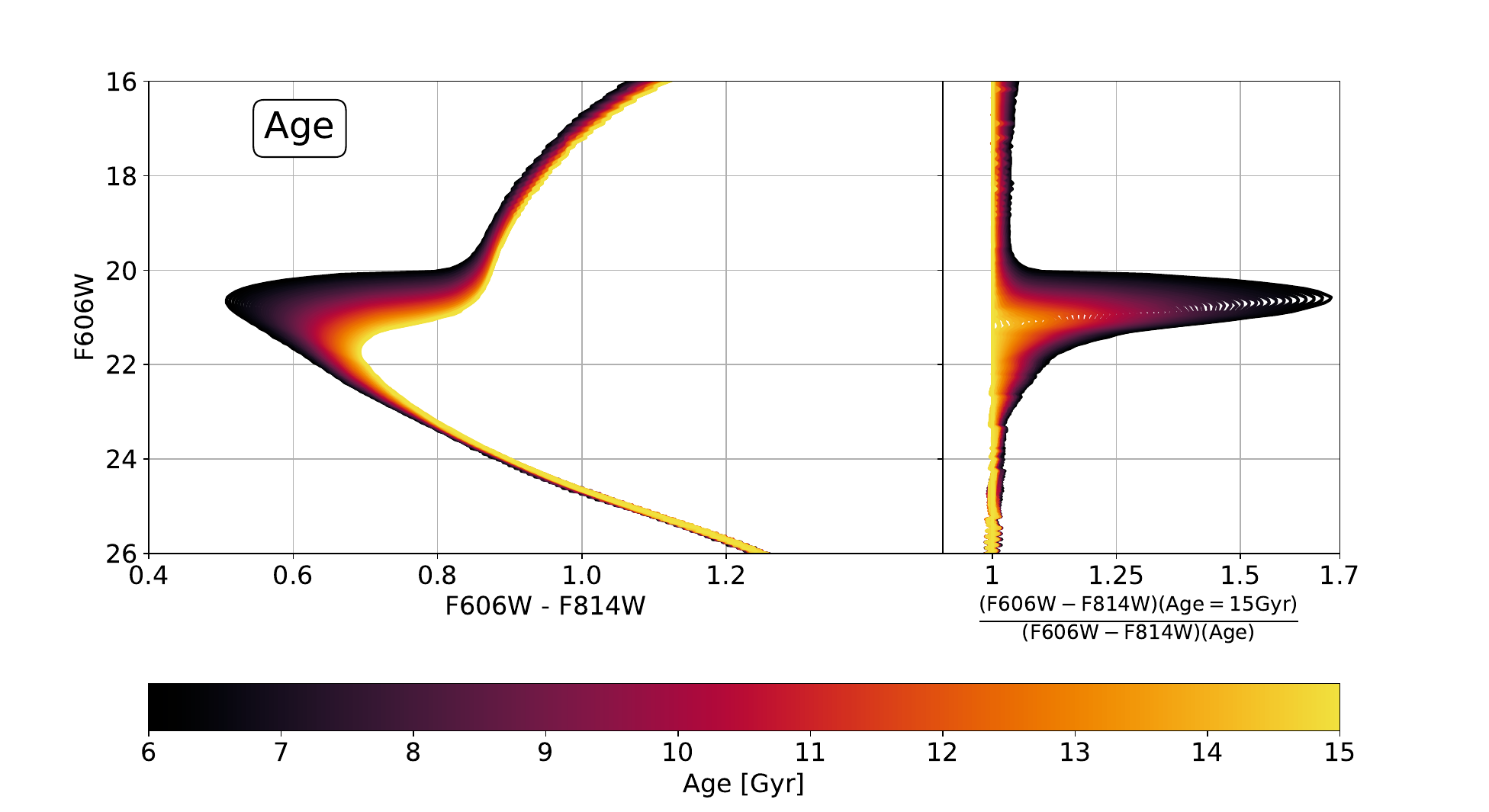}
    \includegraphics[width=0.9\textwidth]{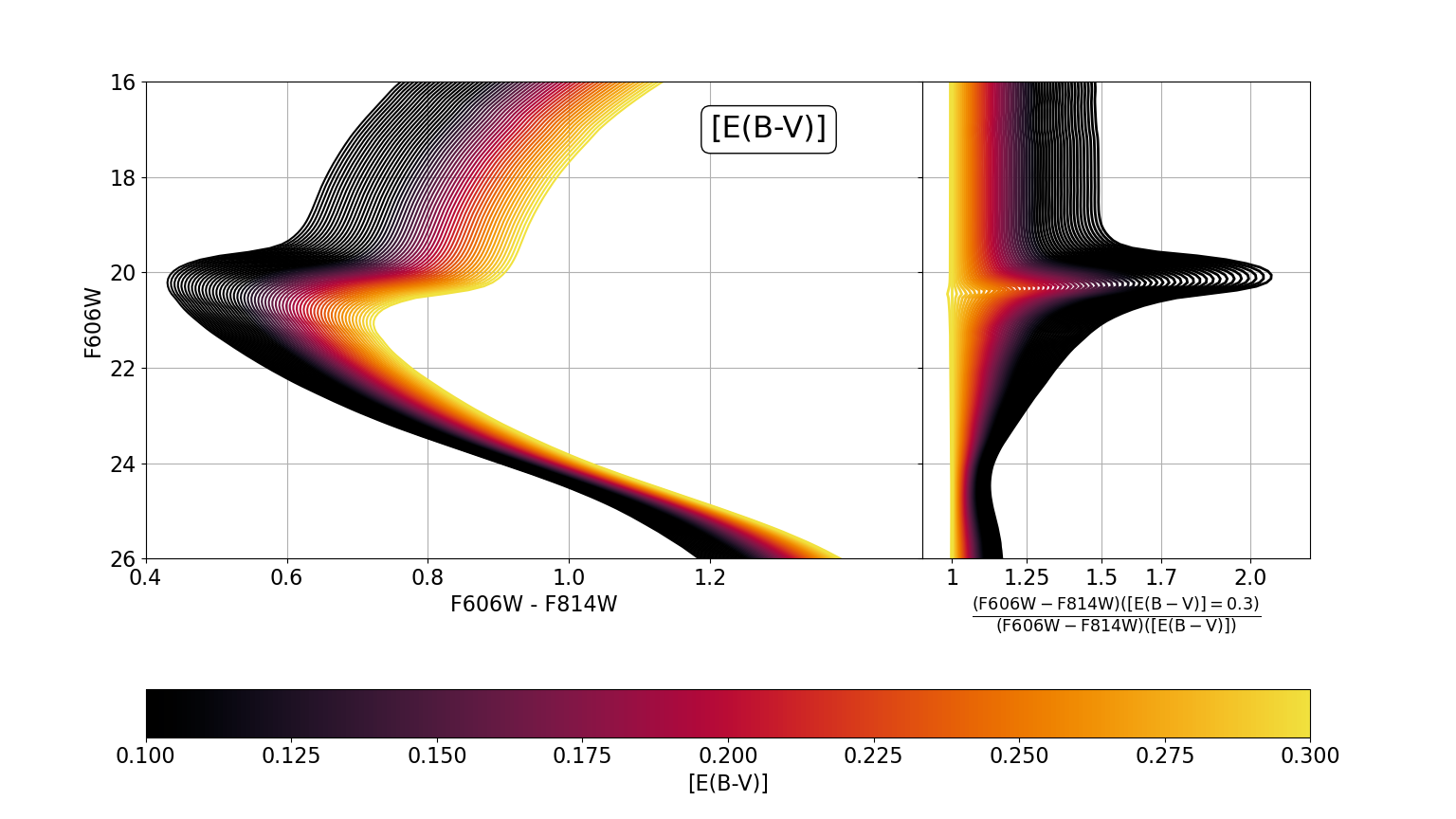}
    \caption{Theoretical curves showing the main sensitivity for age (top) and absorption (bottom). Note that without the full CMD absorption and MSTO luminosity are fully degenerate. GC number is ic4499. E(B-V) estimate value is 0.23.}
    \label{fig:deg_age_abs}
\end{figure}

The absorption correction plays a major role in the inference of the age of every cluster because of the high  absorption-age degeneracy, second only to the age degeneracy with metallicity. In Figure \ref{fig:deg_age_abs} we show the isochrone variations in color as a function of age (top panel) and absorption (bottom panel). The age sensitivity, as it is well known,  is mainly concentrated around the turn off point of the main sequence, which also happens to be a region highly sensitive to changes in absorption. This is why a careful treatment of dust is important.

In  V20,  available extinction coefficients from the literature e.g., \cite{Harris, Dutra} were related  to absorption using the Cardelli \cite{Cardelli} law with constant extinction ratio $R_V=3.1$. 
The standard correction factor\footnote{Recall that  absorption  $A(V)$ and extinction or reddening $E(B-V)$ are related by $A(V)=R_V E(B-V)$} ($R_V = 3.1$), designed to create a uniform extinction law, does not fully encompass the complexity and diversity of galactic environments~\cite{Casagrande,Mathis, Legnardi}.

\begin{figure}
    \centering
    \includegraphics[width=0.9\textwidth]{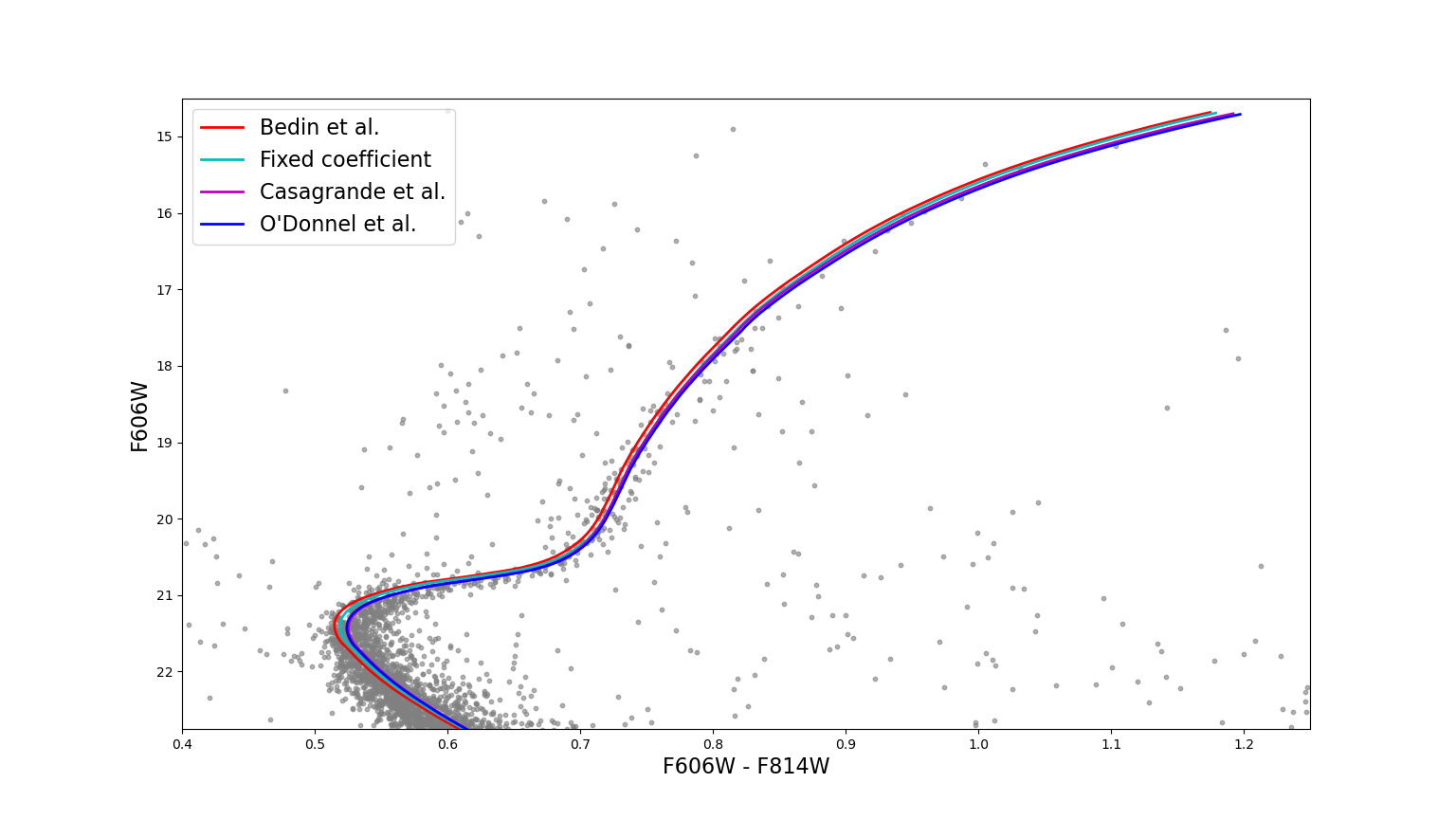}
    \includegraphics[width=0.9\textwidth]{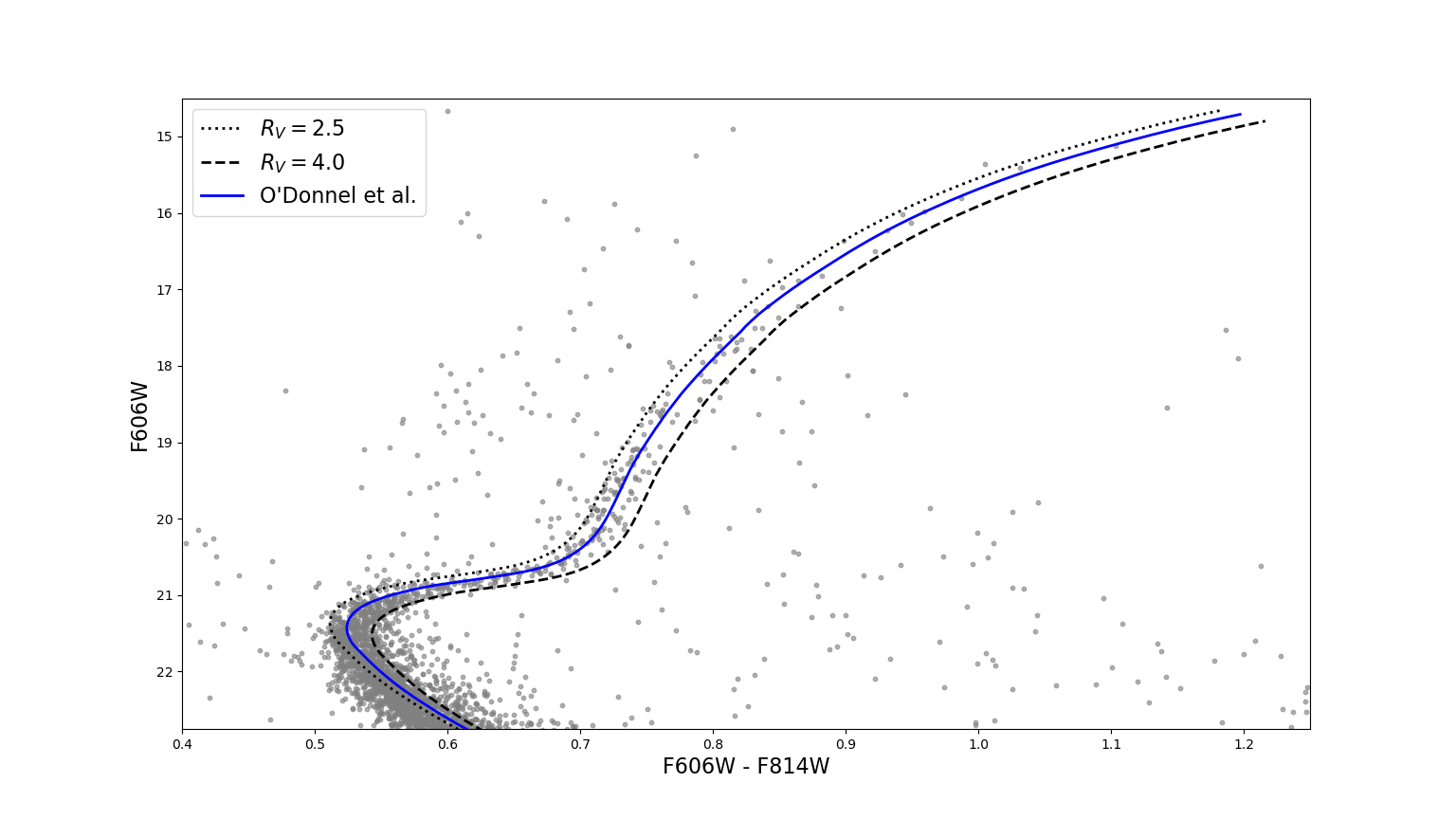}
    \caption{Comparison between isochrones where absorption is converted from V band to HST filters either using a fixed coefficient or a fitting formula. The dots correspond to photometry from the GC Arp 2.}
    \label{fig:fitting_formula}
\end{figure}

There are several formulas available in the literature (Bedin \cite{Bedin}, Cardelli \cite{Cardelli}, Fitzpatrick \cite{Fitzpatrick}, O'Donnell \cite{O'Donnell}, Casagrande \cite{Casagrande}, etc.) that provide a correction factor based on the wavelength. 
In the left panel of Figure \ref{fig:fitting_formula}, we compare a few examples of isochrones corrected with different extinction law with the one from V20 (labeled as "fixed coefficient").

Here we adopt the O'Donnell formula \cite{O'Donnell} for the correction, as the dependence on the $R_V$ parameter is explicit, so this parameter can be varied in the analysis in a simple and transparent way. We thus promote $R_V$ to a free parameter which corrects for possible deviations from universality of the extinction law. Hence, in our analysis we now simultaneously fit the extinction E(B-V) and the new parameter  $R_V$. 
 
The right panel of Figure \ref{fig:fitting_formula} shows 3 isochrones corrected with O'Donnell law but with different value of $R_V$. The standard value of 3.1 is shown in blue, $R_V = 2.5$ in dotted black and $R_V = 4.0$ in dash black. As we can see, the effect on the color is very similar to changes in extinction and a high correlation between E(B-V) and $R_V$ is expected.

\subsection{The helium fraction: an additional free parameter.}
\label{sec:helium}

The helium fraction $Y$ is highly degenerate with metallicity and  [$\alpha$/Fe] abundance. Fixing it to an incorrect value might bias the inference of metallicity and therefore age.
A  vast literature agrees that the helium fraction is one of the most important parameters to highlight the presence of multiple populations in globular clusters, along with CNO abundance and s-processes (see excellent reviews~\cite{Bastian_Lardo} and~\cite{Milone_Marino}).
Most GCs have multiple populations, where the younger populations are helium enhanced.
While there are techniques to disentangle the different populations, these are not applicable here because we work only with two filters. 
When fixing $Y$ at a value close to the primordial one, to accommodate the helium enhancement of multiple populations in the fitting process,   the recovered  metallicity is biased low, resulting in higher recovered ages.
However,  the effect of the presence of multiple populations, can be absorbed by increasing $Y$; hence 
 adopting  $Y$ as a free parameter in the analysis reduces the  sensitivity to  multiple populations \cite{Bastian_Lardo, Milone_Marino, Wagner-Kaiser, Stenning}.

\begin{figure}
    \centering
    \includegraphics[width=\textwidth]{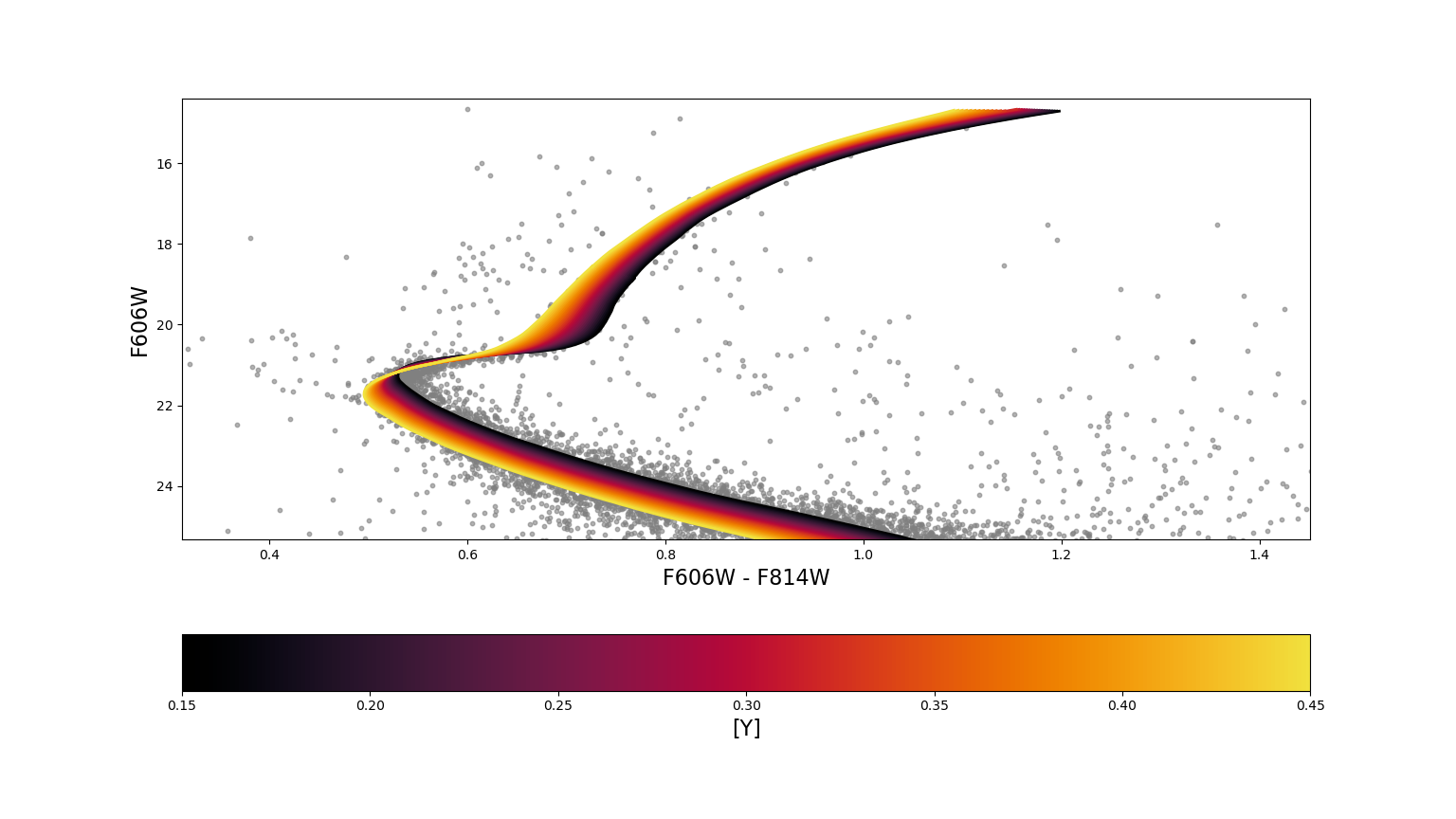}
    \caption{Boundaries of helium uniform prior versus main sequence scatter. The dots correspond to photometry from the GC Arp 2.}
    \label{fig:helium_spread}
\end{figure}

In  V21, the helium dependence on metallicity  was implicitly encoded in the isochrones used:  the helium fraction $Y$ was defined to be proportional to the metallicity ($Y = Y_{\rm init} + 1.5 Z$ where $Z$ is the metal mass fraction and $Y_{\rm init}$ is the initial value). As presented in Table \ref{tab:Table1}, the Dartmouth stellar model also offers isochrones with two fixed initial helium values ($Y_{\rm init} = 0.33$ and $Y_{\rm init} = 0.4$), making it possible to interpolate over a helium range.  In Figure \ref{fig:helium_spread}, we show theoretical isochrones interpolated from $Y=0.15$ to $Y=0.45$\footnote{The Helium range reported here is slightly wider than the prior adopted  in the analysis (see Tab. \ref{tab:priors}). This is to better control the behavior of the interpolation.} 
The highest color sensitivity to helium is located at the bottom of the red giant branch and along the main sequence; helium affects the main sequence turn off (which is very sensitive to age) minimally, indicating that the two parameters should not be degenerate.  However, we expect a degeneracy with absorption (see Figure \ref{fig:deg_age_abs}).

\subsection{Data cleaning, selection and compression}
\label{sec:Dataset selection}
The most notable improvement compared to previous work is a much more sophisticated data compression than the combination of adaptive binning  and cuts of V20.
Data-compression speeds up the analysis significantly. In addition, it makes it easier to compare the data to the model equivalent evolutionary points (EEPs,  see  V20 for more details).

Note that from  the two available filters (F606W and F814W), two color magnitude diagram can be made, one using  F606W magnitudes and one using F814W. Here we perform the analysis on both to improve statistics and  control the stability of the results. 

\subsubsection{Photometry cleaning and Reddening correction} V20 \cite{Valcin2020} used the same prescriptions as Wagner-Kaiser et al. \cite{Wagner-Kaiser} to clean the data of stars with poorly determined photometry. Here, we opt instead for  the methodology of  Milone et al. \cite{Milone2012} for a more streamlined treatment of  the reddening correction. For a detailed explanation of the cleaning and the reddening correction, we refer the reader to the original paper~\cite{Milone2012} which we follow closely. Here, we will briefly summarize the procedures in a few key points:
\begin{itemize}
    \item  after binning stars in bins of 0.4 magnitudes along the reddening line, apply a 4-sigma clipping for the $rms$ of x and y positions  and for PSF fit residuals.
    
    \item Remove stars for which the  total flux in a 0.5 arcsec aperture is dominated by  neighboring stars. 
    
    \item  Remove stars for which photometric errors in both filters fall into the outer 5\% tail  of both filters.
\end{itemize}
This defines the ``well measured" sample.  We then follow the reddening correction  steps detailed in \cite{Milone2012}. The correction is applied to each star individually, and the procedure is done iteratively until it converges.
The two top panels of Figure \ref{fig:data_sel} illustrate the results of photometry cleaning and reddening steps for NGC2298, a cluster with high enough extinction to highlight the effect of the reddening correction.

\subsubsection{Data cuts}

\label{sec:Subsampling}
\paragraph{Magnitude cut:}
We expect the quality of the measurements to be less robust at very low magnitudes. This motivated V20 to define a ``functional'' magnitude interval between the lowest apparent magnitude of the brightest stars and a magnitude cut arbitrarily defined at $m_{cut} = m_{\rm MSTOP}+ 5$ in the F606W filter, where $m_{\rm MSTOP}$ refers to the magnitude corresponding to the Main Sequence Turn off Point. 
Here, considering the increased number of free parameters in the fit,  we move the magnitude cut to slightly lower magnitudes $m_{cut} = m_{\rm MSTOP}+ 7$ in F606W and to the nearest corresponding EEP point for F814W to help break degeneracies.

\begin{figure}
\hspace*{-0.7cm}
    \includegraphics[width=1\textwidth]{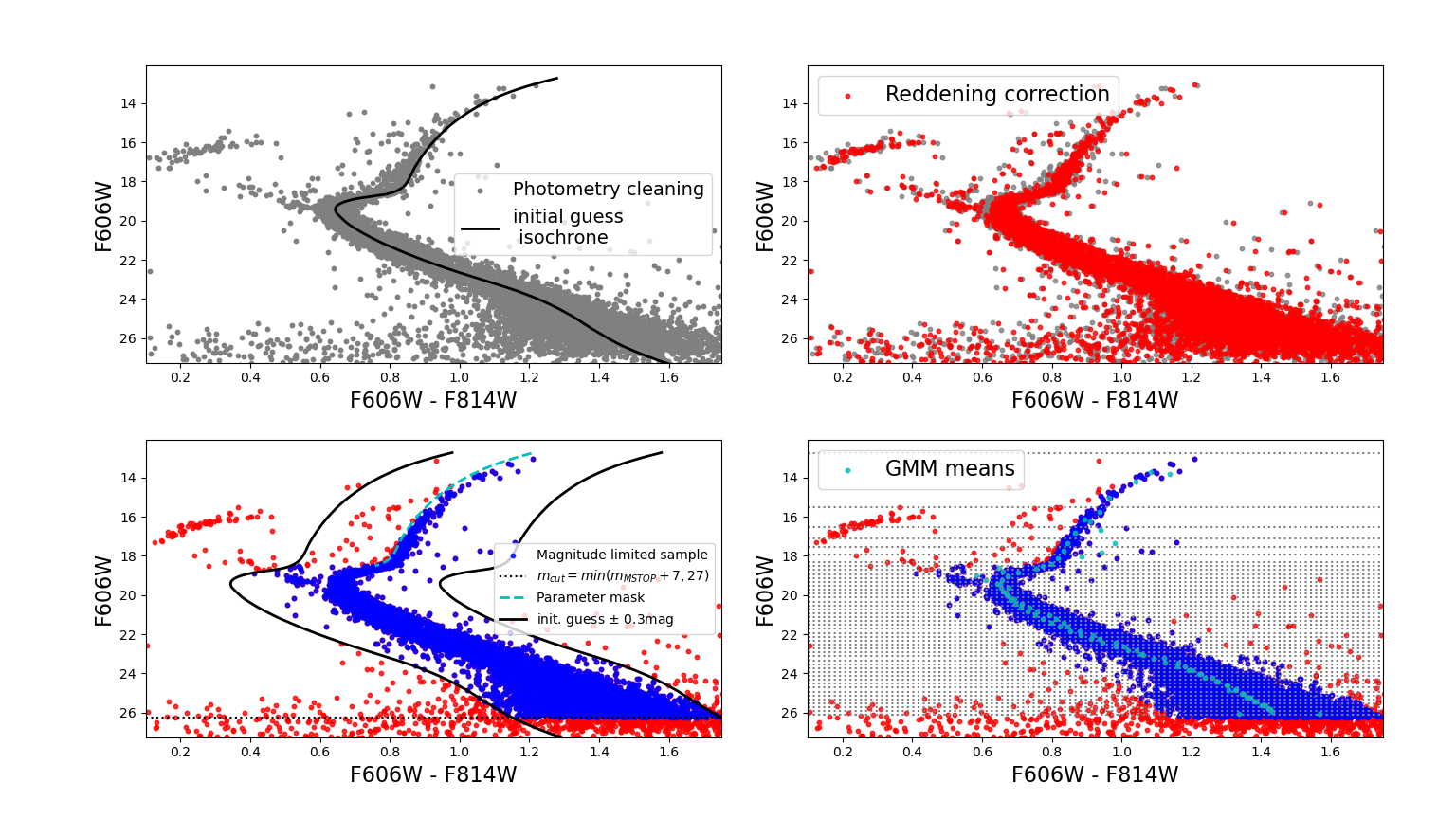}
    \caption{Illustration of the different steps of  data  cleaning, and compression for  NGC2298 as a  representative GC  example.} 
    \label{fig:data_sel}
\end{figure}

Following similar considerations as in V20,  a further cut $m_{cut} = 27$ is enforced to exclude noisy data points.  
Thus, our magnitude cut is as follows:\footnote{c.f. V20 adopted $m_{cut} = min(m_{\rm MSTOP}+ 5 , 26)$.}
\begin{equation}
    m_{cut} = min(m_{\rm MSTOP}+ 7 , 27).
\end{equation}
this is applied in the  F606W filter and to the nearest corresponding  EEP point for F814W.
Our findings are not sensitive to the details of this choice as long as the noisy, dim part of the color-magnitude diagram is removed, and enough
EEPs in the main sequence are retained.
The magnitude cut is illustrated in the bottom left panel of Figure \ref{fig:data_sel}.

\paragraph{Color cut:}
The isochrones of the DSED stellar model only consist of main sequence and RGB stars. To isolate these regions of the CMD we perform color cuts in the following way.  
First, we compute a theoretical isochrone based on estimates from the literature and use that as an initial guess (black line in the top left panel of \ref{fig:data_sel}). Then the cut is defined as a color band ($c_{cut}$) centered around the initial guess with a width of 0.3 magnitude, chosen to safely include the width of the main sequence for all the GCs in our sample.
Thus, 
\begin{equation}
    c_{cut} = c_{initial} \pm 0.3,
    \label{eq:colorcut}
\end{equation}
where $c_{initial}$ refers to the color of the initial guess isochrone, as illustrated in the bottom left panel of Figure \ref{fig:data_sel} as solid black lines.

\paragraph{Parameter mask:}
The color cut removes most of the stars belonging to the horizontal branch, white dwarfs and some of the blue stragglers. However, we can see in Figure \ref{fig:data_sel} that some of the AGB stars are still present. AGB stars are not included in the stellar models  but, due to their proximity to the RGB, their presence alters (biases and worsen the goodness of fit) the parameter inference. 

A possible solution would be to use very narrow priors or even to fix some of the parameters.  
We prefer to proceed by creating what we call a ``parameter mask".

The idea behind it is simple: if few points in the CMD  force the recovered parameters to be outside a reasonable prior range, these should be excluded.
A detailed explanation of the mask is provided in Appendix \ref{app:Parameter mask}. The mask creates a color band around the initial guess isochrone (see black lines in Figure \ref{fig:param_mask}). Since we only want to  exclude  AGB stars that are typically bluer than RGB stars,  the mask is applied only in the blue direction.

\paragraph{Influence of the initial guess:}
Some of the choices made to calibrate the color cut, and the parameter mask depend on  an initial guess,  which choice is based on estimates from the literature. 
We have empirically verified that even if the isochrone chosen for the initial guess is a  bad fit to the data,  the cuts still include the majority of the main sequence and RGB stars and do not affect the final parameter inference.
As an example, the case of NGC 2298 is shown in Figure \ref{fig:data_sel} and \ref{fig:param_mask}. Its initial isochrone is not a good fit, but an offset of 0.3 mag in color (see eq.~\ref{eq:colorcut}) combined with the parameter mask is enough to include most of the stars of interest in our sample.

\subsubsection{Gaussian mixture model}
\label{sec:Gaussian_mixture_model}
The key step in our data compression is the transformation of our star sample into a group of representative points that correspond to the EEP's for the stellar model. As explained in V20, the isochrone profile in the color-magnitude diagram is sampled by EEPs (which are ``universal'' across different isochrones), which can be obtained for different adopted values of the parameters of interest. 
Indeed, interpolating the stellar models in the EEP space allows us to remove the mass from the fitting parameters, thus allowing a massive reduction in  parameter dimensionality (from $O(10^2 - 10^4)$ to  to  $O(10)$). 

We assume that the scatter in color of stars follows a Gaussian distribution centered on the  (true) underlying isochrone. So in the presence of multiple populations, there will be as many  underlying isochrones as populations.
Stars are then binned in magnitude bins of default size of 0.2 mag when the number of stars per bin is greater than $n_{\rm stars} = 30$. \footnote{This is usually the minimum sample size quoted for the central limit theorem to hold.}  
 If a default bin doesn't have enough stars, which is usually the case for the brightest stars, we increase its size in 0.2 mag increments until the $n_{\rm stars}$ limit is satisfied. 
 The binning for NGC2298 is shown in the bottom right panel of Figure \ref{fig:data_sel} as gray dotted lines.
  
V20 discarded bins where the color distribution could not be fit by a unimodal Gaussian distribution as it could be an  indication of the presence of multiple populations.
Here we adopt a Gaussian mixture model (GMM)  which  estimates the probability of each data point to belong to a Gaussian distribution,  accounting  for the effect of multiple populations. We adopt the GMM algorithm distributed with the scikit-learn (\texttt{sklearn}) python library. 
Our choices for the set up  of the Gaussian mixture model are reported in Table~\ref{tab:gmm}.

 \begin{table}
\centering
\begin{threeparttable}
\resizebox{\textwidth}{!}{%
\begin{tabular}{|c|c|c|}
\hline
Parameters & Calibration & Motivation \\ \hline\hline
covariance\_type & full & Each component has its own cov. matrix. 
More precise \\ \hline
fit dimension & 2d & 
Also gains information from the magnitude distribution 
\\ \hline
random\_state & fixed to 42& For reproducibility \\ \hline
max\_iteration & 10000 & Enough to ensure convergence \\ \hline
\# initializations & n\_init=30 & Central limit theorem. Enough for convergence \\ \hline
Convergence threshold. 
& tol=$1e^{-3}$ & default value \\ \hline
Initialize weights, means \& precision & init\_params = 'random' &  
Performs better when means overlap
\\ \hline
\end{tabular}%
}%
\end{threeparttable}
\caption{Set up of the  Gaussian Mixture model available in the \texttt{sklearn} package.}
\label{tab:gmm}
\end{table}
In addition to the parameters of  Table~\ref{tab:gmm}, the code requires to input a fixed number of  Gaussian components for the fit, which can be motivated by the physics of the problem. In our case, there is no definite way to know the number of stellar populations within a cluster. Instead of resorting to previous estimates in the literature, we proceed as follows.

\paragraph{Optimal number of Gaussian components}
We limit the maximum number of components to 5. 
The cluster with the highest number of stellar populations, Omega Centauri, is thought to be the host of 17 of them. However, for our approach (only two filters and variable $Y$) using more than 5 components only makes the numerical analysis more complex, unstable and time-consuming. 
The number of components is then varied  as a hyperparameter and chosen by balancing the  improvement in goodness of fit vs overfitting.  The adopted package provides several quantities to do so: the  Akaike information criterion (AIC),   the Bayesian information criterion (BIC) and  the ``Score" which is  the per-sample average log-likelihood. In practice,  rather than selecting the number of component by minimization (AIC, BIC) /maximization (Score) of one of these quantities, it is more robust  to use the biggest inflection point.
We use the \texttt{kneed}\footnote{https://github.com/arvkevi/kneed} python package, which implements the kneedle algorithm to determine the knee point as a function of the convexity and direction. In well-behaved bins -- those that are well populated-- all three estimates agree. When they disagree, we adopt the following  hierarchy to determine the optimal number of components. We give priority to the ``Score", if \texttt{kneed} can't find the inflection point  (this happens in some poorly populated  bins in the RGB, for a total of $\sim$ 10\% of the bins) we move to the BIC, if that also fails (below 5\% of the times) use the AIC. Only  in the few cases (0.1 \% of the cases) where  \texttt{kneed} can't find the inflection in any of the above, we resort to use the extreme value (maximization/minimization) for AIC.
This order is motivated by the fact that   BIC is less susceptible to overfitting than AIC, but the Score has a more direct Bayesian interpretation.

\begin{figure}
    \centering    \includegraphics[width=\textwidth]{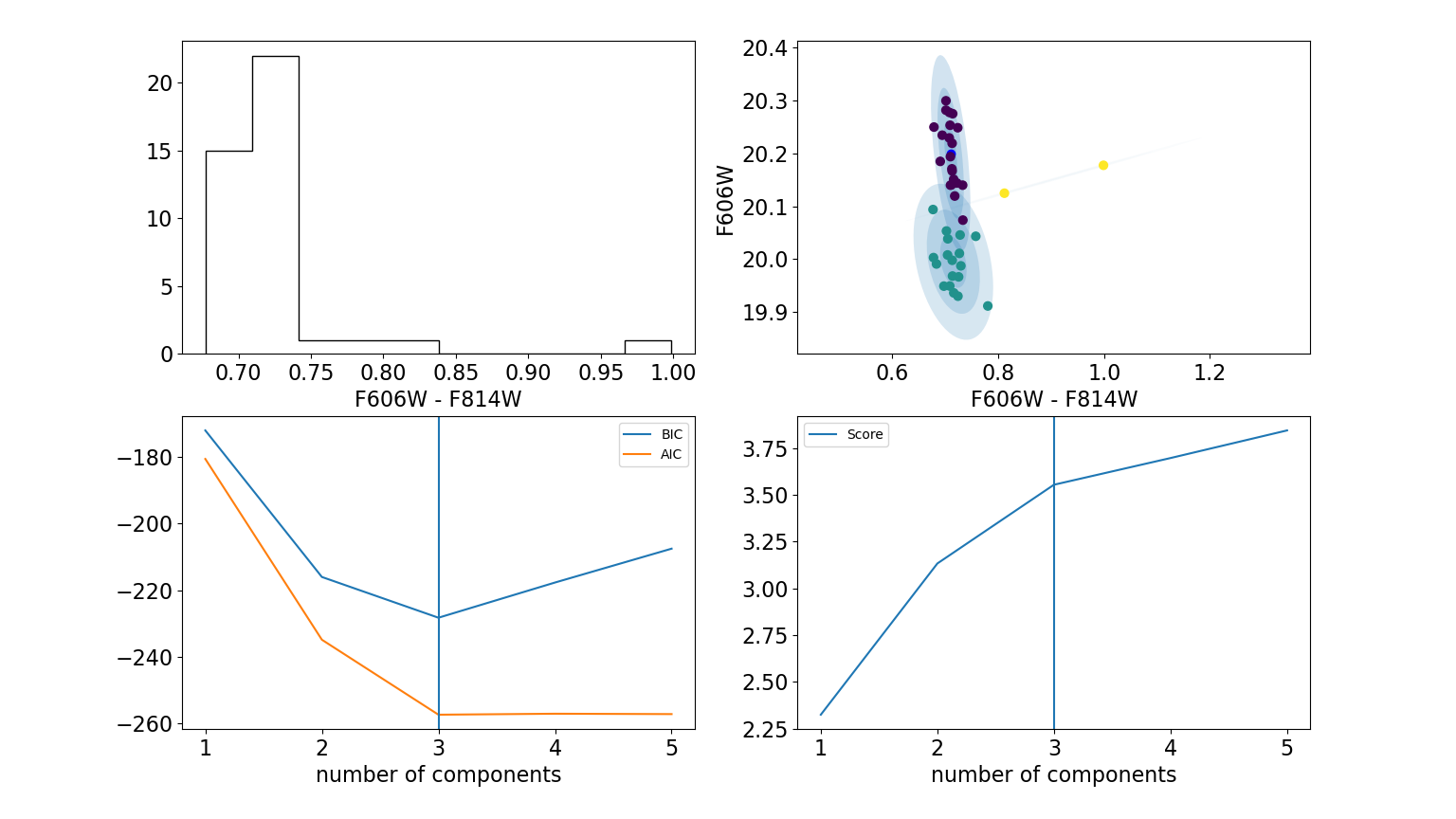}
    \caption{Illustration of the steps in the GMM pipeline. Top left: the profile of a representative bin in the CMD of GC Arp2. Top right, the corresponding points on the CMD color coded based on the Gaussian component they belong to. The three Gaussian component estimated by the GGM are also shown. Bottom panels: the Score, BIC and AIC as a function of the number of components. In this  case, all the estimates prefer three Gaussian components.}
    \label{fig:gmm_calib}
\end{figure}

We show an example of the choice of components in Figure \ref{fig:gmm_calib} for a given bin. In the top row we show the color-magnitude distribution  of the stars within the bin (right) and the associated color histogram (left) along with the three gaussian components. In the bottom row, the optimal number of components (3) found with \texttt{kneed} package for the score (right panel) and AIC/BIC (left panel). This is a well-behaved bin, as all three estimates agree.

\paragraph{Additional considerations:  singularities, filter dependence and interpretation of GMM fit}

The GMM fit can sometimes return singularities: when a single star is identified as a component\footnote{It happens because the GMM algorithm need to use all the data points within the sample.}, that component has associated an  infinite variance. 
The minimum sample size per bin $n_{\rm stars}$ ensures 
that such singularities are very rare. 
They can be easily removed from the dataset, but we choose to keep them, as they are accounted for in our nuisance likelihood introduced in the next section. 

We apply the pipeline described above to both the  ``F606W - F606W-F814W" and ``F814W - F606W-F814W" color magnitude diagrams. The binnings are slightly different in the two cases, and so  are the errors. Unsurprisingly, the GMM shows some sensitivity to the choice of the CMD. 
To mitigate this, we fit both CMD simultaneously.

\begin{figure}
    \includegraphics[width=\textwidth]{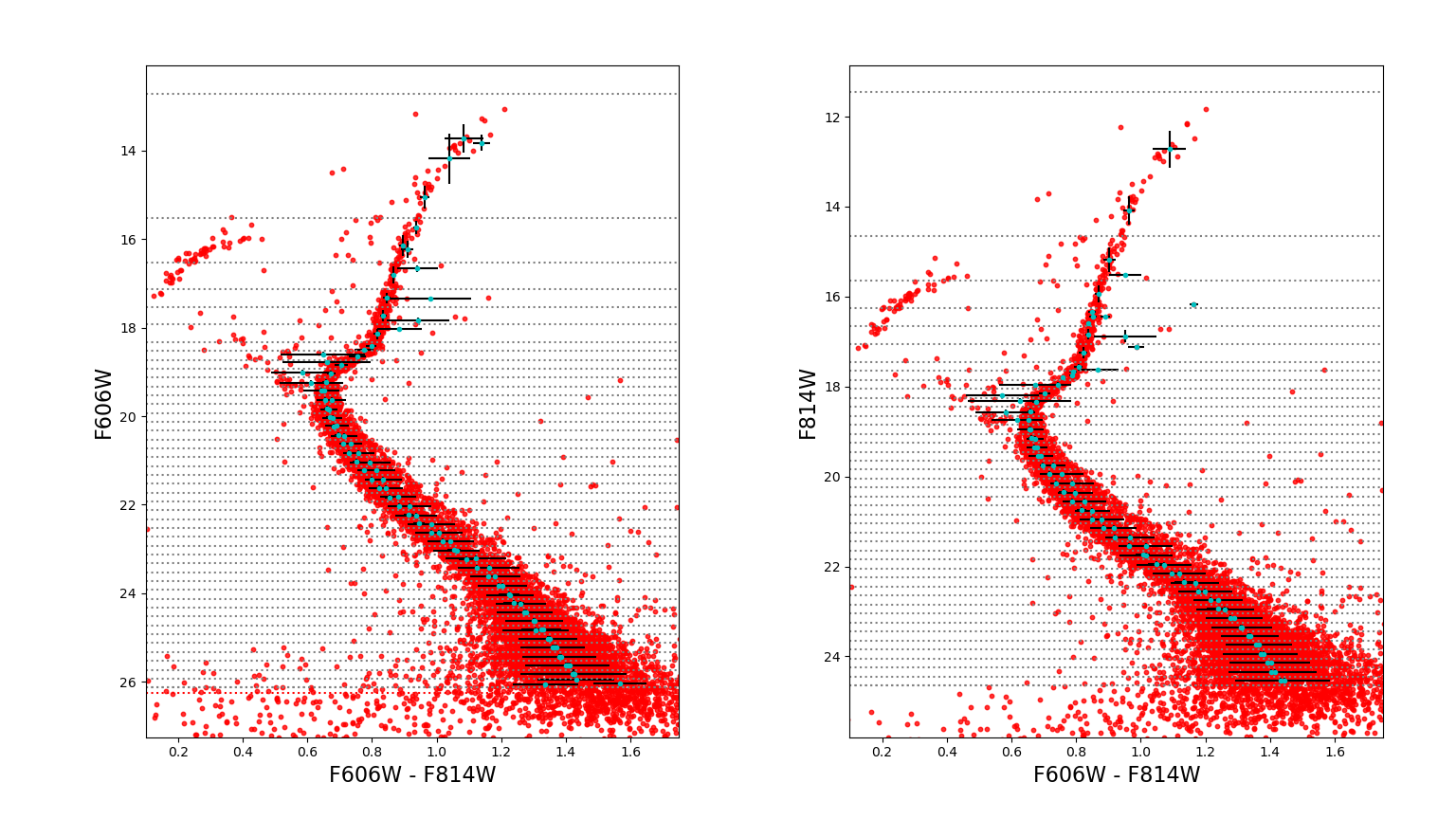}
    \caption{Influence of the choice of CMD on the Gaussian mixture model fit. The red points are the  reddening corrected stellar sample, gray dotted lines indicate the binning and cyan points with error bars denote the bin mean and errors recovered by the Gaussian Mixture Model.} 
    \label{fig:data_sel2}
\end{figure}
Figure \ref{fig:data_sel2} highlights the differences between the two CMD. Even though the two datasets are highly redundant, the means, errors and number of components per bin  are different in the two cases.\\

Finally, we emphasize that the GMM is not used as a mean to estimate true underlying distribution and stellar populations, but rather as a flexible density estimator in the CMD which allows us  to accurately represent the stellar sample while considerably decreasing the number of  parameters in the fit.

\section{Parameter inference}
\label{sec:inference}
\subsection{Mixture likelihood}

The sequence of points in the CMD produced by the GMM can be directly compared to the model's output based on EEP interpolation.
 Our dataset  $\mathcal{D}$ is then the sequence points in the CMD produced by the GMM concatenated for the two CMD $\mathcal{D}_{\rm F606W}, \mathcal{D}_{\rm F814W}$. 
We use the (simplified) mixture likelihood of Stennings et al. \cite{Stenning} already marginalized over the cluster membership probability distribution:

\begin{equation}
    L(\theta|{\mathcal{D}}, \Sigma) = \prod_{i=1}^N [Q \times P(\mathcal{D}_i|\Sigma_i,\theta_i, {\cal Z}_i = 1) + (1-Q) \times P(\mathcal{D}_i| {\cal Z}_i = 0)],
\label{eq:likelihood}
\end{equation}
where $\mathcal{D}$ denotes the dataset, $\theta$  the model's parameters defining the EEPs, $N$  the number of elements in the dataset, ${\cal Z}$ is the cluster membership (${\cal Z}=1$ means member, ${\cal Z}=0$ not member). Here $P$ denotes the Gaussian probability distribution.   
$Q$ is a constant resulting from  marginalization over cluster membership. While  after some tests, Ref.~\cite{Stenning} opted for a fixed  value of $Q = 0.95$,  we leave it as a free parameter in our analysis.

Our pipeline is designed to work in color-magnitude space, while 
the likelihood in Eq.~\ref{eq:likelihood} was originally written for the magnitudes in $n$ filters. Since colors are linear combinations of magnitudes, we can apply Eq.~\ref{eq:likelihood} as long as correlations are correctly accounted for.
Although in Ref.~\cite{Stenning} the covariance matrix is  diagonal, here  the GMM provides a full covariance accounting for the cross-correlation between the two filters. 

\subsection{Priors}

\begin{table}
\centering
\begin{tabular}{|c|c|c|}
\hline
Parameter & Hard boundaries  & Gaussian prior: \\
per GC & of uniform prior & central value,  width \\\hline\hline
Age & 0-15 Gyr & --  \\ \hline
{[}Fe/H{]} & -2.5-0.5 dex &Ref.\cite{Harris}, $\sigma =0.2$ dex \\ \hline
Distance & 0-$\infty$& Ref.~\cite{Baumgardt}\\ \hline
[$\alpha$/Fe] & 0-0.4  & Ref.~\cite{Recio-Blanco}, $\sigma=0.1$ \\ \hline
Extinction & 0-3 & Tab.~\ref{tab:input-extinction}, $\sigma_{E(B-V)}$=0.02 \\ \hline
$R_V$ & 1.5-5 & -- \\ \hline
Y & 0.2-0.4  & --\\ \hline
Q & 0-1  & --\\ \hline
 \end{tabular}
\caption{Adopted priors on the model's parameters. There are uniform priors with hard boundaries for all parameters.  Further, for some parameters additional Gaussian priors are included, based on published literature, as reported. The first five parameters are the parameters of interest, but note that $A_V$ is a derived parameter.}
\label{tab:priors}
\end{table}

The parameters  to be constrained by the data are: age, metallicity [Fe/H], distance, alpha enhancement [$\alpha$/Fe], extinction E(B-V), extinction correction $R_V$,  and Helium fraction $Y$. $Q$ is treated as a nuisance parameter and marginalized over. Note that the parameters of interest are:  age, metallicity [Fe/H], distance, alpha enhancement [$\alpha$/Fe] and absorption $A_V$, the latter being a derived parameter from  E(B-V) and $R_V$.
To ensure that we remain inside the interpolation domain of the stellar model, we use uniform priors as follows:
[1,15] Gyr for age, [-2.5,0.5] dex for metallicity, [0,3] for extinction, [0,$\infty$] for distance,  [0.2,0.4] for Helium  abundance $Y$, [0, 0.4] for [$\alpha$/Fe].\footnote{Although the maximum range for [$\alpha$/Fe] is [-0.2,0.8], the isochrones with fixed Helium fraction are only available for [$\alpha$/Fe] = 0 and 0.4. We then use a uniform prior [0, 0.4] to limit the interpolation range.}  For the extinction correction $R_V$, we use a very conservative prior [1.5, 5.0] because there is no real prior information available.   
The cluster membership parameter ${\cal Z}$ takes  discrete values of either 0 or 1. Its marginalized counterpart $Q$ can take any values in the continuous interval [0,1].

 As mentioned above, the uniform priors are chosen to ensure that the sampling of the parameter space stays within reasonable boundaries. However, most of them are very wide, making it more difficult for the MCMC chains to converge and break the degeneracies. This is why we include additional Gaussian priors to some of the free parameters. We adopt $\sigma_{\rm [Fe/H]} = 0.2$ dex for the width of the Gaussian priors for the metallicity, based on spectroscopic measurements \cite{Harris},  corresponding to twice the typical errors reported in Ref.~\cite{Bolte+}).\footnote{In principle, this prior could be more stringent,  following  Ref.~\cite{Carretta}. However, we decide not to do this here and explore a wider range in metallicity.} The central values and width adopted for the distance prior is taken from Ref.~\cite{Baumgardt} database\footnote{https://people.smp.uq.edu.au/HolgerBaumgardt/globular/}. For  [$\alpha$/Fe] we adopt a prior centered on the values of \cite{Recio-Blanco} with a width of $\sigma_{\alpha} = 0.1$ which is equivalent to half the sampling step of the DSED stellar grid. 
We assume central values for the reddening as reported in Tab.~\ref{tab:input-extinction} and a dispersion on the reddening $\sigma_{E(B-V)} = 0.02$, in agreement with Ref.~\cite{OMalley}, which translates into Gaussian priors on absorption with $\sigma_{\rm abs} = 0.06$ following the Cardelli \cite{Cardelli} relation.
This is summarized in table~\ref{tab:priors}.
Unlike the priors on metallicity or distance, which are conservative compared to recent literature, the prior on extinction needs to be restrictive to reduce the degeneracy between age and extinction. Although it may seem narrow, it should be noted that this parameter is usually kept fixed in other analyses in the literature or kept free while the metallicity is fixed.\footnote{We have also explored relaxing the metallicity prior by increasing the width of the Gaussian by a factor few. We find that this more conservative choice does not affect the final results of the inferred age ($t_{\rm GC}$, $t_{\rm U}$) as the statistical errors remain below the systematic ones.} 

\subsection{Sampling and parameter inference}
Our posterior distribution is complex and often multi-modal.
V20 
used the Affine Invariant Markov Chain Monte Carlo (MCMC) Ensemble sampler EMCEE for parameter inference. In this paper, we used instead \texttt{pocoMC}\footnote{https://github.com/minaskar/pocomc},  which is a Python package designed for the rapid estimation of posterior Bayesian and model evidence using the preconditioned Monte Carlo algorithm. It offers speed advantages over traditional methods such as Markov Chain Monte Carlo (MCMC) and Nested Sampling, making it suitable for complex scientific problems with costly likelihood evaluations, nonlinear correlations, and multi-modal distributions, such as in this application.

 The \texttt{pocoMC} set up was chosen to be consistent with the EMCEE sampling of our previous analysis and includes:  number of effective particles  $n_{\rm effective} = 256$,
 number of active particles $n_{\rm active} = 128$, preconditioned MCMC (precondition=True), 
number of processes to use for parallelization pool=4,\footnote{With this set the entire sample can be analyzed 
simultaneously on our available computing facilities} and total number of effectively independent samples to be collected $n_{\rm total}=5000$.
Several convergence tests on a subset of clusters, varying the number of effective particles and increasing the total number of effectively independent samples, have shown  that this does not change the results.

\subsection{Hierarchical modeling and age distribution}
In V20, to determine the age limit of the cluster,  metallicity cuts were applied  and inverse variance weighting was used for the error. As the global age inference can be correlated with the error associated with individual clusters, here we increase sophistication and  opt for a hierarchical model. Hierarchical models assume that the posterior (the age  distribution  in this case) is inherited from an underlying ``true" distribution. 
Here we  use PyMC\footnote{https://www.pymc.io/welcome.html}, a Python-based probabilistic programming library that enables users to construct Bayesian models via an easy-to-use Python API and apply Markov Chain Monte Carlo (MCMC) techniques for fitting.

Our tests (see appendix \ref{app:PyMC}) show that given several realizations of  distributions drawn  from an underlying (true) distribution that is either a single Gaussian or a mixture of Gaussians, PyMC recovers correctly the underlying true distribution, with quantifiable errors.
Therefore,  we find PyMC well suited to infer what could be the underlying  age distribution given the age posteriors of each of the 69 GCs.

\subsection{Systematic uncertainties}
\label{sec:syserr}
 There is no change in the  quantification and treatment of systematic  errors from V21,  which we adopt here unchanged.
 In brief, we  estimate the remaining systematic uncertainties in the ages of GCs induced by the theoretical stellar model using the recipe in  Table~2 of Ref.~\cite{OMalley}. The dominant systematic uncertainty is related to reaction rates and opacities.\footnote{Rotation is another source of systematic uncertainty, as the rotation speed of stars in GCs is unknown. However, the main effect of rotation is to alter the depth of the convection zone. Given that we have explored a wide range of values of the mixing length parameters, the effect of rotation is effectively included in our systematic budget estimation.} Everything else, (including  the remaining HST photometry systematic uncertainty) is subdominant, thus we argue that the combined effect of these two components captures well the extent of systematic errors. We follow the prescription in V21 (\cite{Valcin2021}) to estimate these systematic uncertainties, which propagate into ages errors  at the level of $0.23$ (conservatively $0.33$) Gyr with (conservatively without) using spectroscopic metallicity determinations to constrain the allowed range of mixing length parameter variations.

\section{Results}
\label{sec:results}

We apply the methodology presented in the previous sections to our catalog of  69 GCs.  The two-dimensional marginalized posteriors of all pairs of parameters for  two   representative clusters  (Arp2 and NGC2298) are shown in Fig.~\ref{fig:cornerplots}
(in  Appendix~\ref{app:GCtable-params}).
Across the full sample, the recovered parameter distributions indicate that a Gaussian assumption in interpreting the confidence regions is reasonable,
so we report the mean and standard deviation of the posterior for each parameter. The inference for the whole sample is reported in
table \ref{table:param_constraints} of Appendix~\ref{app:GCtable-params}.

\begin{figure}[ht]
    \centering
    \includegraphics[width=0.49\linewidth]{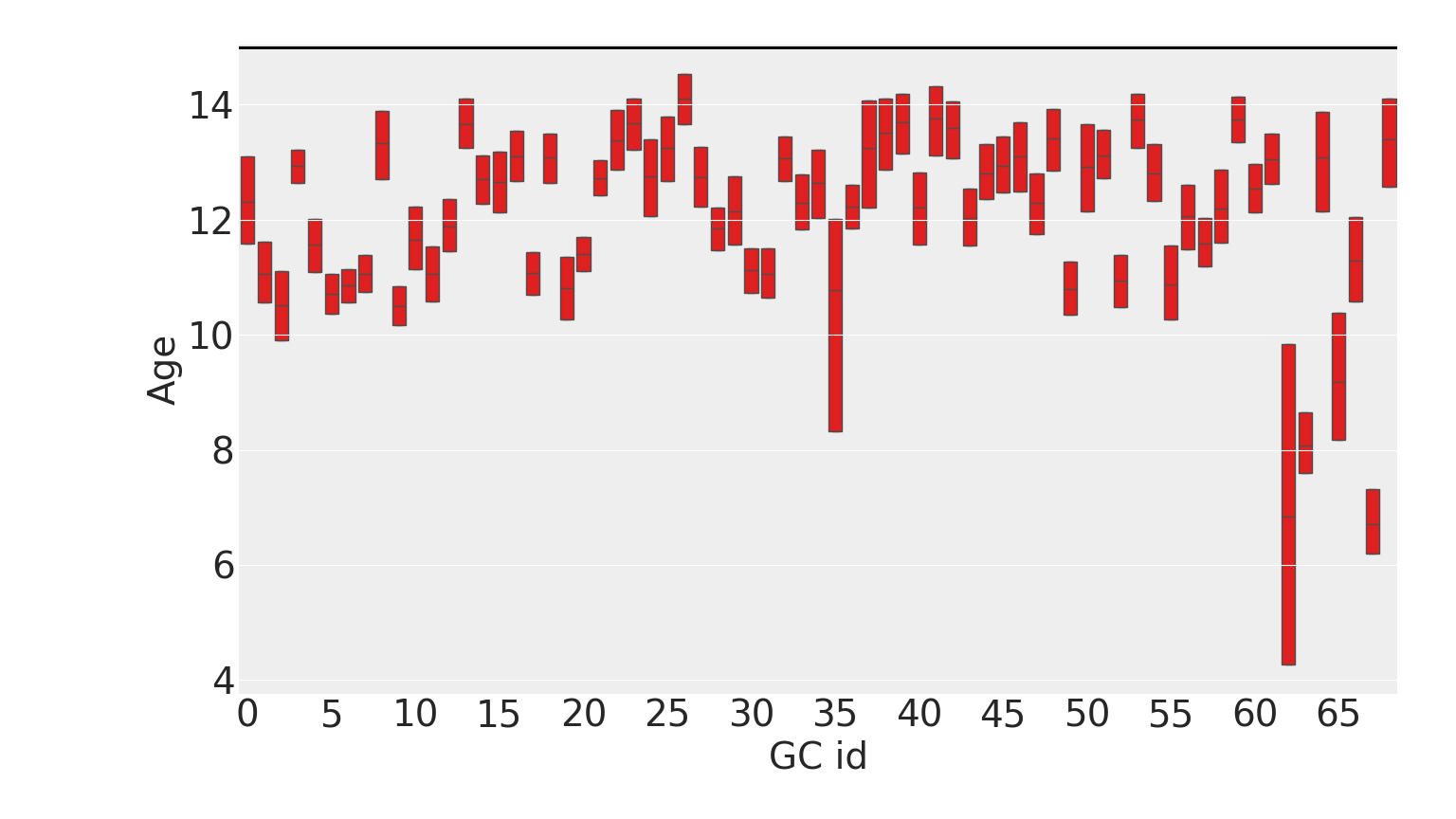}
    \includegraphics[width=0.49\linewidth]{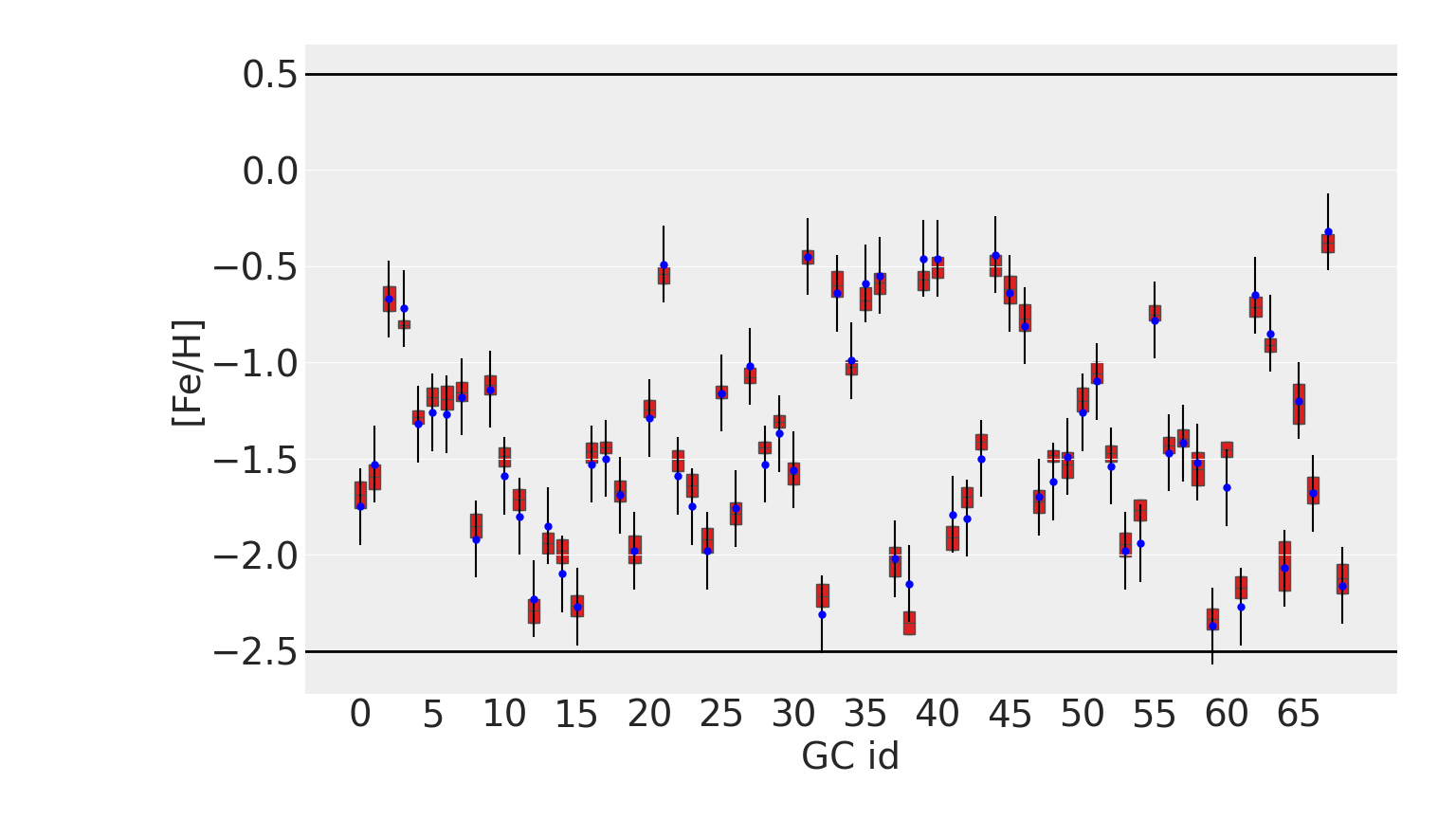}
    \includegraphics[width=0.49\linewidth]{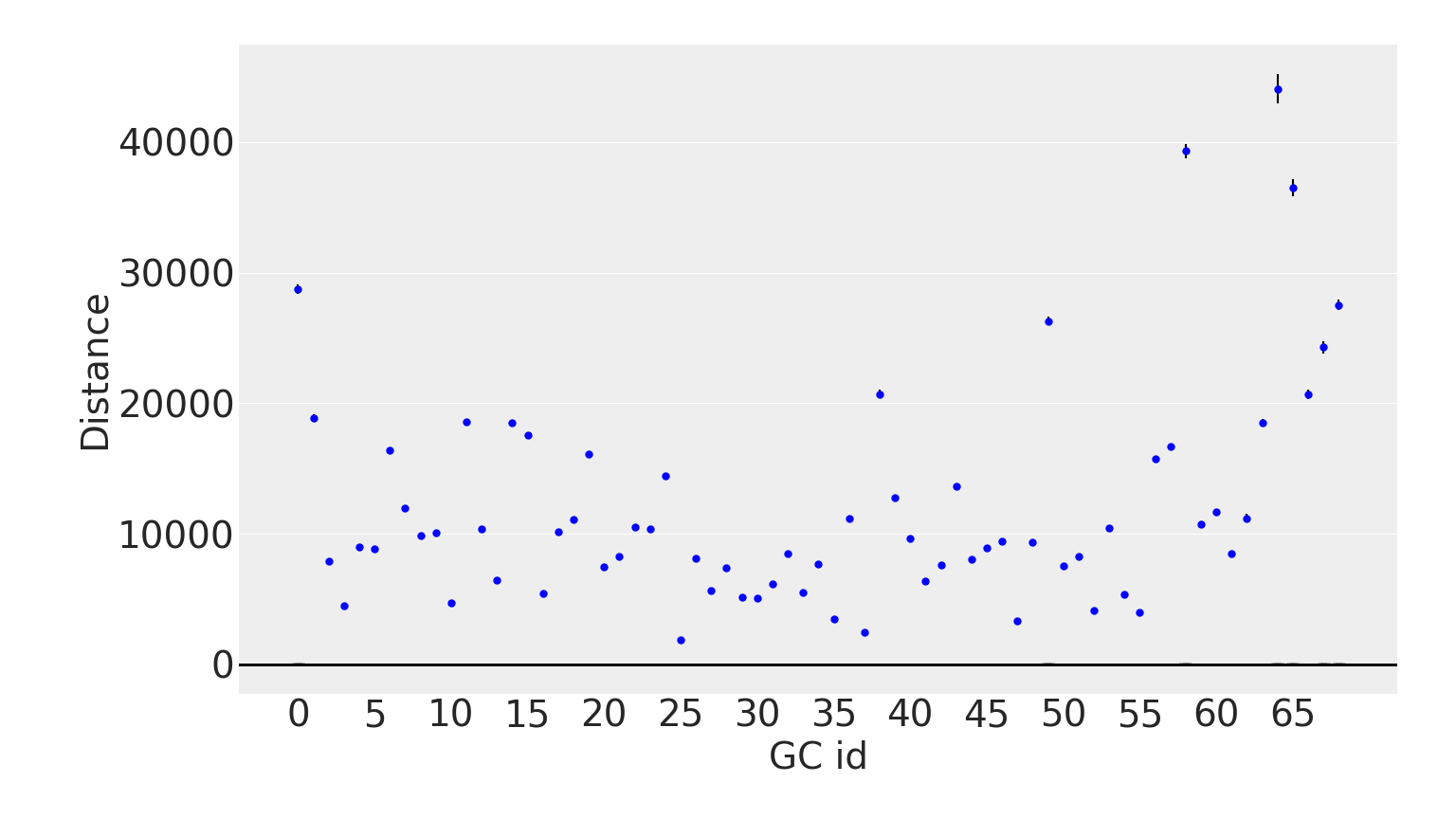}
    \includegraphics[width=0.49\linewidth]{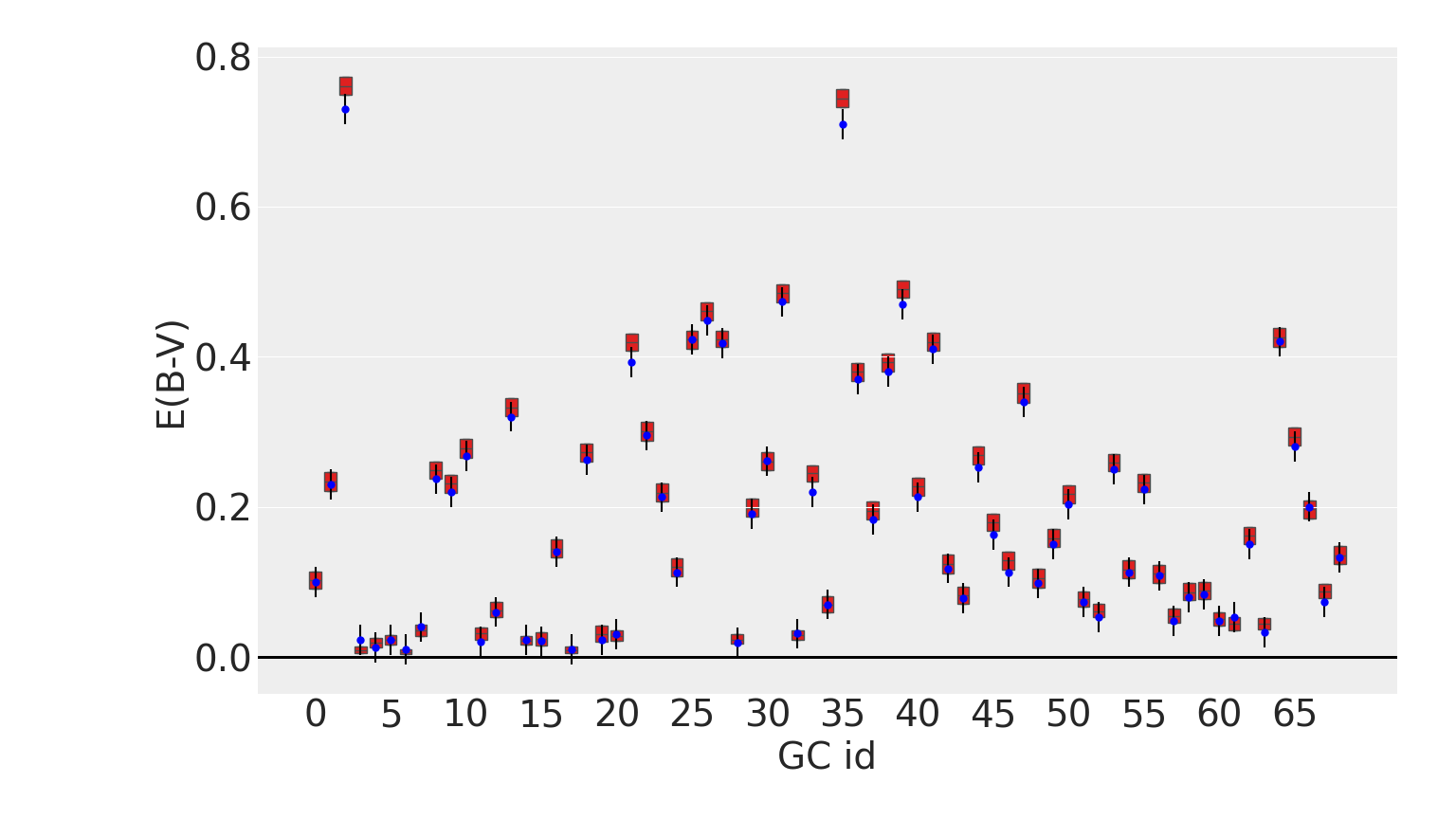}
    \includegraphics[width=0.49\linewidth]{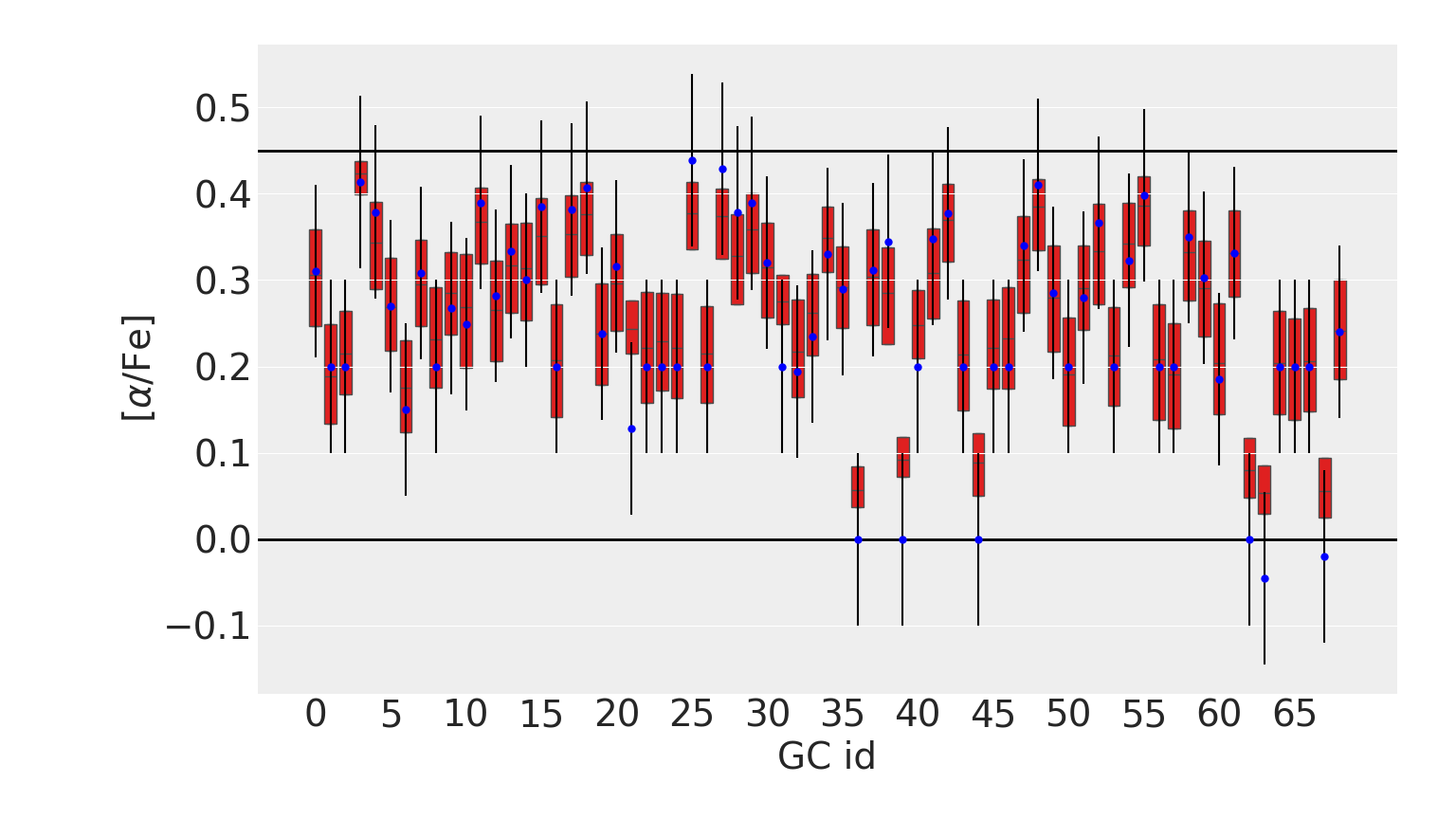}
    \includegraphics[width=0.49\linewidth]{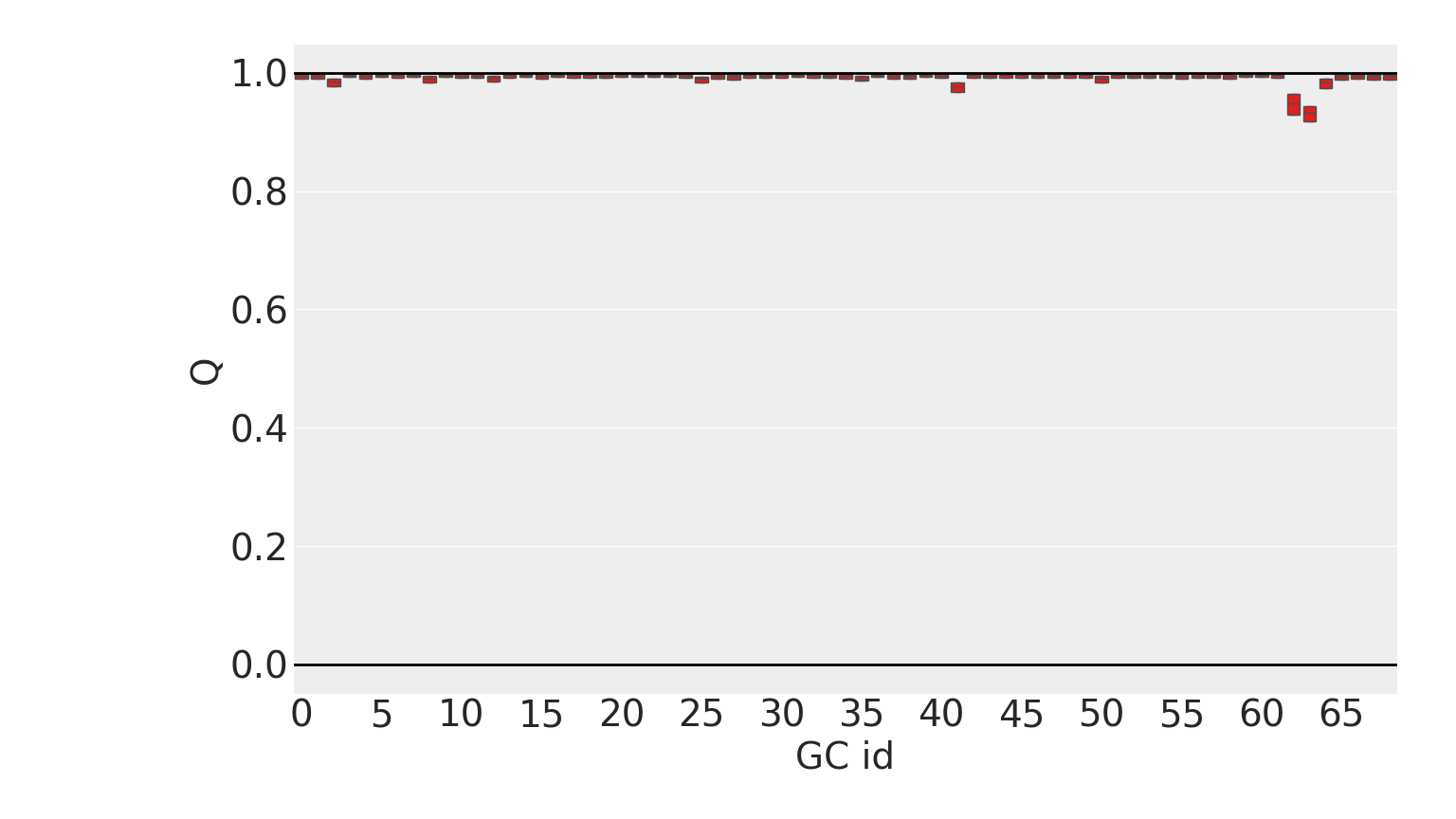}
    \includegraphics[width=0.49\linewidth]{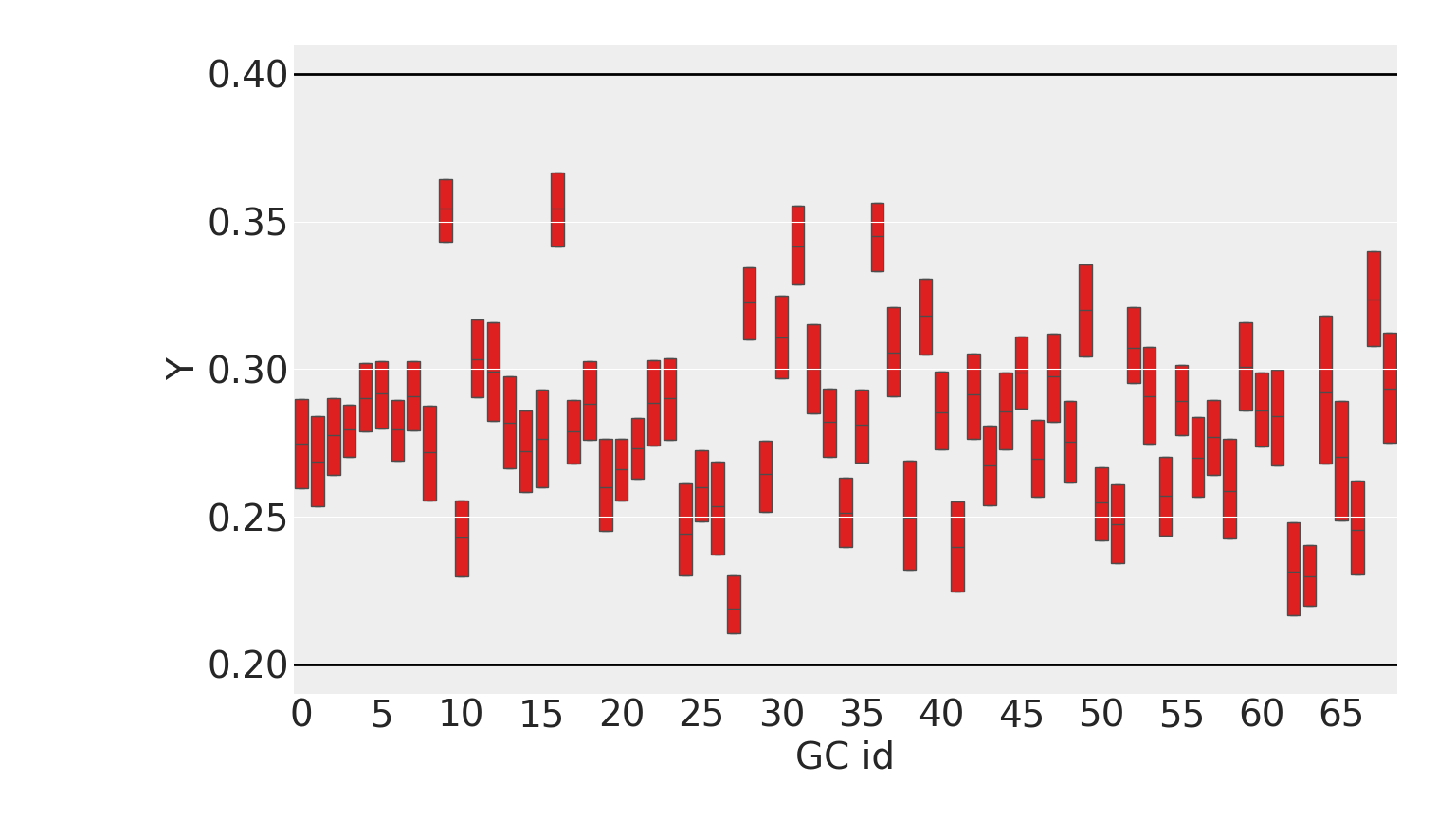}
    \includegraphics[width=0.49\linewidth]{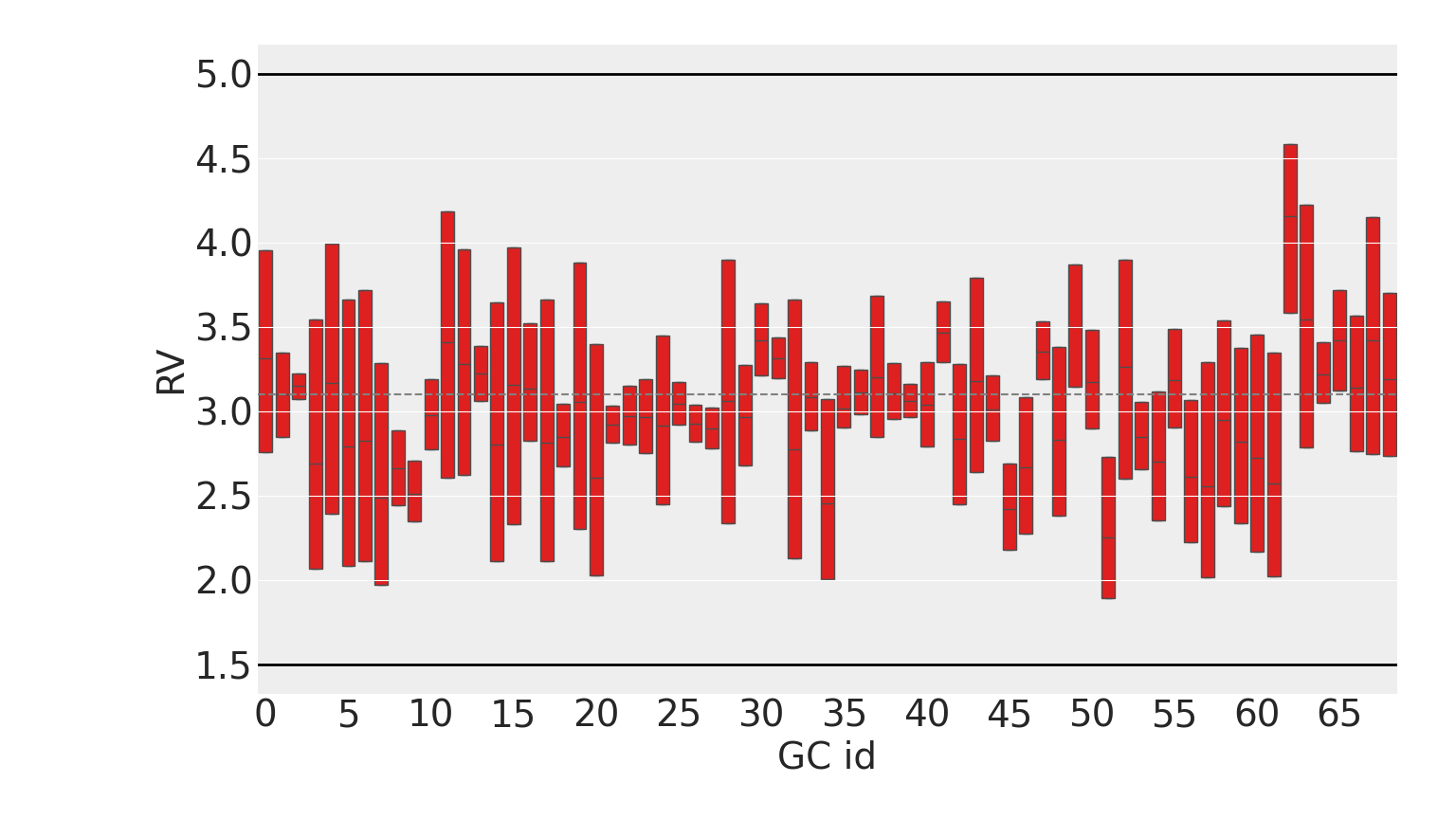}
   
    \caption{Marginalized best fit values for each of the 69 GC in the sample, with corresponding 1D uncertainty bounds (as red bars) and the priors used for each parameter, horizontal thick black  lines are the hard priors, thin vertical lines are the Gaussian priors. The distance determinations are dominated by the prior, see discussion around Fig.~\ref{fig:V20vsV25} and sec.~\ref{sec:V20V25}. Note how the $Q$ distribution is localized, as expected,  around the upper prior edge, how  the Y distribution scatters round the standard primordial value  and how the $R_v$ distribution  is centered around the universal value of 3.1.}   \label{fig:posterior_distrib}
\end{figure}

Figure~\ref{fig:posterior_distrib} shows the marginalized constraints  for each of the parameters for  all the GCs in the sample. The thick horizontal black lines mark the edges of the hard priors imposed. 

The Gaussian priors imposed are indicated by the blue dots (central values) and the vertical black lines (width), see Table \ref{tab:priors}. The red bars denote the standard deviations around the central (mean) values.

Before focussing on the recovered ages (sec.\ref{sec:ages} and below), several considerations are in order.  
In general, for the clusters in this sample we find no significant  correlation between age   and  distances,  absorption  or [$\alpha$/Fe]. On an individual cluster-basis the constraints on  [$\alpha$/Fe] are weak, however values of  [$\alpha$/Fe]$>0.4$ and smaller than $0.1$ are typically disfavored. The obtained range for [$\alpha$/Fe] might reflect the actual spread in chemical enrichment of these elements. 
The use of the full color-magnitude diagram, along with the adoption of the priors motivated in sec.~\ref{sec:inference},  enables us to  break the age--distance--metallicity--redenning degeneracy.

Unlike in V20, helium is a free parameter with a wide prior ($0.2 < Y < 0.4$)  in this study. The bottom left panel of Fig.~\ref{fig:posterior_distrib} shows the recovered  $Y$ values for the GCs. The medium value, $Y = 0.275$, is in excellent agreement with estimations of the primordial Helium abundance from BBN and Planck18~\cite{Planck18} and the Helium-metallicity relation (e.g. Ref.~\cite{Helium,Helium2}).

\begin{figure}
    \centering    \includegraphics[width=0.8\linewidth]{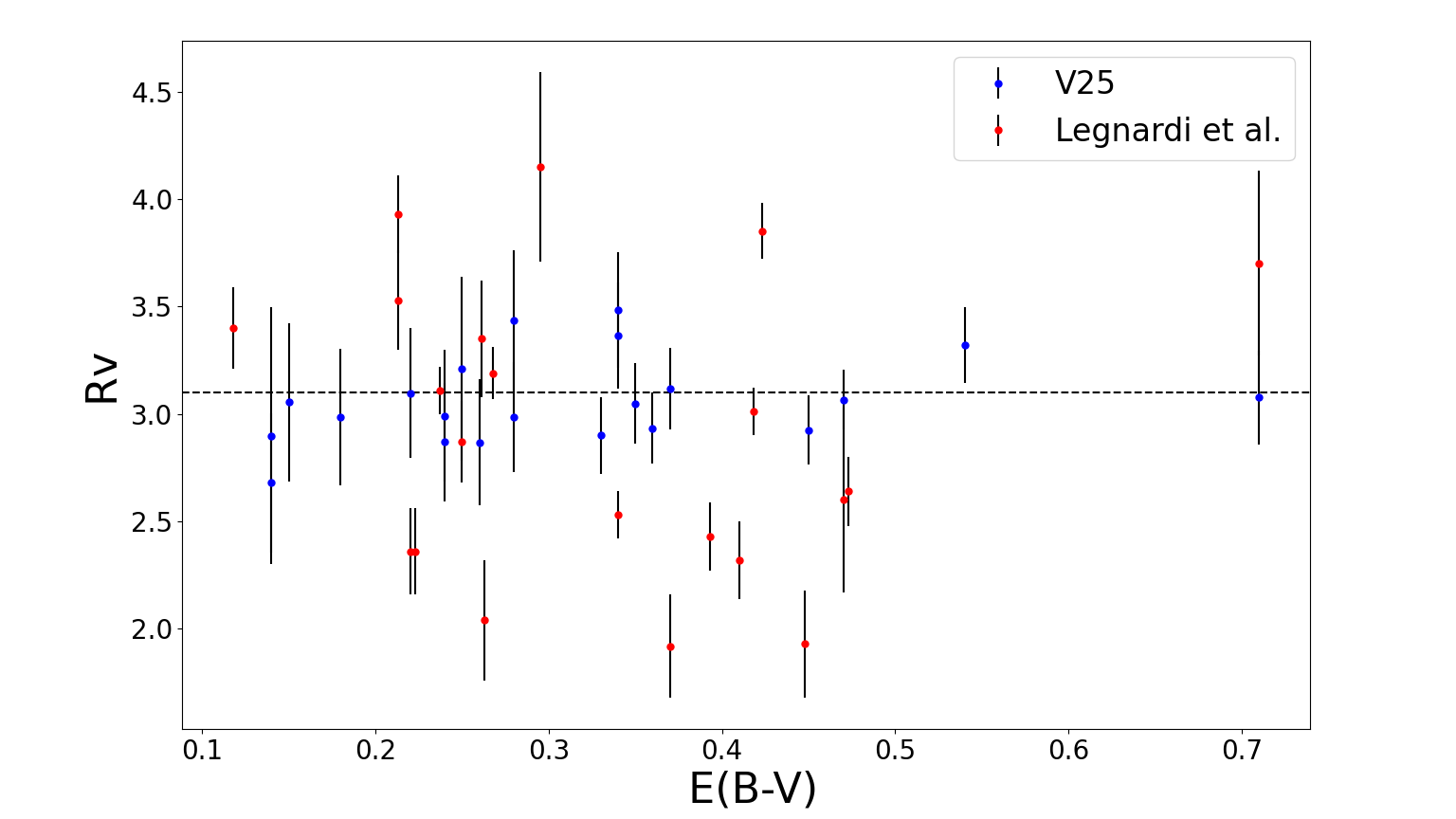}
    \caption{Recovered  $R_v$ value compared to the one from~\cite{Legnardi} as a function of extinction. Our recovered values show less scatter around the universal value of $3.1$.}
    \label{fig:rv_comp}
\end{figure}

The recovered $R_v$  values (also kept fixed in previous analyses) can be compared  to the ones from ref.~\cite{Legnardi} who use the older Harris \cite{Harris} catalog for extinction.  Fig.~\ref{fig:rv_comp} shows this comparison as a function  of extinction. 
Our recovered $R_v$ values show less scatter and are closer to the  universal value of $3.1$ (dashed horizontal line in the figure).
This could be due to the fact that our updated compilation of extinctions is likely more precise; when the adopted extinction has a larger scatter around the true value, the recovered $R_v$ will also show a larger scatter.

\subsection{Comparison with V20}
\label{sec:V20V25}
\begin{figure}
    \centering
    \includegraphics[width=0.45\linewidth]{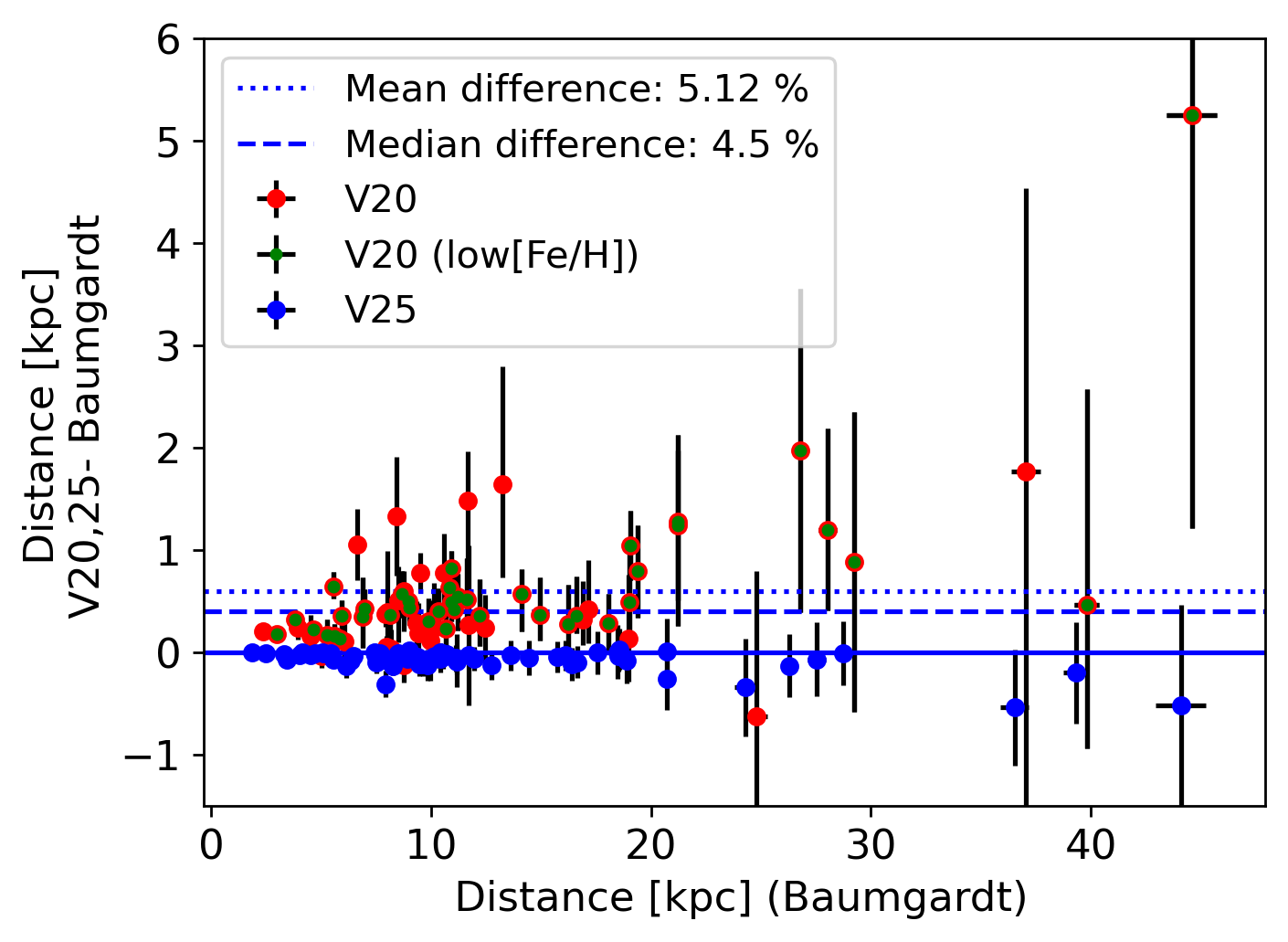} 
    \includegraphics[width=0.45\linewidth]{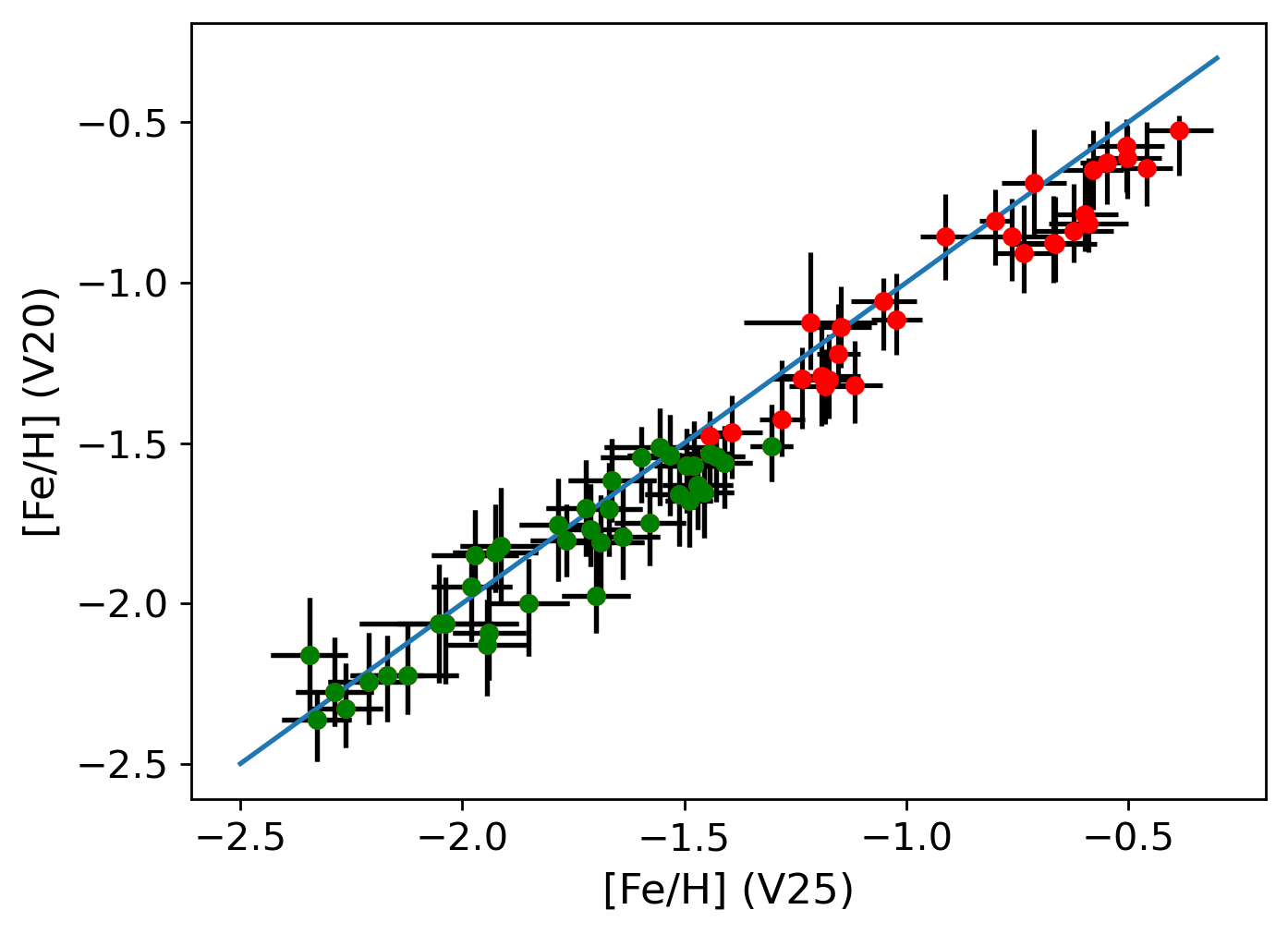}   \includegraphics[width=0.45\linewidth]{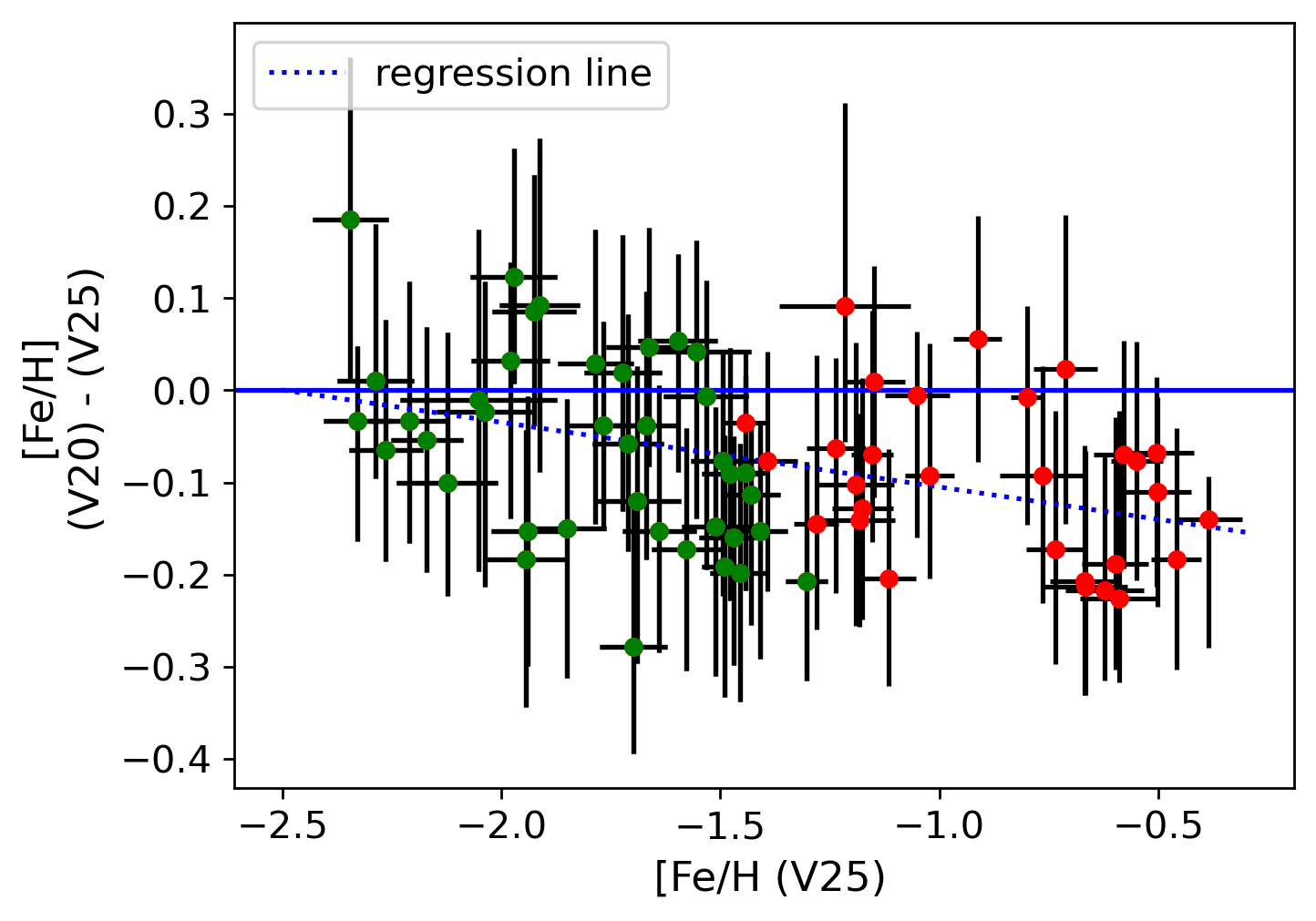}
\includegraphics[width=0.45\linewidth]{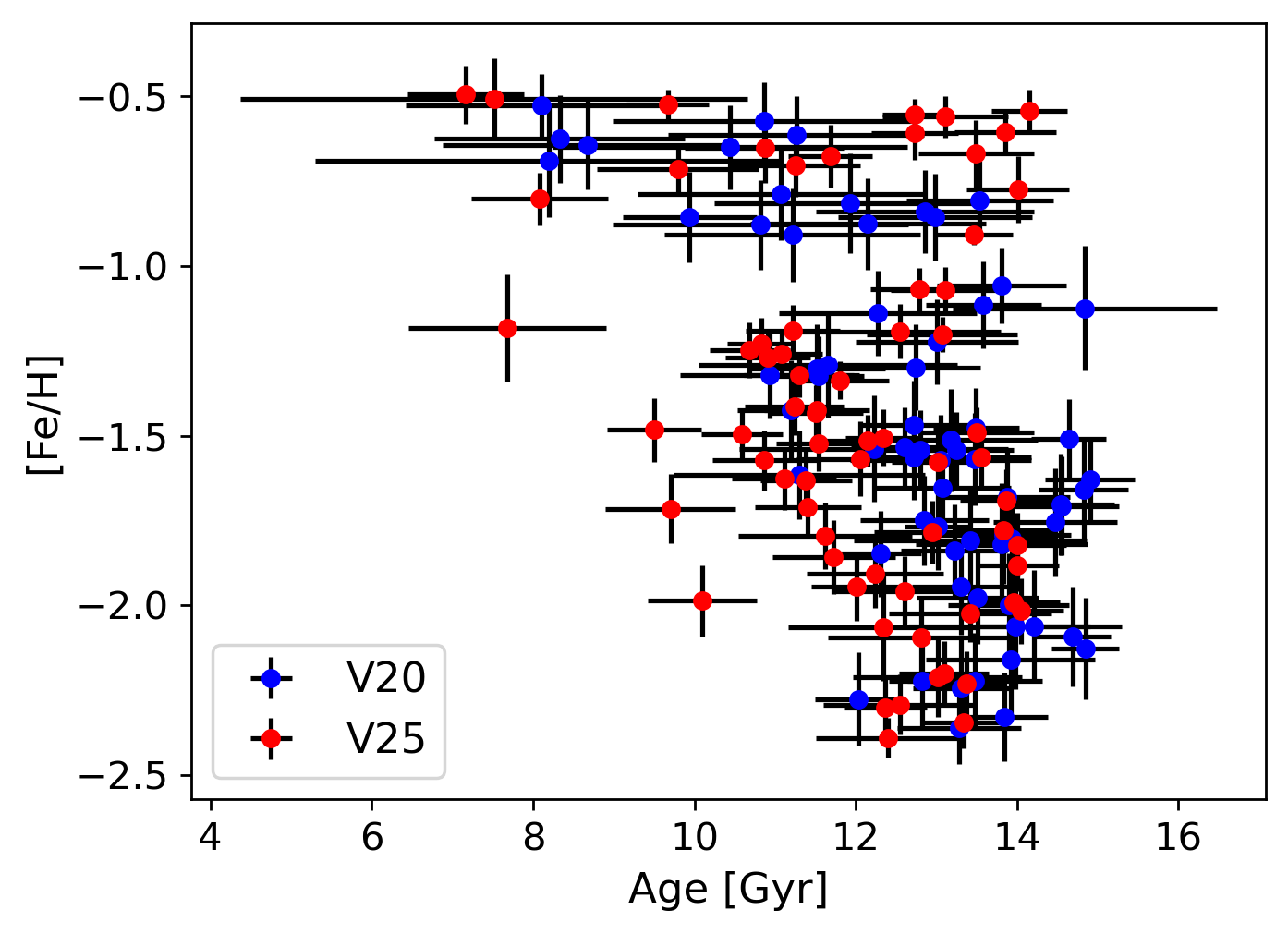}
\includegraphics[width=0.45\linewidth]{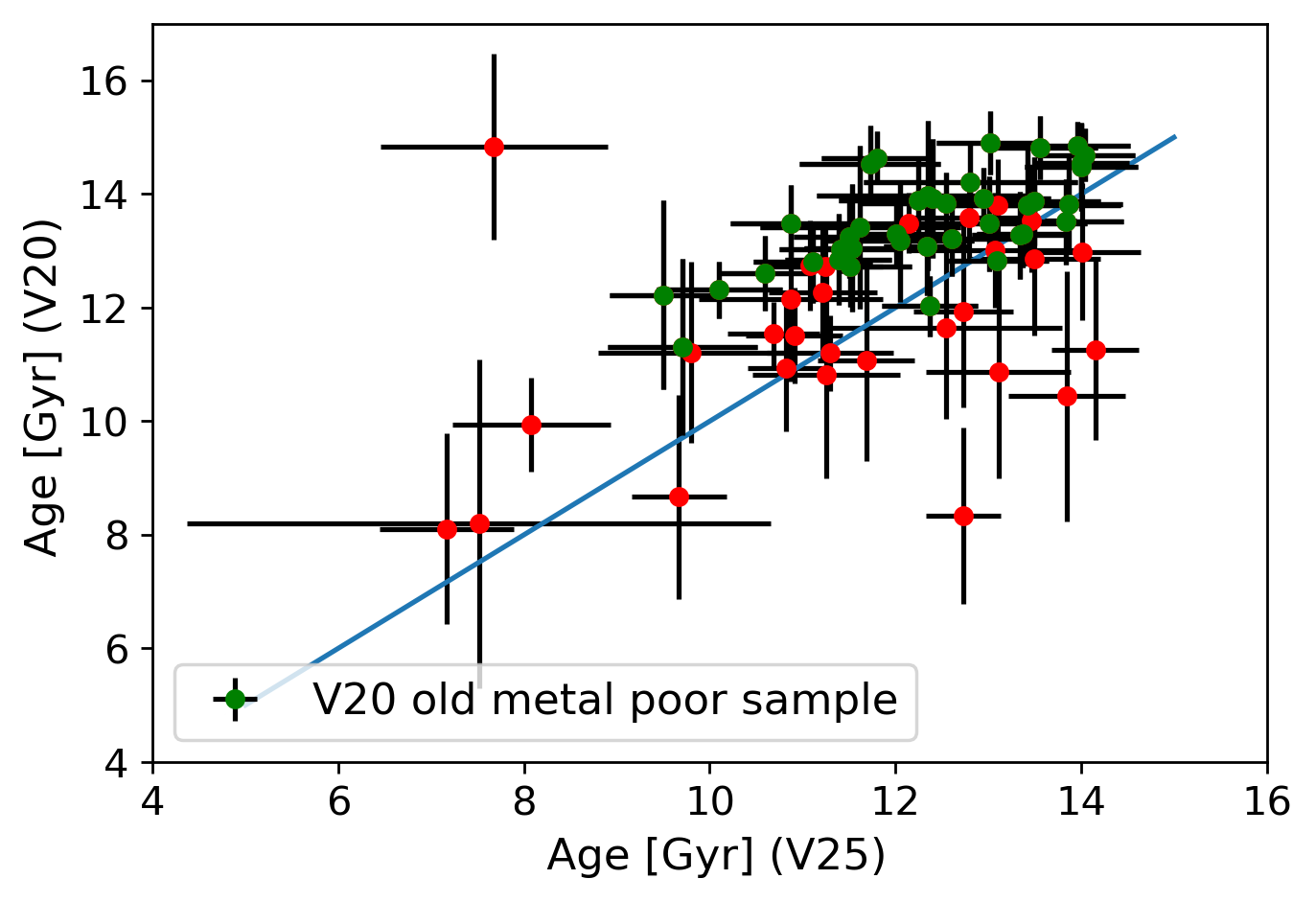}
\includegraphics[width=0.45\linewidth]{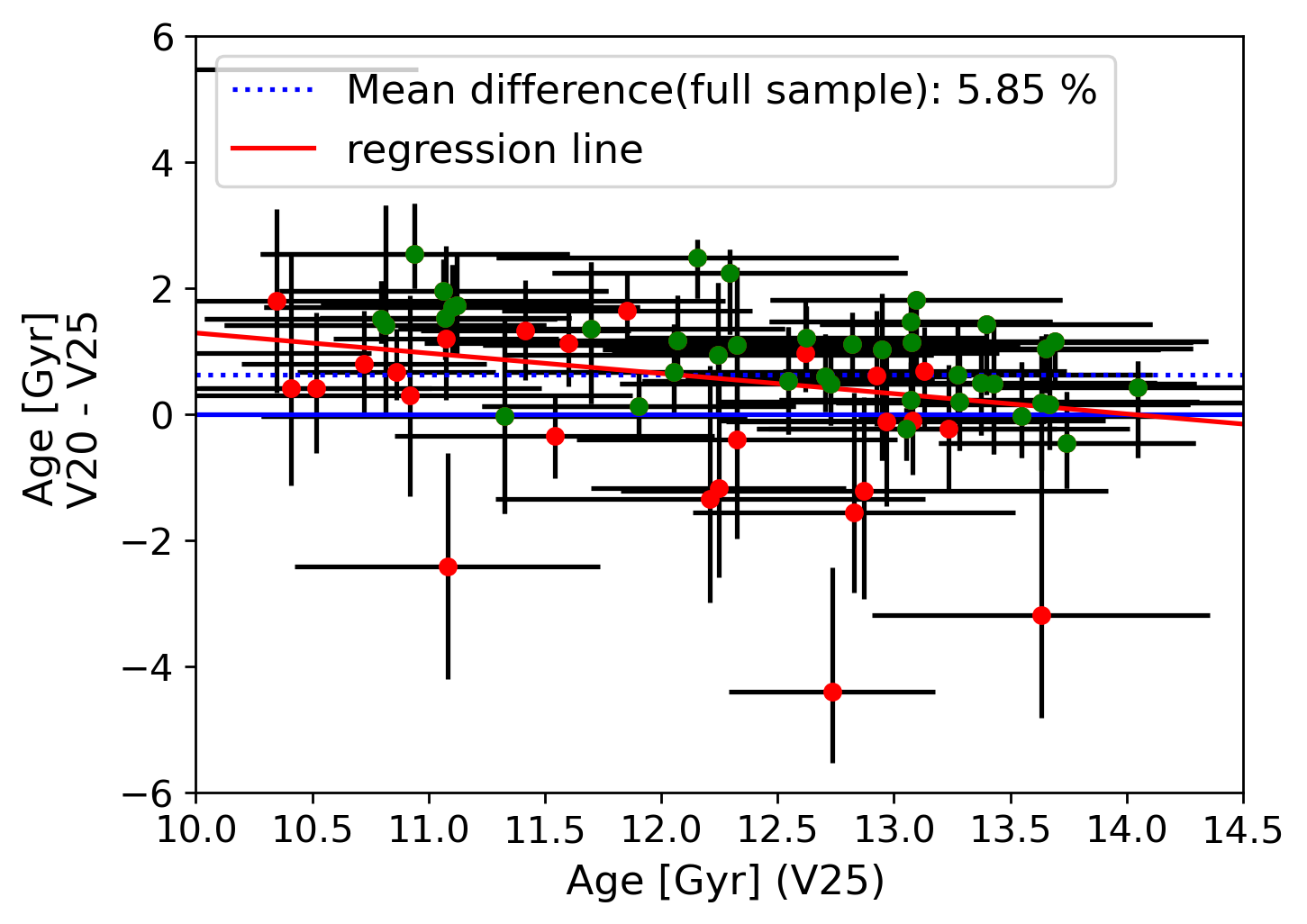}
    
    \caption{Comparison of V20 GC parameters with these results (V25). In all panels, the green points indicate  the old, low metallicity sample used in V20 to estimate the age of the Universe. The top left panel can be compared directly with fig.~\ref{fig:comp_gaia}. For clarity, the V20 points have been displaced slightly to the right. The top-right and middle left panels compares the inferred metallicities. While there is a trend for the full sample (indicated by the dotted line) the agreement is particularly good at low metallicities. The middle right  panel compares the age-metallicity distributions. Recall that the V20 old sample had [Fe/H]$<-1.5$.  Finally  the bottom  panels show the recovered ages.  While for the full sample V20 ages tend to be older than V25 by $\sim 6\%$, the agreement improves substantially for older ages, as shown in the bottom right panel that zooms in into ages older than 10 Gyr and shown the regression line computed from the full sample. For ages above 12 Gyr there is no evidence of any  significant difference on average. 
    }
    \label{fig:V20vsV25}
\end{figure}

An important  difference with  V20 is that in this study we have adopted Gaia~\cite{Baumgardt} distances as our prior. 
  Figure~\ref{fig:V20vsV25} (top left panel) shows the comparison of distances between V20, Baumgardt (who used Gaia)  and the present work (V25). Not surprisingly, current distances are closer to the Gaia ones. We will see below that this mismatch has a small effect on the recovered ages and does not affect the determination of the older ages in any significant way.

\begin{figure}[ht]
    \centering \includegraphics[width=\linewidth]{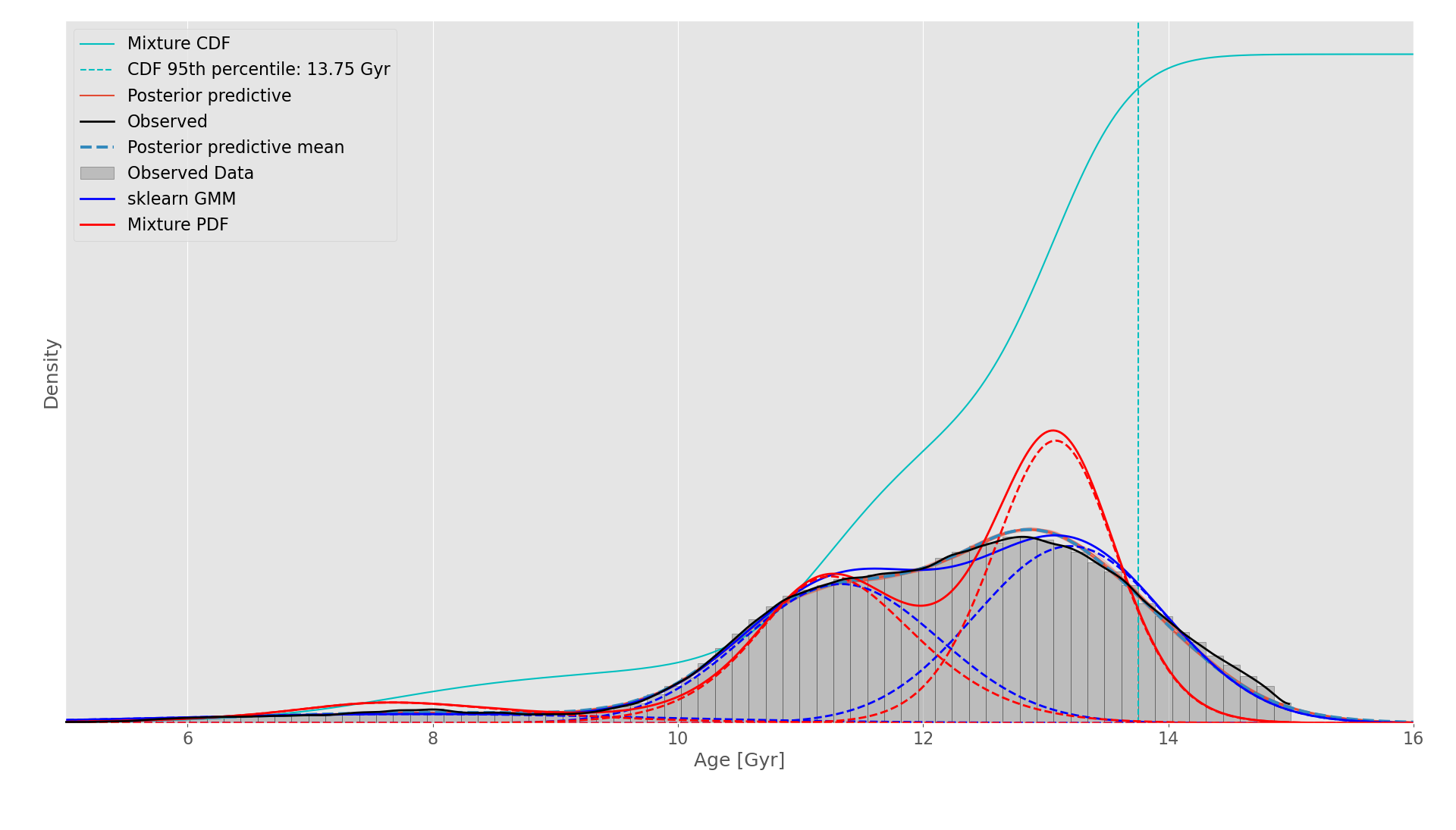}
  \includegraphics[width=\linewidth]{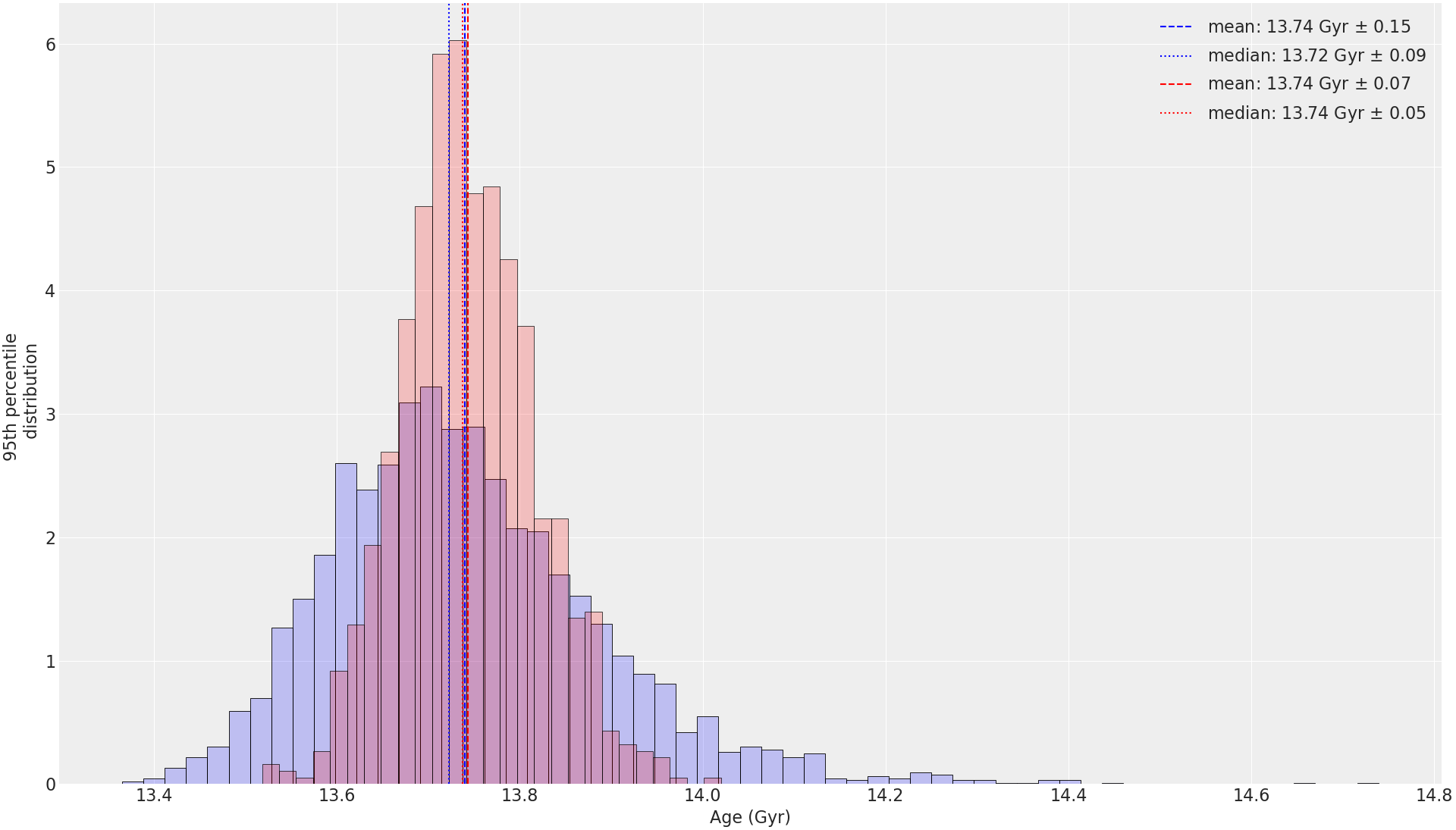}  
\caption{Posterior distribution of the ages of the GC in the sample and interpretation in terms of Gaussian mixture. Three components can be recognized with respective means  and standard deviations of $t_{GC}=\{8.80, 11.27, 13.30\}\pm \{ 1.48, 0.61,0.56\}$. The bottom panels show the age distribution for the 95\% percentile.}
\label{fig:final_ages}
  \end{figure}

Metallicity determinations (top right and middle left panels)  between V20 and the present work are in 
good  agreement, especially for the metal poor objects, which are those relevant here (see middle left panel). The most metal poor GCs are also the oldest as shown in the middle right panel.
For lower metallicity, the absorption lines in the spectrum are much weaker. Thus, they affect the integrated photometry less: the photometry becomes fully dominated by the continuum, which has information about both age and metallicity. The weaker absorption lines  therefore yield
less degeneracy in the age-metallicity plane from the integrated photometry. 
The bottom right panel  compares the recovered ages, highlighting the old, metal poor sample used in V20 to estimate the age of the Universe. Although there is a large scatter, especially for the objects with larger error-bars,  V20 ages  are on average $\sim$ 6\% older than in  this work. However,  the shift reduces significantly for the older, metal-poor population as shown in   the bottom right panel, where the regression line to the full sample is also shown. For objects older than $\sim 11-12$ Gyr there is no evidence of any age difference.

\subsection{Ages}
\label{sec:ages}
Fig.~\ref{fig:final_ages} shows the posterior distribution of the GC ages, PyMC finds three Gaussian components: one (subdominant) centered around $9$ Gy: few objects have a recovered mean age around this value, and they tend to have large age uncertainty.   A second component is centered around 11 Gyr and a third around 13.5 Gyr. The width of these distributions encompasses the measurements errors and the intrinsic age scatter of the sample, which theoretical models predicts to be of the order of 200 Myr, see e.g.~\cite{TrentiGC} and references therein.

\begin{figure}
    \centering
    \includegraphics[width=0.5 \linewidth]{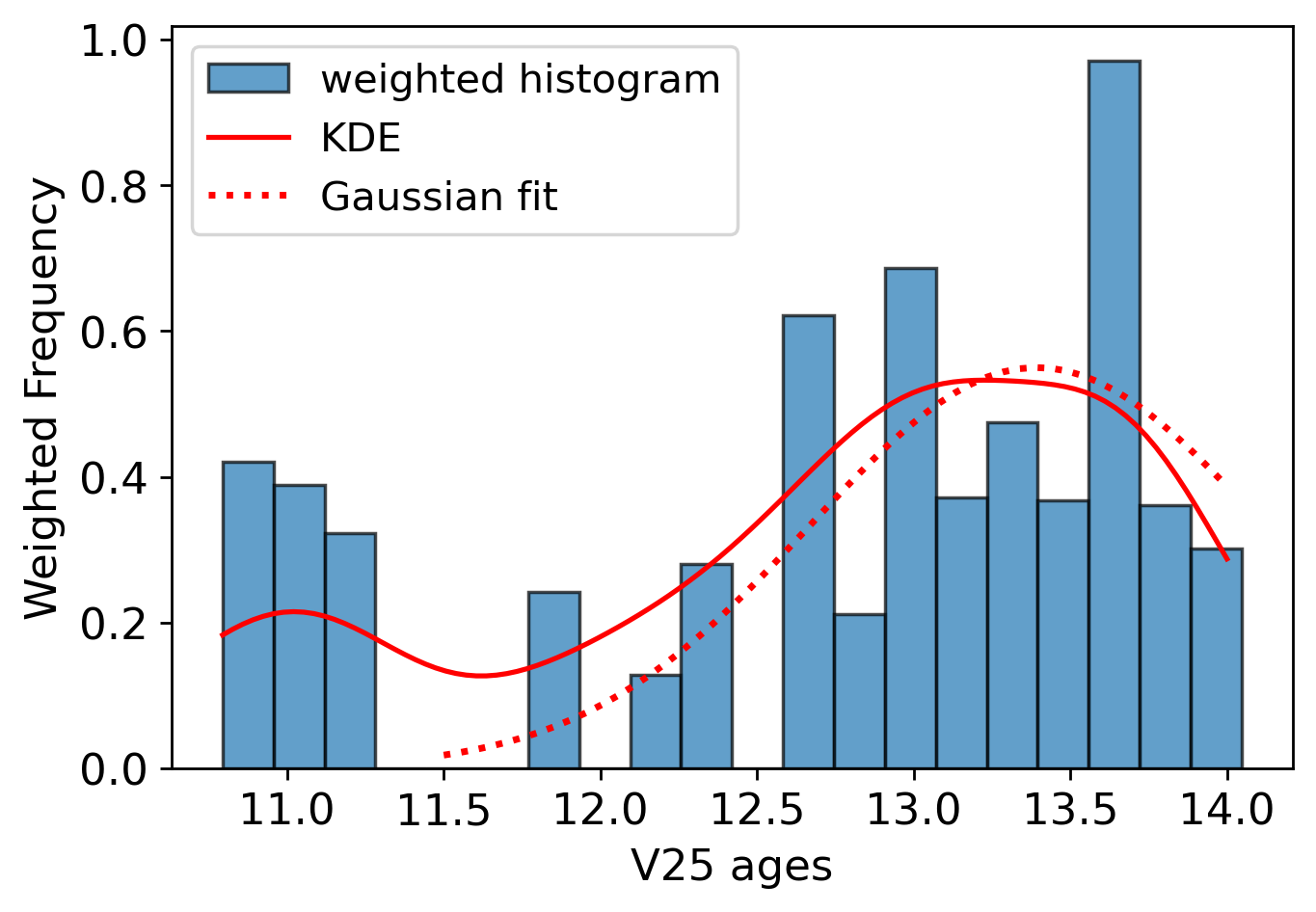}
    \caption{Age distribution for GCs with metallicity $< -1.5$ [Fe/H]. There is an isolated distribution above 11.5 Gyr, and a  sharp cut  at 14 Gyr; the age prior is at 15 Gyr. In blue the (inverse variance) weighted histogram, the solid red line is a kernel density estimator of the distribution where the kernel width is given by the error of each object's mean age determination.  The dotted red line is a  Gaussian fit to the oldest ages peak.}
    \label{fig:2G_lowmet}
\end{figure}

We wish to estimate the age of the {\em oldest} stars from our GC sample, $t_{GC}$. The intrinsic formation spread may represent a non-negligible contribution to  the width of the Gaussian component,  the mean age of the oldest group found by the  mixture model-PyMC algorithm is $t_{GC}^{GM}=13.30\pm 0.10$. 
A robust upper limit can be obtained from the 95\% quantile of the full age distribution,  which  PyMC returns to be $t_{GC}^{\leq}= 13.75 \pm 0.05\,{\rm (stat)}$ (see Fig.~\ref{fig:final-ages}).
An estimate of $t_{GC}$ is to follow V20 and cut by metallicity, as the least metal rich GCs are expected to be those that have formed earlier. Indeed, if we select GCs with, ${\rm [Fe/H]} < -1.5$ we see  an isolated distribution above 11.5 Gyr.  A non-parametric (a kernel density estimator)  fit finds it centered around 13.30 Gyr  while a (truncated) Gaussian fit finds the maximum at 13.39\ Gyr and  68\% limits of $\pm 0.52$ which corresponds to an error on the mean of 0.11 Gyr.
(see Fig.~\ref{fig:2G_lowmet}). The 
PyMC for this sample returns an estimate of $13.39\pm 0.10$. These results are robust to  reasonable  changes in the cutoff value for metallicity. 

Considering the scatter among the different ways to extract, $t_{GC}$ we conclude that there is no need to increase the errors to account for uncertainty  due to the adopted methodology, and we take the most sophisticated value obtained by PyMC of  $t_{GC}=13.39\pm 0.10 $ for our determination for the age of the oldest stellar population of GCs.

\subsection{From globular cluster ages to the age of the Universe}
\label{sec:ageuni}

The age of the oldest stars in the GC sample, $t_{GC}$, which were formed at redshift $z_{\rm f}$, sets a lower limit for the age of the Universe, $t_{\rm U}$: $t_{\rm U}=t_{\rm GC}+\Delta t$, where  the formation time  $\Delta t$ corresponding to the look back time at $z_{\rm f}$.

As shown in Ref.~\cite{JimGC},  it is possible to estimate  the probability distribution of $\Delta t$ by adopting a  well-motivated value for $z_{\rm f}$. Observations indicate very early star  and galaxy formation ~\cite{2018Natur.557..392H,2020ApJ...888..124S,2019MNRAS.489.3827B} including the direct detection of GCs~\cite{GEMS}, thus ~\cite{JimGC, Valcin2020,BernalTriangles21} assume $z_{\rm f}=11$, although it is shown there that  $\Delta t$ is largely  insensitive to this choice.

Ref.~\cite{JimGC} provides  the probability distribution of  $\Delta t$ marginalizing over $H_0$, $\Omega_{m,0}$ and $z_{\rm f}$, with $z_{\rm f}$ varying between $z_{\rm f, min} = 11$ and $z_{\rm f, max}$. The resulting distribution depends very weakly on cosmology for reasonable values of the cosmological parameters, weakly on $z_{\rm f, min}$ and very weakly on the choice of  $z_{\rm f, max}$ provided $z_{\rm f, max}> 20$. Ref.~\cite{Valcin2020} provides an analytical  fitting formula to the distribution, to simplify marginalization over $\Delta t$. Here, we estimate the full probability distribution of $t_{\rm U}=t_{\rm GC}+\Delta_t$ by performing a convolution of the posterior  probability distribution for $t_{\rm GC}$ and the probability distribution of $\Delta_t$. 
Keeping the systematic error budget of V21, we obtain $t_{\rm U}=13.57^{+ 0.16}_{-0.14} ({\rm stat})\pm 0.23 ({\rm sys.})$ and a robust 95\% confidence upper limit of $t_U^{\leq}=13.92_{-0.10}^{+0.13}({\rm stat}) \pm{0.23} ({\rm sys})$. 
 
\section{Summary and Conclusions}
\label{sec:summary}

We have built upon our previous results \cite{Valcin2020,Valcin2021} with improvements in the methodology to estimate the ages (and other properties) of GCs. In particular, we have increased the flexibility of the model, including two additional free parameters, the helium fraction and reddening, that were previously fixed.
We have incorporated a new Bayesian hierarchical model to robustly and precisely estimate the GC parameters. It includes a Gaussian mixture model to reduce possible biases arising from presence of multiple populations. This model allows us to  analyze  each of the  GC, varying simultaneously their age, distance, metallicity, [$\alpha$/Fe], helium fraction, dust absorption and reddening  adopting physically-motivated and wide priors based on independent measurements of distances, metallicities and extinctions found in recent literature.

Despite the additional freedom, we find that the use of the full color-magnitude diagram, specifically,  the main sequence and red giant branch, is sufficient to break degeneracies among all the parameters.  Except for distances (where the  adopted prior is given by the Gaia measurements of \cite{Baumgardt}),  our posteriors are not dominated by the adopted priors, but by the data.

We found the average age of the oldest (and most metal poor)  GCs is $t_{GC}= 13.33 \pm 0.11 ({\rm stat})\pm 0.23 ({\rm sys.})$  and a robust (95\% confidence) upper limit $t^{\leq}_{\rm GC}=13.75 \pm 0.05 {\rm (stat.)} \pm 0.23 {\rm (sys.)}$  Gyr.
We note that with the updated modeling presented here,  despite the bigger parameter space explored, the statistical uncertainty remains well under control.
The systematic errors are dominated by   theoretical stellar model  uncertainties in nuclear reaction rates and opacities. These remain the dominating systematic uncertainties, as distances and mixing length parameter are determined from data. Systematic errors  are  bigger than  the statistical errors,  once constraints from several objects are combined. Hence, to make  further progress,   uncertainties in stellar model-building should be addressed.

 This age measurement can be used to estimate the Universe absolute age by taking into account the look back time to the big bang  at the likely redshift of formation of these objects. We find  the age of the Universe as determined from stellar objects to be $t_U^{\leq}=13.92_{-0.1}^{+0.13}({\rm stat}) \pm{0.23} ({\rm sys})$. The total uncertainty is 1.86\%. This is now sufficiently small  to warrant comparison to the CMB model-dependent inferred age, which is one of the most accurately quantities inferred from the CMB~\cite{Jaffe,Knox}. Thus, comparing the CMB-derived value to independent astrophysical estimates can yield precious insights into possible new physics, or provide "guardrails" around the  $\Lambda$CDM model. We leave this to future work, here we limit ourselves to note that our determined value of $t_{\rm U}$ is fully compatible with the inferred value from the Planck mission observations assuming the $\Lambda$CDM model.

\begin{acknowledgments}
Funding for the work of RJ and LV  was partially provided by
project PID2022-141125NB-I00, and the “Center of Excellence Maria de Maeztu 2020-2023” award to the
ICCUB (CEX2019- 000918-M) funded by MCIN/AEI/10.13039/501100011033.  Based on data products from observations made with ESO Telescopes at the La Silla Paranal Observatory under ESO programme ID179.A-2005 and on data products produced by TERAPIX and the Cambridge Astronomy Survey Unit on behalf  of the UltraVISTA consortium.

\end{acknowledgments}

\bibliographystyle{JHEP}

\providecommand{\href}[2]{#2}\begingroup\raggedright\endgroup

\clearpage

\appendix

\newpage

\section{Parameter mask}
\label{app:Parameter mask}
Here we detail the procedure for making a parameter mask.
\begin{enumerate}
\item Start with a initial guess isochrone based on the initial guess  for the parameters ($p$) taken from the literature (as described in the main text). The initial guess for the parameters is defined by a pair of vectors  ($\mu_p, \sigma_p$) where $\mu_p$ represents the initial guess for the parameters   and $\sigma_p$ their uncertainties, and we refer to such isochrone as $I(\mu_p)$.

    \item From this initial isochrone $I$ we compute a suite of isochrones $I_i$ by varying one by one
   the parameters metallicity, distance, alpha enhancement  and extinction so that $\mu_i\longrightarrow L_i$ as follows 

    \begin{itemize}
        \item If the prior on the parameter (see table~\ref{tab:priors} is Gaussian, $L_i = \mu_i \pm n \times \sigma_i$. The value of $n$ is a trade-off to make sure that we include most of the width of the red giant branch while removing most of the AGB stars.
        \item If the prior is uniform, $L_i = E_i$, where $E_i$ is the boundary of the uniform prior. 
        \item  If both priors are applied to a parameter, $L_i = {\rm max}(\mu_i \pm n \times \sigma_i, E_i)$ as we want the extreme value to be conservative.
    \end{itemize}

    \item   The mask limit is given by the envelope of all the isochrones $I_i$.
    \item Age is not included because isochrones are almost not sensitive to age above the sub giant branch.  We also limit the mask to a magnitude of,  $mp_{cut} = m_{\rm MSTOP}-1$ which roughly corresponds to the bottom of the RGB.
    \end{enumerate}

\begin{figure}
    \centering
    \includegraphics[width=\textwidth]{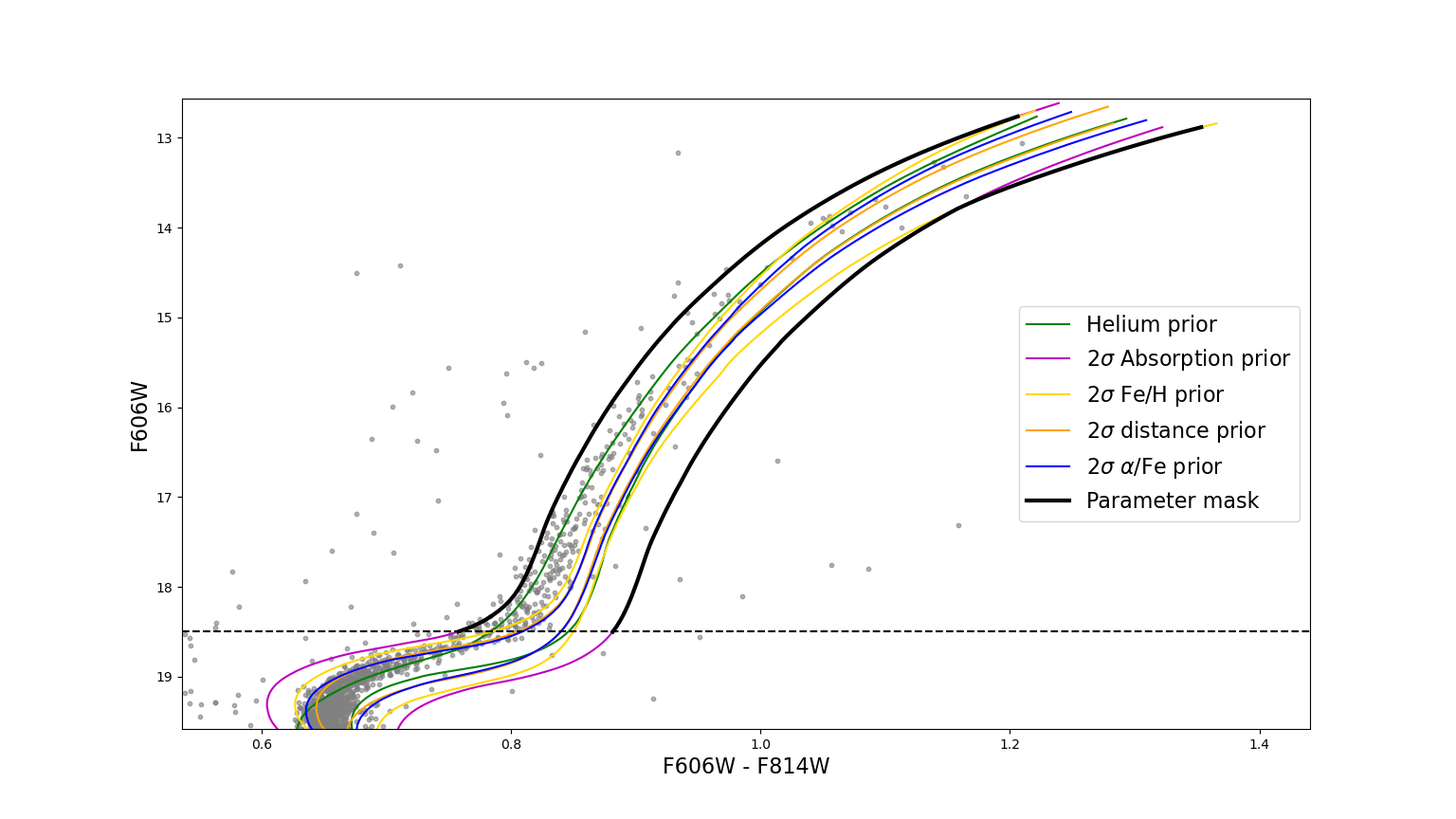}
    \caption{Parameter mask compared to isochrones for extreme values $L_i$ for each parameter}
    \label{fig:param_mask}
\end{figure}

A visual example is shown in Figure \ref{fig:param_mask} where we plot the  initial guess isochrone for NGC 2298 with 2 $\sigma$ ($n=2$) variation in 5 of the most impacting free parameters ([Fe/H], Absorption, distance, $\alpha$/Fe and $Y$). NGC 2298 is an interesting example, as its initial guess is not very good (the mask is centered at the right of the RGB). Even with a bad initial guess, a variation of 2$\sigma$ for our prior choices is sufficient to include most of the RGB while discarding stars belonging to the AGB. We can also see that for this cluster, the mask is completely dominated by the ``absorption" isochrone (magenta).

\section{PyMC inference tests}
\label{app:PyMC}

Each of the following tests are run with the NUTS sampler for 4 chains. Each chain runs for 1000 steps after a tune period of 1000 steps.
To make sure that we can recover the underlying distribution, we perform a series of tests. 
First, we simulate 69 normal distributions with their own means and standard deviations. Each distribution is drawn from an underlying Gaussian. The resulting trace is shown in appendix \ref{sec:Simple Gaussian}. We can see that we both recover very well the properties of every individual distributions and the overall one.
For the second test, we also simulated 69 normal distribution. Contrary to the first test, these distributions are drawn from a  mixture of Gaussian. The trace of the inferred parameters is also shown in appendix \ref{sec:Gaussian mixture}. PyMC does a good job recovering the mean and standard deviations of every normal distribution. More impressive, the sampled means, standard deviations and weights of each component of the mixture are also in good agreement with the true values.
The last test is what we call here a null test. As the number of components inside the mixture is subjective and can be approximated from different methods. We checked whether the software recovers artificial components. For this we sampled 69 normal distributions coming from a single Gaussian  but this time we fit a normal mixture as the underlying distribution. The results are shown in appendix \ref{Mixture null test}.

In this section, we will explore a series of tests to assess the effectiveness and functionality of PyMC. These tests include evaluating model convergence, performing posterior predictive checks, and assessing the accuracy of the parameter estimation. For consistency, we  use the same calibration for each of the tests.\\

Configuration of PyMC for the different tests:
\begin{itemize}
\item The number of samples to draw = 1000
\item Number of iterations to tune = 1000
\item Number of chains = 4
\item Initialization method to use for auto-assigned NUTS samplers = advi+adapt\_diag
\item target\_accept = 0.99
\end{itemize}

The last parameter (not that we adopt a high value of the target\_accept parameter, the default value being 0.9) is not necessary for the fit of the simple Gaussian model. However, it improves the convergence of the mixture model. We next briefly specify our choice of prior for the inference of the model parameters. For the underlying ``true" mean \textit{global\_mean} and the mean of each cluster \textit{gc\_means} we use a normal prior. For their associated error \textit{global\_sigma} and \textit{sigma\_y} we use a half-normal prior. For the mixture model, we also define a Dirichlet prior for the weights of the mixture.\\ 

For each model tested, we also add a representation of the components to help the reader easily visualize the hierarchical structure.

\subsection{Simple Gaussian}
\label{sec:Simple Gaussian}
\begin{figure}
    \centering
    \includegraphics[width=0.7\textwidth]{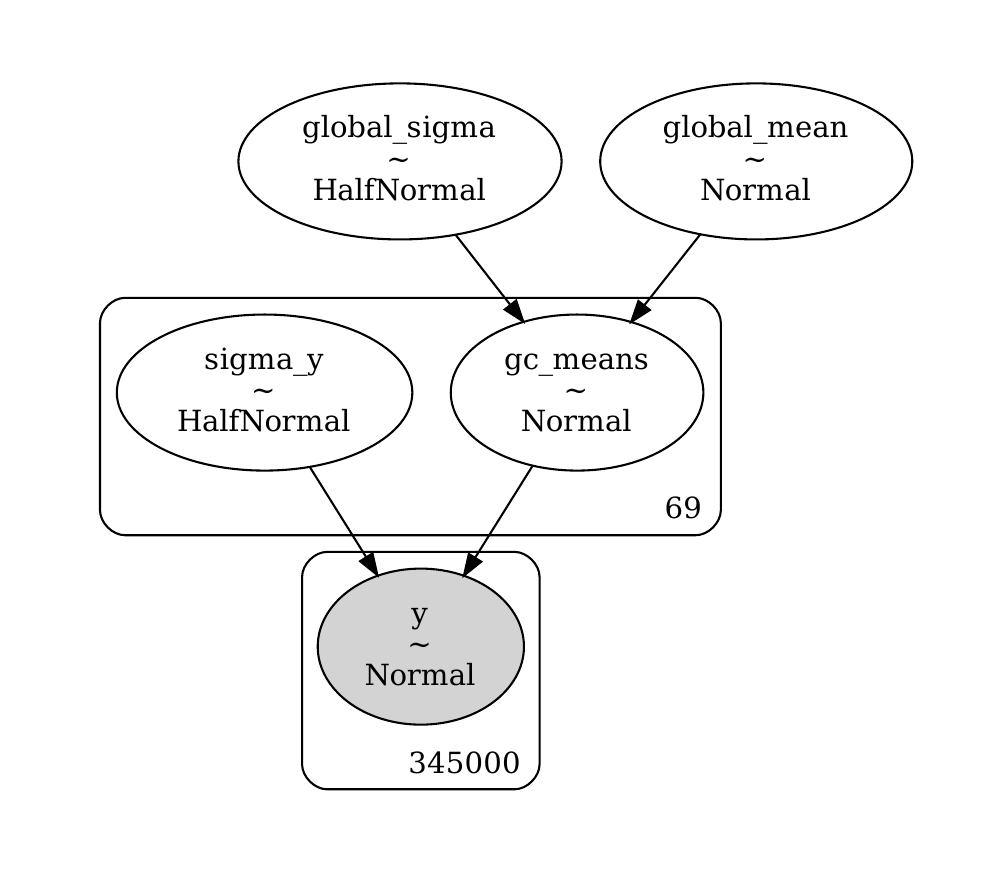}
    \caption{Diagrammatic representation of the structure of  of the hierarchical Bayesian model. The gray cell represents the likelihood. In our case, the total marginalized age posterior. The middle panel shows the prior at the cluster level for the mean and standard deviation. The top row shows the prior applied for the mean and standard deviation of the underlying distribution.}
    \label{fig:simple-structure}
\end{figure}

First, we would like to test if PyMC is able to recover the parameters from samples drawn from a simple Gaussian (as opposed to  mixture Gaussian see below). For this, we define a normal distribution with an arbitrary mean \textit{global\_mean} and variance \textit{global\_sigma}. From this underlying distribution, we draw 69 values (similar to the number of clusters in our HST sample). These values represent the mean ages \textit{gc\_means} of our test clusters. For each cluster, we produce a posterior distribution of 5000 draws centered around the mean ages. To add a bit a complexity to our model, each cluster has its own standard deviation \textit{sigma\_y}. The structure can be seen in Figure \ref{fig:simple-structure}. The \textit{y} value represents the likelihood, where the observable to fit is the total posterior distribution (hence the $69*4000 = 345000$ data points).

\begin{figure}
    \includegraphics[width=\textwidth]{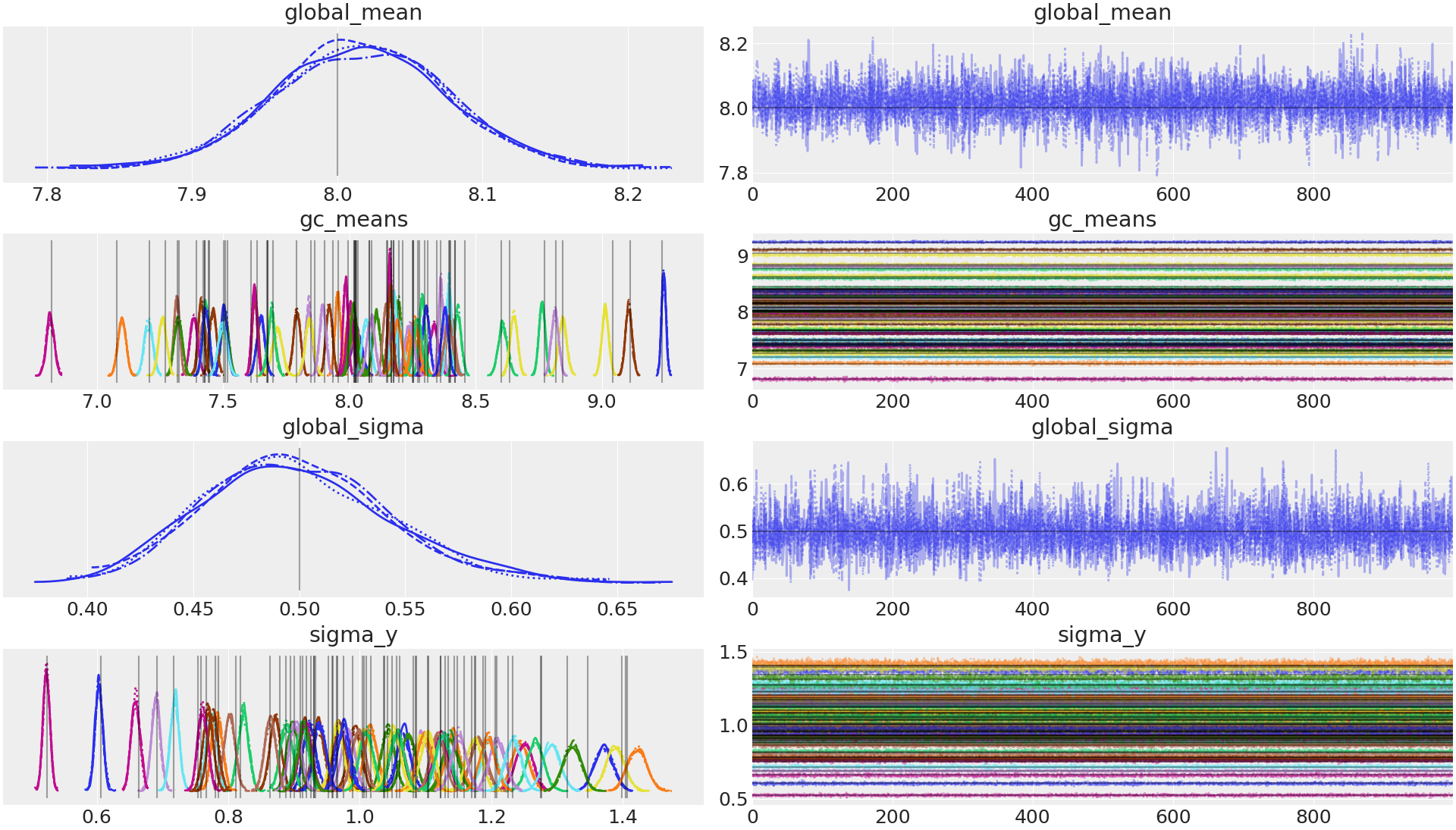}
    \includegraphics[width=\textwidth]{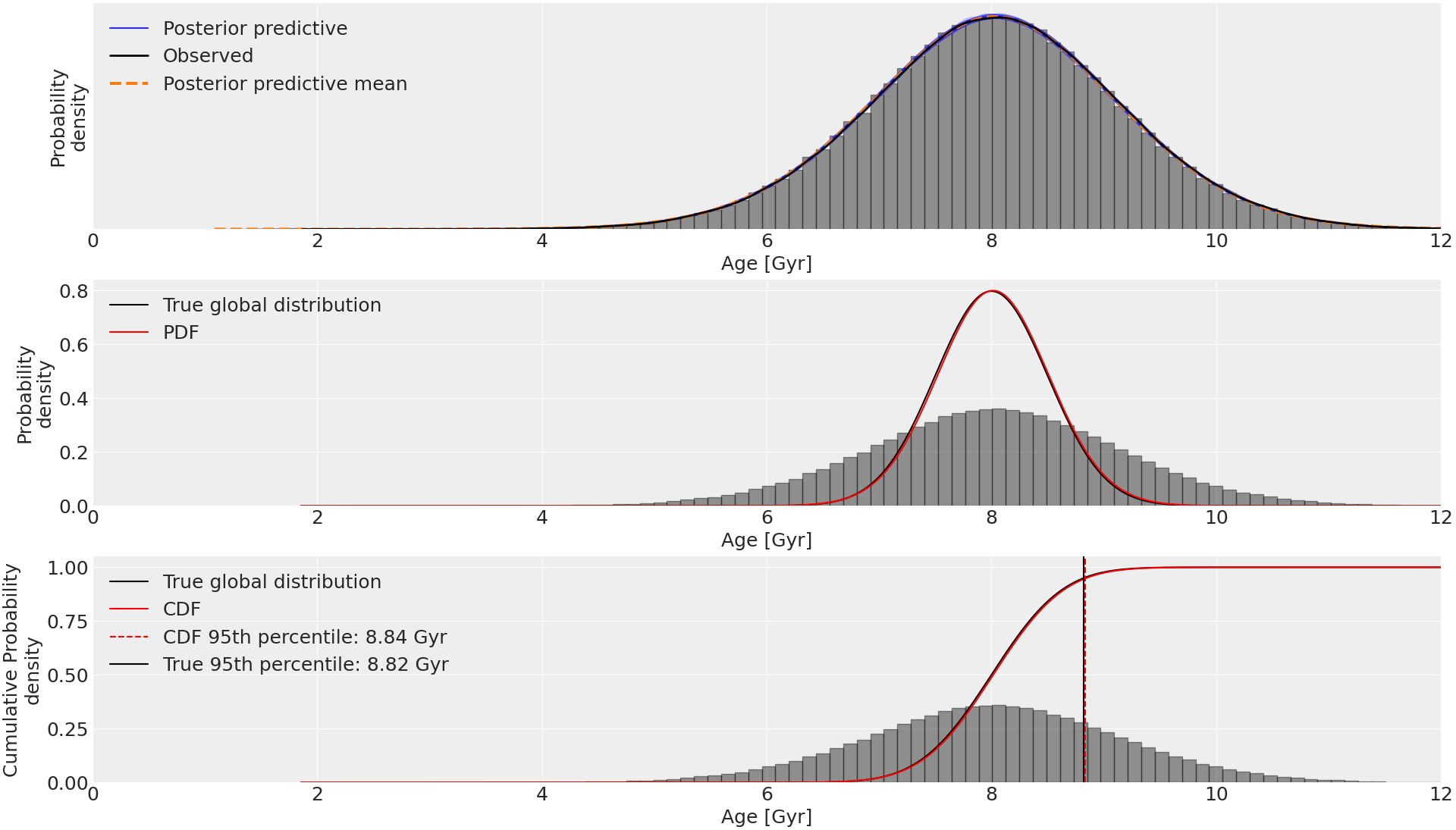}
    \caption{Trace of PyMC fit for the 4 parameters of interest. The grey line is true parameters and the black lines represent the mean and standard deviation of the simulated sample.}
    \label{fig:trace_single}
\end{figure}

In Figure \ref{fig:trace_single}, we show the results of the PyMC fit. The left column shows the distribution of the different parameters for the 4 chains, as well as the true value (gray vertical lines). The right column shows the evolution of the chains as a function of the number of steps (1000 in our case). The chains appear to be well mixed and converged. The parameters of the individual ``clusters" are well recovered (the second row shows the mean and the last row the standard deviations). The parameters of the underlying distribution are also well recovered, especially the mean, where the recovered peaks are located close to the true value (first row). The standard deviation is slightly underestimated. We experimented with several numbers of clusters for the test sample (see appendix \ref{Convergence test}) and argue that the convergence improves as the number of clusters increases--as expected. However, since our HST sample is limited to 69, we might not perfectly recover the true deviation, but the discrepancy here is less than $1\sigma$. This can be seen in the third row of Figure \ref{fig:trace_single}  where we show the total posterior distribution of our 69 test clusters (gray histogram). The solid black line represents the global distribution that we used to create the sample (with \textit{global\_mean} and \textit{global\_sigma}), and the dashed green line represents the distribution using the mean of the sampled parameters. Even though the location of the two normal distributions is in good agreement, the sampled one is slightly more peaked, indicative of a smaller variance.

The last step is to check if we can recover the total posterior distribution (gray histogram) using PyMC predictive checks. The top panel of Figure \ref{fig:trace_single} presents the results of this test. The solid black line follows the edge of the histogram of the total posterior distribution (our observable), the blue lines are 50 posteriors drawn from the parameter distributions\footnote{For each posterior, a random value is drawn from each of the parameter distribution (\textit{global\_mean}, \textit{global\_sigma}, \textit{gc\_means} and \textit{sigma\_y}.}, and the red dashed line is the posterior drawn from the mean values of the parameters. We can see that the predictive posteriors perfectly recover the total posterior of our simulated sample. 

\subsection{Gaussian mixture}
\label{sec:Gaussian mixture}
\begin{figure}
    \centering
     \includegraphics[width=\linewidth]{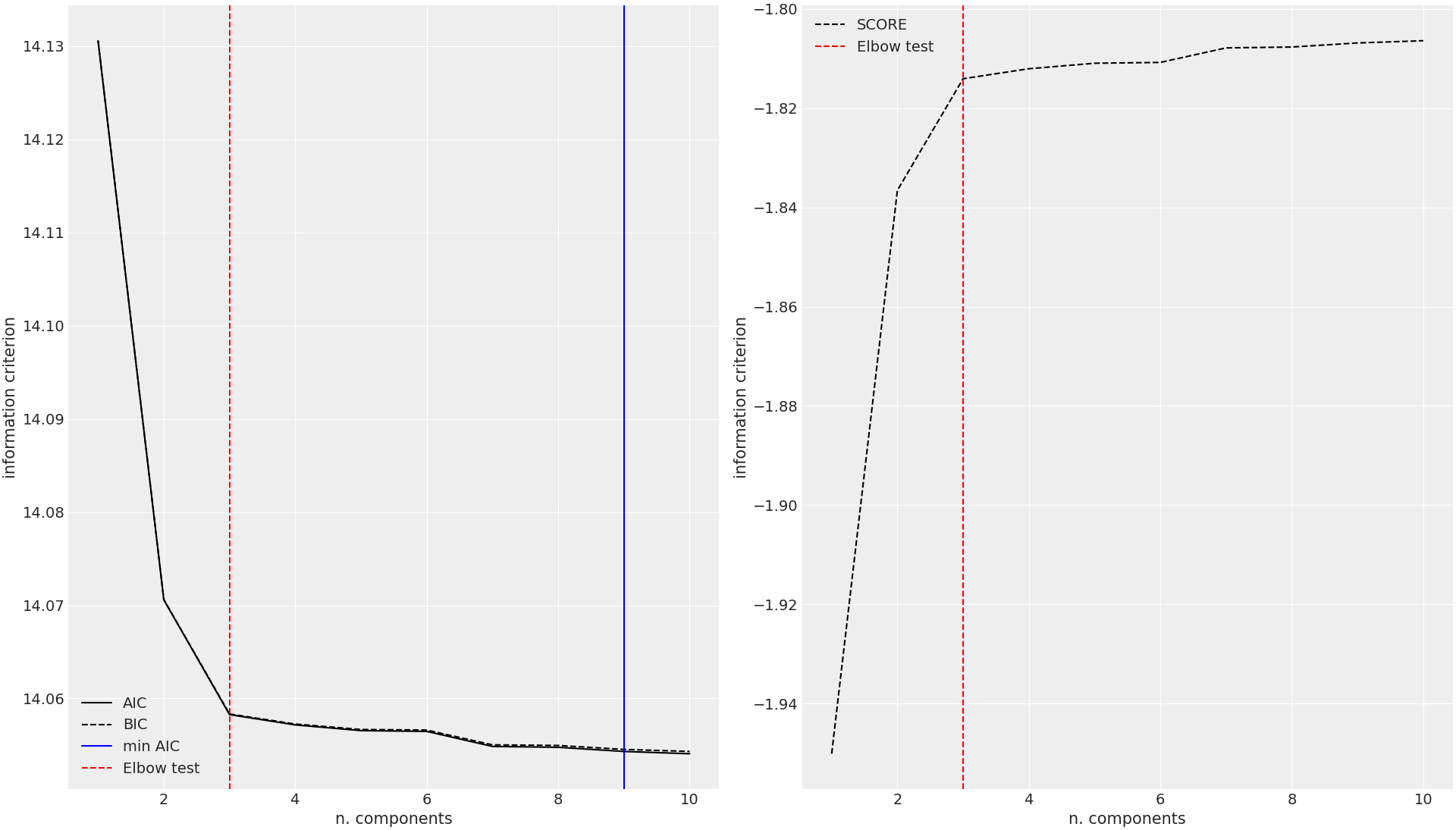}
       \includegraphics[width=\linewidth]{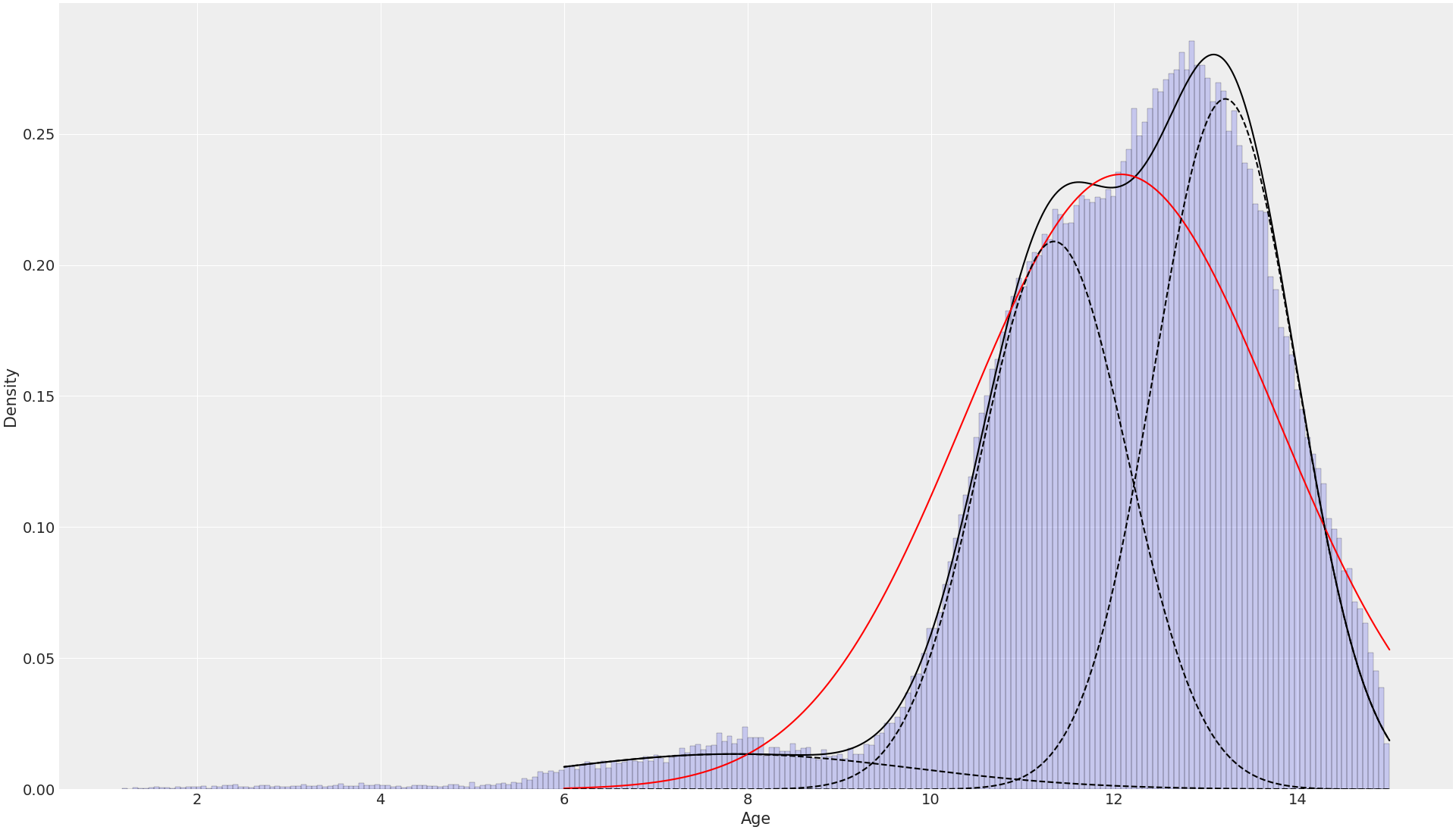}
    \caption{GMM fit of the total age posterior}
    \label{fig:totpostcomp}
\end{figure}

The second test we want to perform is to check whether PyMC can recover the parameters of clusters drawn from a more complex distribution. In this case we opt for a Gaussian mixture model, what could be thought of as several periods of cluster formation. \\
As the mixture model requires a fixed number of parameters, and we want to apply our model to the real data eventually,  we use the sklearn GMM library to fit the total age posterior that we got from pocoMC and apply the elbow method to obtain the optimal number of components. We found that both the score, AIC and BIC gave a preference for a mixture of 3 components (see Figure \ref{fig:totpostcomp}). The lower panel of the figure shows each of the component, the mixture, and a single Gaussian as a comparison (red). The total age posterior is also plotted (blue histogram). \\

To define our model, we create a mixture of 3 Gaussians, with their individual mean \textit{mu}, standard deviation \textit{sigma\_y} and weights  \textit{w}. Similar to the first test, we draw 69 values (similar to the number of clusters in our HST sample) from our underlying distribution (here a mixture). These values represent the mean ages \textit{gc\_means} of our test clusters. For each cluster, we produce a posterior distribution of 5000 draws centered around the mean ages. As before, to add complexity to our model, each cluster has its own standard deviation \textit{sigma\_gc}. 

\begin{figure}
    \includegraphics[width=\textwidth]{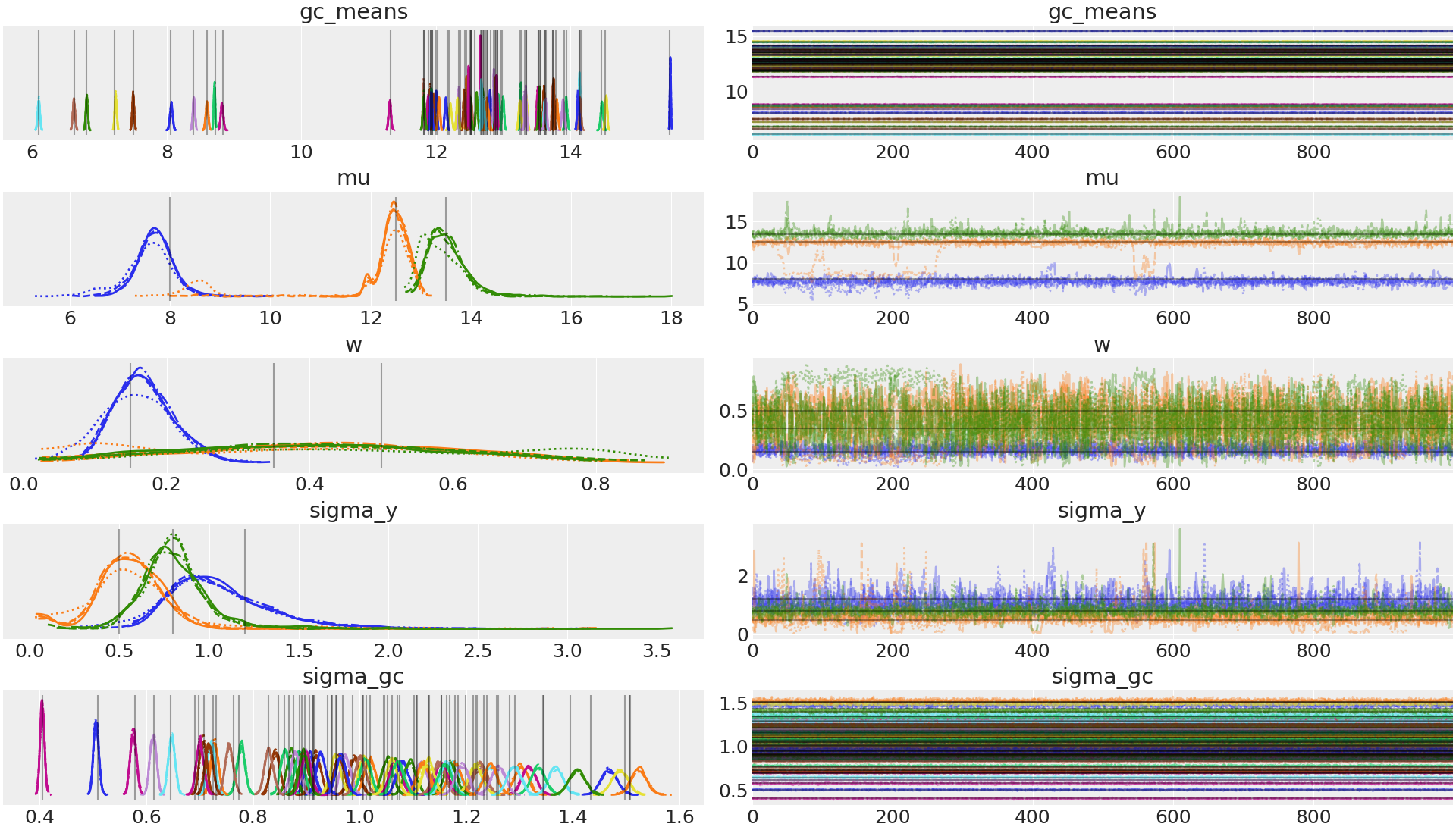}
    \includegraphics[width=\textwidth]{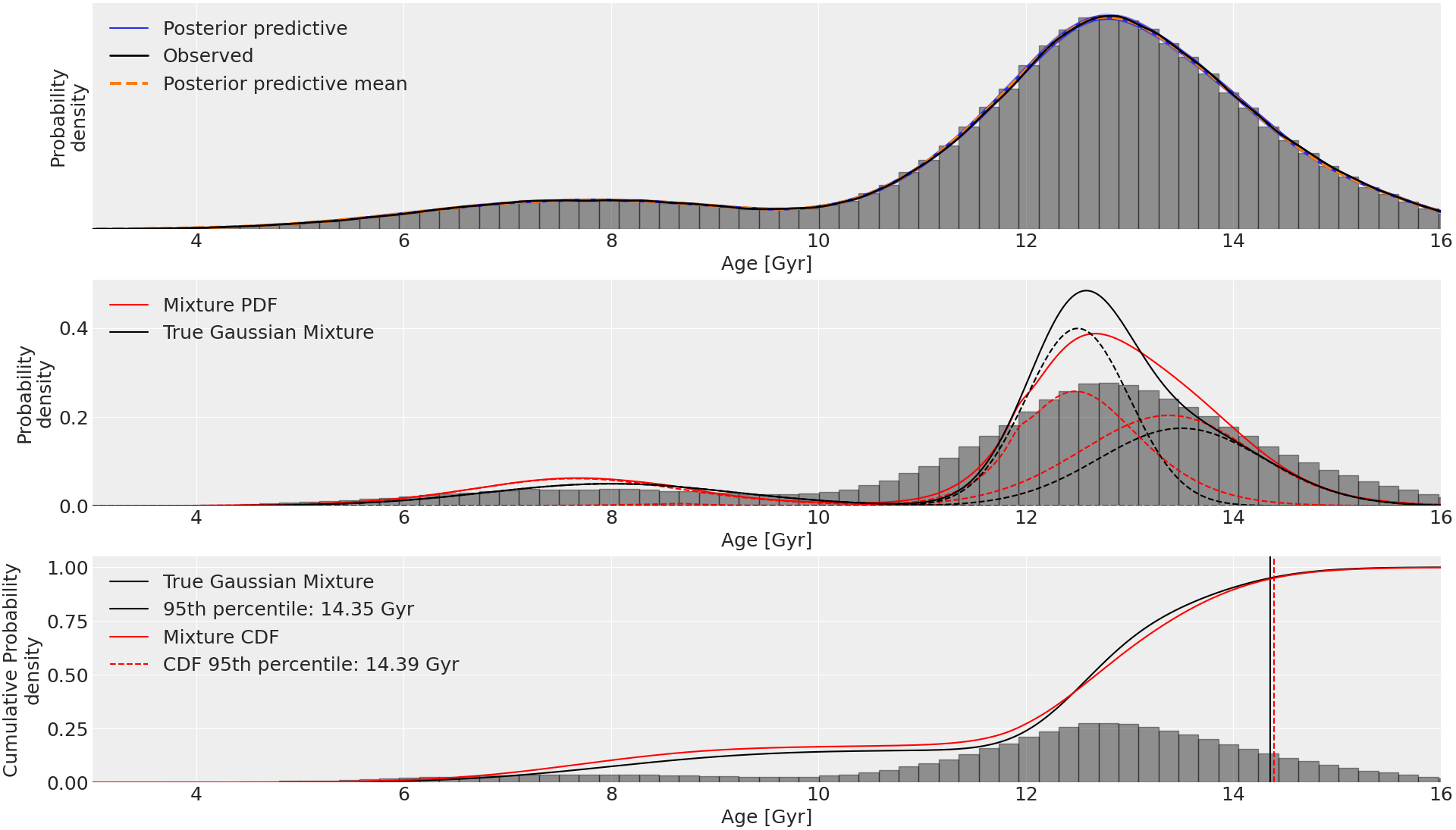}
    \caption{Trace of PyMC fit for the 4 parameters of interest. The gray line is true parameters, and the black lines represent the mean and standard deviation of the simulated sample.}
    \label{fig:trace_mixture}
\end{figure}

In Figure \ref{fig:trace_mixture}, we show the results of the PyMC fit. The parameters of the individual "clusters" are well recovered. The mean and the variance of the underlying distribution are also well recovered, especially the mean, where the recovered peaks are located close to the true value. Two out of three components have standard deviations in very good agreement with the true values. The third is slightly underestimated. Two of the components weights have large distributions. It is very complicated to recover the true weights because of label switching\footnote{The components of the mixture are interchangeable, and one way to break the degeneracy is to order the components \cite{Stephens}.Even then there is still switching between the chains.}, especially when the means of the mixture are close. The last two panels are very interesting. Even though the PDF appears to be off, because the means are correct but the variance and weights are different, the CDF is in very good agreement and the 95th percentile is close to the true value.

\subsection{Mixture null test}
\label{Mixture null test}

The last test to perform is what we called here a null. Indeed, some of the readers might argue that the number of components chosen for the fit will alter the inference of the parameters or that the formation process can be explained by a single Gaussian. In the latter case, the introduction of extra components can be similar to overfitting. Both points are valid and therefore worth investigating.

\begin{figure}
    \includegraphics[width=\textwidth]{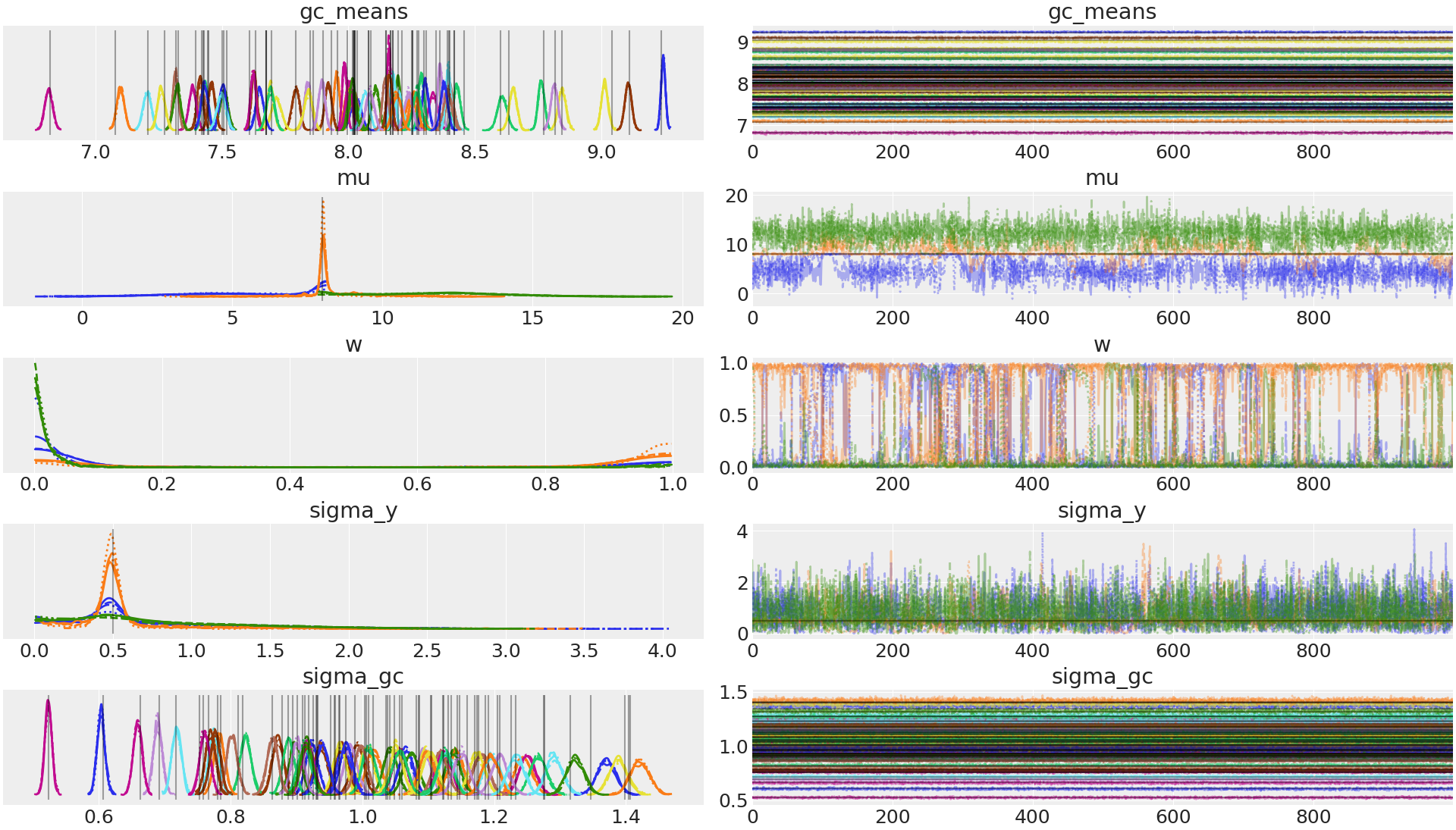}
    \includegraphics[width=\textwidth]{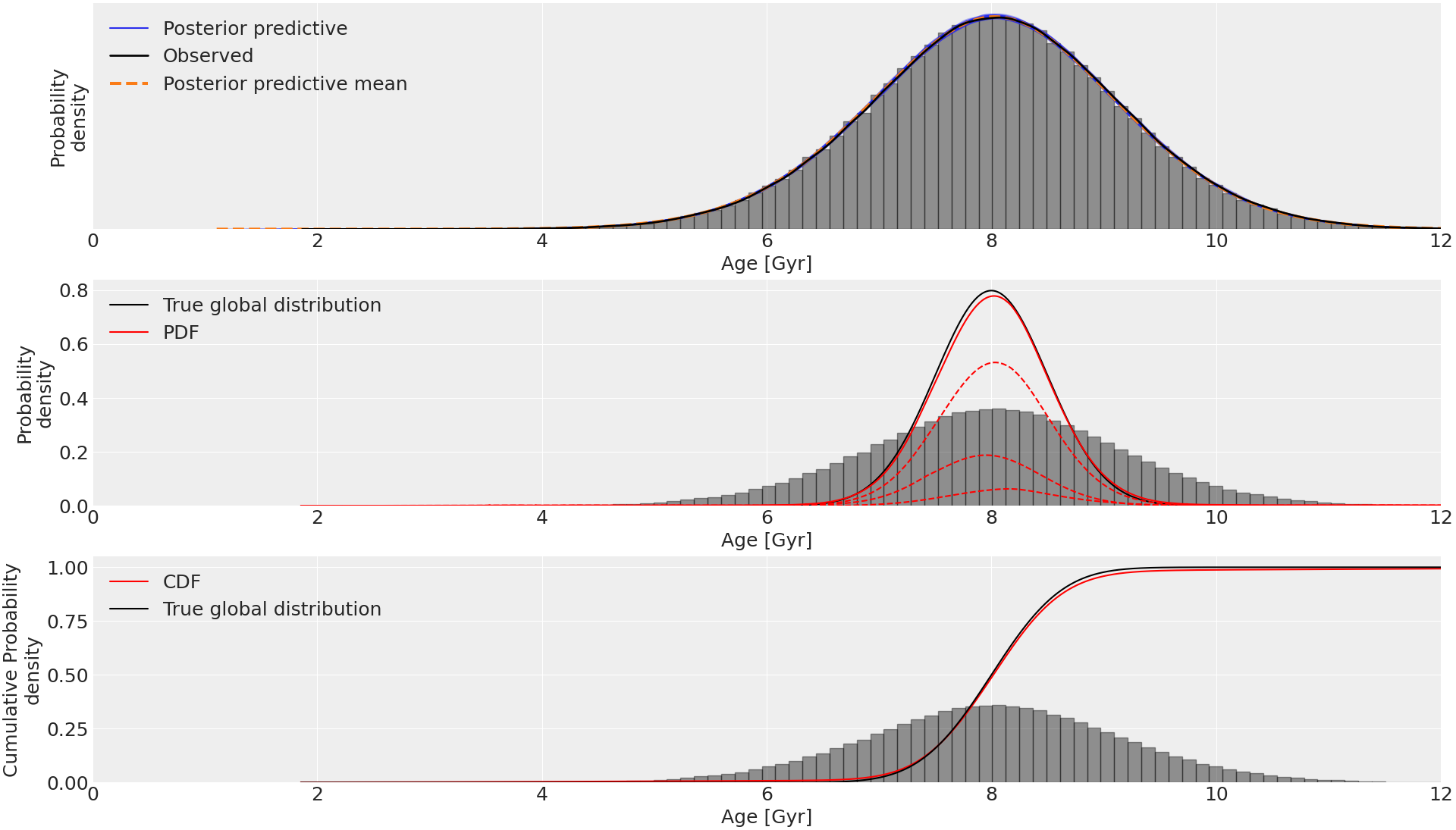}
    \caption{Trace of PyMC fit for the 4 parameters of interest. The grey line is true parameters and the black lines represent the mean and standard deviation of the simulated sample.}
    \label{fig:trace_null}
\end{figure}

Ultimately, we are interesting by the underlying distribution of the sample and not the individual components of a possible mixture. Can we recover the underlying posterior by wrongfully fitting a sample generated by a single Gaussian with a mixture. In order to avoid redundancy in our explanation, we can summarize the null test this way. The clusters are created with a procedure similar to the simple Gaussian case (see appendix \ref{sec:Simple Gaussian}. However, the fitting procedure matches the mixture case \ref{sec:Gaussian mixture}. The results can be seen in Figure \ref{fig:trace_null}. The cluster parameters are still well recovered. The 3 mixture components are peaked at the true value for the mean and the standard deviation. The weights are peaked at either 0 or 1, indicating overfitting and the likelihood of a single Gaussian. It is also reassuring to see that the cumulative distribution is almost perfectly recovered.

\subsection{Convergence test}
\label{Convergence test}

As the inference of the parameters might be affected by the number of clusters in our simulated sample, we perform convergence tests for the simple and mixture hierarchical models. 
We arbitrarily selected 7 sample sizes. To also test the randomness of the PyMC fit, for each sample size, we start the sampling from 7 different seeds. Here are the values tested:
\begin{itemize}
    \item sample size = 5,15,40,69,100,150,250
    \item Random seeds = 4,8,42,88,159,222,428
\end{itemize}

\begin{figure}
    \centering
\includegraphics[width=0.9\textwidth]{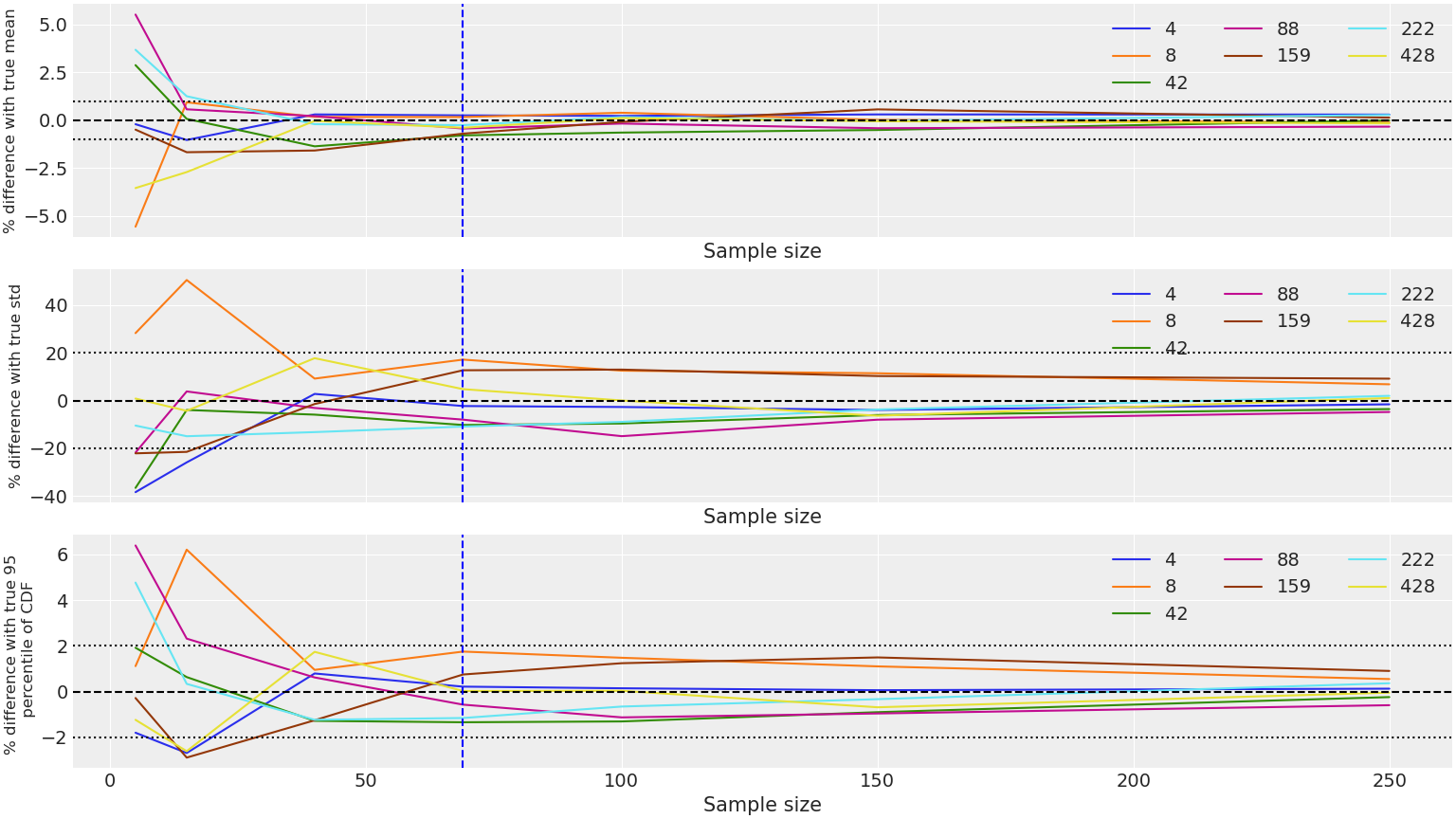}
    \caption{Convergence test for single Gaussian model.}
    \label{fig:conv_test_simple}
\end{figure}

The results for the mixture model are shown in figure \ref{fig:conv_test_simple}. We can see that the sampled mean converges rapidly towards the true mean of the distribution (less than 1\% for a sample size of 69). As expected, the standard deviation is very sensitive to the sample size (with a difference of 20\% for a sample size of 69). The interesting part here is the last row of figure \ref{fig:conv_test_simple} where we show the difference on the recovered 95 percentile of the cumulative probability distribution (CDF). Indeed, as you are interested in the age of the oldest cluster, it is important to check if we can recover the tail of the distribution. For a sample of 69 clusters, the difference is below 2\% for all random seeds \footnote{Here, the 95 percentile used to compare to the true distribution is taken from the mean of the 95 distribution. That is, for each step of each mcmc chain, a CDF is computed with its 95 percentile. We used the mean of all these percentiles 4 * 1000 for the comparison, which introduces a bit of scatter}. It is worth noting that the 95 percentile seems to mirror the standard deviation difference.

\begin{figure}
    \centering
\includegraphics[width=0.9\textwidth]{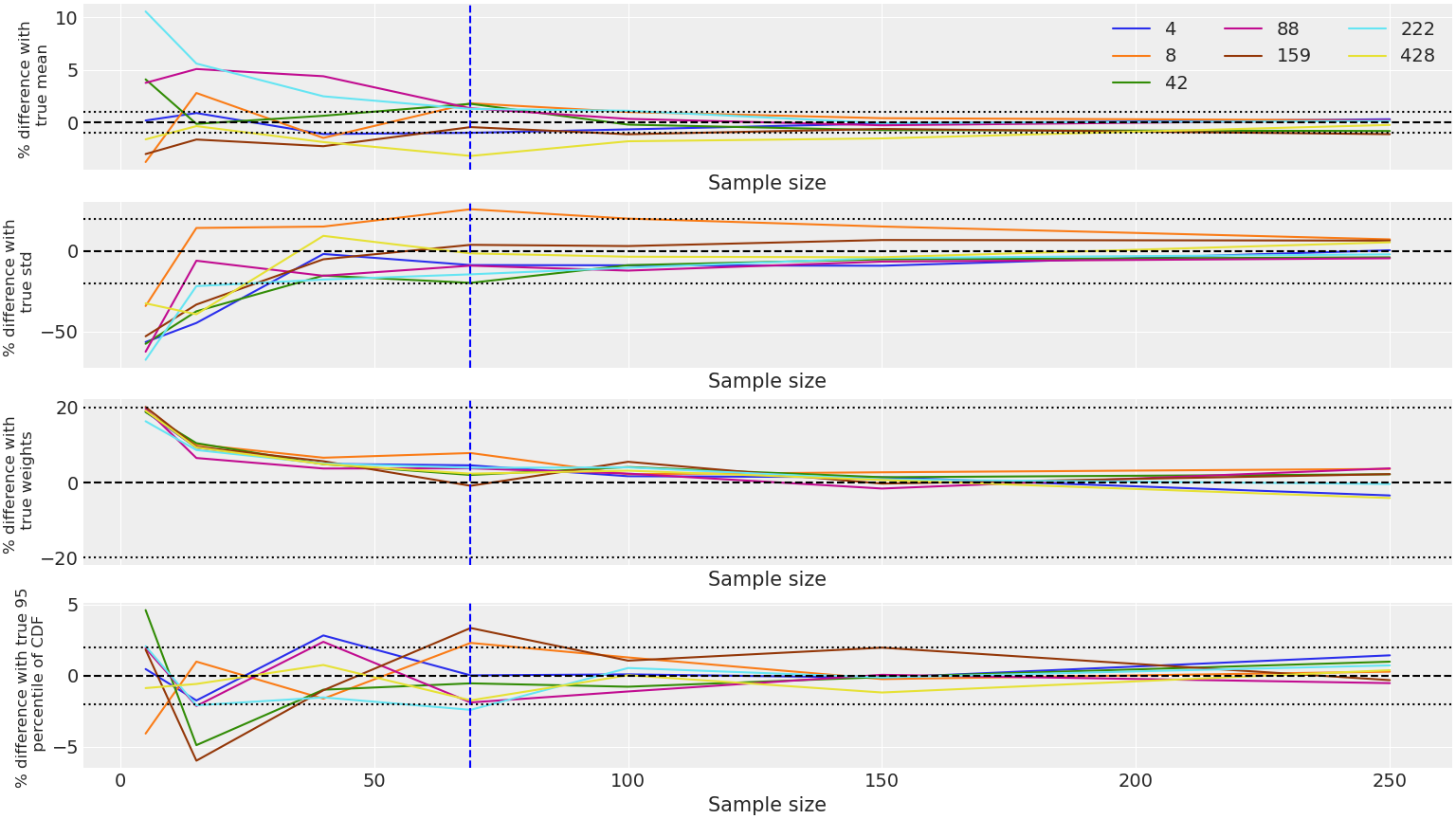}
    \caption{Convergence test for Gaussian mixture model}
    \label{fig:conv_test_mixture}
\end{figure}
\subsection{Application to our sample}

\begin{figure}[ht]
    \centering
    \includegraphics[width=\linewidth]{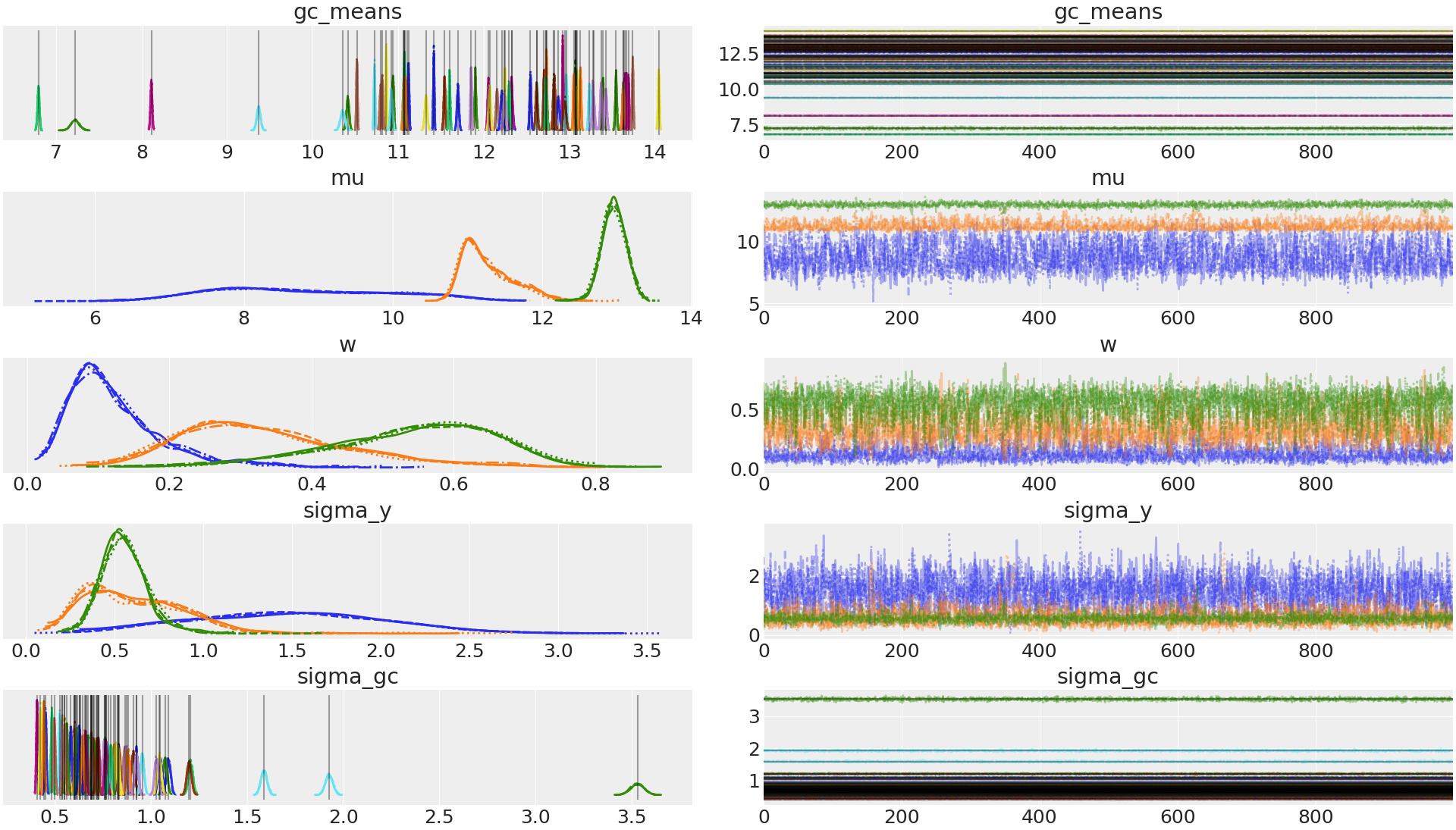}
    \caption{Trace of PyMC fit for the  dataset. Grey lines are the mean and standard deviation from pocoMC.}
    \label{fig:final-ages}
\end{figure}

Among the 3 models tested, we chose the hierarchical mixture to fit the total age posterior with PyMC. We already have an optimal number of components $3$ (see Figure \ref{fig:totpostcomp}). The results of the fit are shown in Figure \ref{fig:final-ages}. Instead of the true values displayed as vertical gray lines for the toy models, here we show the mean and standard deviation inferred from pocoMC (see Figure \ref{fig:posterior_distrib}). PyMC is in perfect agreement with the cluster parameters.

\section{Extinction source catalogs}
\label{AppC}
The source catalogs used as central values for the prior on extinction are indicated in table~\ref{tab:input-extinction}. 
\begin{table}
\begin{tabular}[t]{|l|l|l|}
\hline
nb & Cluster\_name & Extinction\_catalog \\ \hline
0  & arp2          & Harris              \\ \hline
1  & ic4499        & Harris              \\ \hline
2  & lynga7        & Harris              \\ \hline
3  & ngc0104       & Dotter              \\ \hline
4  & ngc0288       & Dotter              \\ \hline
5  & ngc0362       & Dotter              \\ \hline
6  & ngc1261       & Harris              \\ \hline
7  & ngc1851       & Dutra               \\ \hline
8  & ngc2298       & Dotter              \\ \hline
9  & ngc2808       & Harris              \\ \hline
10 & ngc3201       & Dotter              \\ \hline
11 & ngc4147       & Harris              \\ \hline
12 & ngc4590       & Dutra               \\ \hline
13 & ngc4833       & Harris              \\ \hline
14 & ngc5024       & Dotter              \\ \hline
15 & ngc5053       & Dotter              \\ \hline
16 & ngc5139       & Dutra               \\ \hline
17 & ngc5272       & Harris              \\ \hline
18 & ngc5286       & Dotter              \\ \hline
19 & ngc5466       & Dotter              \\ \hline
20 & ngc5904       & Harris              \\ \hline
21 & ngc5927       & Dotter              \\ \hline
22 & ngc5986       & Dotter              \\ \hline
23 & ngc6093       & Dotter              \\ \hline
24 & ngc6101       & Dotter              \\ \hline
25 & ngc6121       & Dotter              \\ \hline
26 & ngc6144       & Dotter              \\ \hline
27 & ngc6171       & Dotter              \\ \hline
28 & ngc6205       & Dotter              \\ \hline
29 & ngc6218       & Dotter              \\ \hline
30 & ngc6254       & Dotter              \\ \hline
31 & ngc6304       & Dotter              \\ \hline
32 & ngc6341       & Dotter              \\ \hline
33 & ngc6352       & Harris              \\ \hline
34 & ngc6362       & Dutra               \\ \hline
35 & ngc6366       & Harris              \\ \hline
36 & ngc6388       & Harris              \\ \hline
\end{tabular}
\begin{tabular}[t]{|l|l|l|}
\hline
nb & Cluster\_name & Extinction\_catalog \\ \hline
37 & ngc6397       & Dotter              \\ \hline
38 & ngc6426       & Dutra               \\ \hline
39 & ngc6441       & Harris              \\ \hline
40 & ngc6496       & Dotter              \\ \hline
41 & ngc6535       & Dutra               \\ \hline
42 & ngc6541       & Dotter              \\ \hline
43 & ngc6584       & Dotter              \\ \hline
44 & ngc6624       & Dotter              \\ \hline
45 & ngc6637       & Dotter              \\ \hline
46 & ngc6652       & Dotter              \\ \hline
47 & ngc6656       & Harris              \\ \hline
48 & ngc6681       & Dotter              \\ \hline
49 & ngc6715       & Harris              \\ \hline
50 & ngc6717       & Dotter              \\ \hline
51 & ngc6723       & Dotter              \\ \hline
52 & ngc6752       & Dotter              \\ \hline
53 & ngc6779       & Dutra               \\ \hline
54 & ngc6809       & Dotter              \\ \hline
55 & ngc6838       & Dotter              \\ \hline
56 & ngc6934       & Dotter              \\ \hline
57 & ngc6981       & Dotter              \\ \hline
58 & ngc7006       & Dutra               \\ \hline
59 & ngc7078       & Dotter              \\ \hline
60 & ngc7089       & Dotter              \\ \hline
61 & ngc7099       & Dotter              \\ \hline
62 & palomar1      & Harris              \\ \hline
63 & palomar12     & Dotter              \\ \hline
64 & palomar15     & Dutra               \\ \hline
65 & pyxis         & Dutra               \\ \hline
66 & ruprecht106   & Harris              \\ \hline
67 & terzan7       & Dotter              \\ \hline
68 & terzan8       & Dotter              \\ \hline
\end{tabular}
\caption{The source catalog providing the central value of the prior on extinction for each individual cluster; Ref.~\cite{Harris} is denoted by Harris, Ref.~\cite{Dotter2010} by Dotter, Ref.~\cite{Dutra} by Dutra.}
\label{tab:input-extinction}
\end{table}

\section{Parameter constraints: globular clusters}
\label{app:GCtable-params}
\begin{figure}
    \centering
    \includegraphics[width=\textwidth]{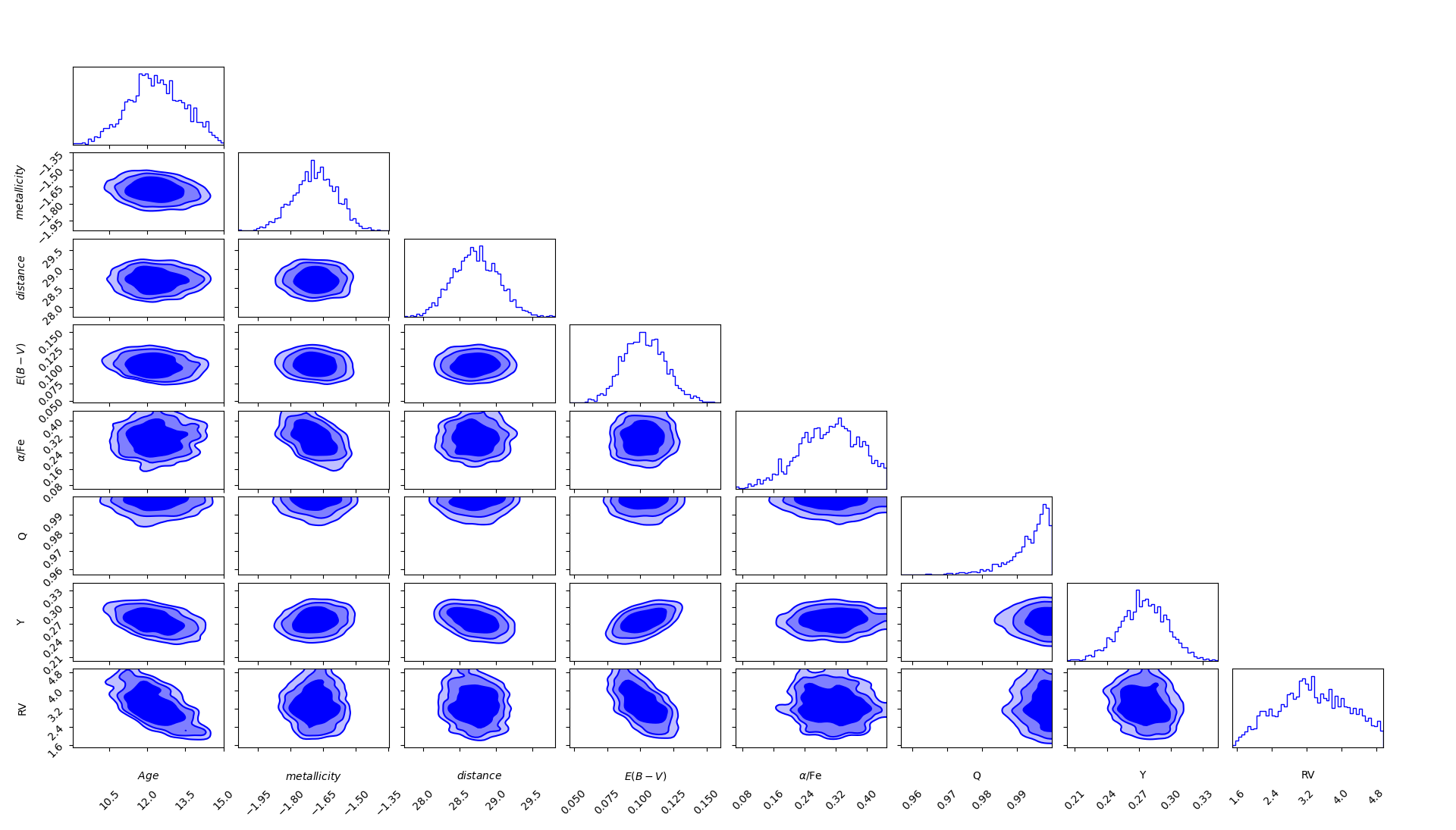}
    \includegraphics[width=\textwidth]{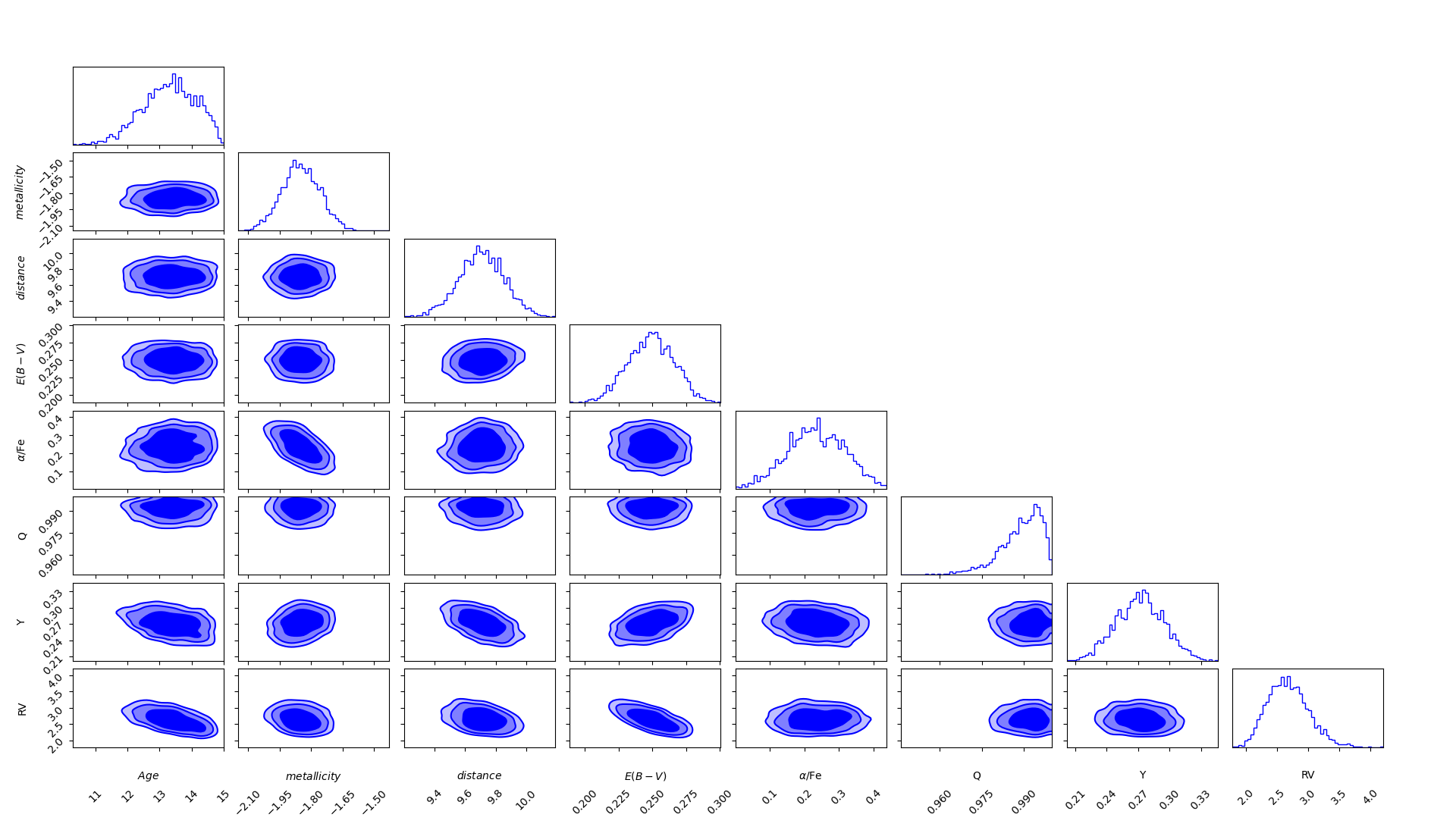}
    \caption{Corner plot of the free parameters. Top for Arp2 and bottom for NGC2298.}
    \label{fig:cornerplots}
\end{figure}

In Fig.~\ref{fig:cornerplots} we show corner plots of the two of the cluster used for calibration, Arp2 and NGC2298. 

Table~\ref{table:param_constraints} reports the best-fit parameters for the GC sample considered in this paper and the  one-dimensional marginalized statistical uncertainties at 68\% confidence level.

\begin{table}
\resizebox{\textwidth}{!}{%
\begin{tabular}{|l|l|l|l|l|l|l|l|l|l|l|l|l|l|l|l|l|}
\hline
Cluster\_name & Age (Gyr) & $\pm$ & {[}Fe/H{]} & $\pm$ & Distance (kpc) & $\pm$ & E(B-V) & $\pm$ & {[}$\alpha$/Fe{]} & $\pm$ & Q     & $\pm$ & Y     & $\pm$ & Rv    & $\pm$ \\ \hline
arp2          & 12.326    & 1.092 & -1.689     & 0.099 & 28.716         & 0.313 & 0.103  & 0.016 & 0.301             & 0.077 & 0.993 & 0.007 & 0.275 & 0.022 & 3.338 & 0.803 \\ \hline
ic4499        & 11.101    & 0.808 & -1.597     & 0.091 & 18.810         & 0.229 & 0.234  & 0.018 & 0.192             & 0.083 & 0.993 & 0.007 & 0.269 & 0.022 & 3.114 & 0.385 \\ \hline
lynga7        & 10.410    & 1.074 & -0.665     & 0.095 & 7.589          & 0.127 & 0.761  & 0.018 & 0.217             & 0.070 & 0.983 & 0.009 & 0.277 & 0.020 & 3.149 & 0.114 \\ \hline
ngc0104       & 12.924    & 0.408 & -0.799     & 0.037 & 4.494          & 0.027 & 0.010  & 0.006 & 0.414             & 0.033 & 0.995 & 0.005 & 0.279 & 0.014 & 2.846 & 0.917 \\ \hline
ngc0288       & 11.542    & 0.688 & -1.280     & 0.051 & 9.009          & 0.078 & 0.020  & 0.010 & 0.337             & 0.065 & 0.994 & 0.006 & 0.290 & 0.017 & 3.208 & 0.949 \\ \hline
ngc0362       & 10.723    & 0.525 & -1.175     & 0.070 & 8.796          & 0.088 & 0.024  & 0.010 & 0.271             & 0.073 & 0.995 & 0.005 & 0.291 & 0.017 & 2.923 & 0.952 \\ \hline
ngc1261       & 10.861    & 0.426 & -1.183     & 0.082 & 16.281         & 0.159 & 0.008  & 0.005 & 0.180             & 0.076 & 0.995 & 0.005 & 0.279 & 0.015 & 2.953 & 0.953 \\ \hline
ngc1851       & 11.074    & 0.485 & -1.148     & 0.070 & 11.884         & 0.117 & 0.036  & 0.011 & 0.295             & 0.071 & 0.995 & 0.005 & 0.291 & 0.017 & 2.690 & 0.879 \\ \hline
ngc2298       & 13.274    & 0.834 & -1.850     & 0.090 & 9.705          & 0.150 & 0.249  & 0.017 & 0.232             & 0.083 & 0.989 & 0.008 & 0.272 & 0.023 & 2.681 & 0.324 \\ \hline
ngc2808       & 10.517    & 0.497 & -1.117     & 0.064 & 10.016         & 0.103 & 0.230  & 0.017 & 0.286             & 0.065 & 0.995 & 0.005 & 0.353 & 0.016 & 2.535 & 0.279 \\ \hline
ngc3201       & 11.699    & 0.831 & -1.494     & 0.073 & 4.728          & 0.038 & 0.278  & 0.018 & 0.263             & 0.088 & 0.994 & 0.006 & 0.243 & 0.018 & 2.992 & 0.305 \\ \hline
ngc4147       & 11.065    & 0.706 & -1.711     & 0.081 & 18.569         & 0.198 & 0.032  & 0.012 & 0.357             & 0.062 & 0.994 & 0.005 & 0.303 & 0.019 & 3.374 & 0.951 \\ \hline
ngc4590       & 11.903    & 0.673 & -2.288     & 0.088 & 10.396         & 0.095 & 0.064  & 0.014 & 0.262             & 0.084 & 0.990 & 0.007 & 0.299 & 0.024 & 3.296 & 0.858 \\ \hline
ngc4833       & 13.652    & 0.631 & -1.939     & 0.083 & 6.446          & 0.072 & 0.333  & 0.017 & 0.311             & 0.072 & 0.994 & 0.005 & 0.282 & 0.023 & 3.229 & 0.233 \\ \hline
ngc5024       & 12.705    & 0.602 & -1.979     & 0.090 & 18.464         & 0.157 & 0.022  & 0.009 & 0.307             & 0.078 & 0.995 & 0.005 & 0.272 & 0.020 & 2.931 & 0.946 \\ \hline
ngc5053       & 12.624    & 0.768 & -2.264     & 0.084 & 17.534         & 0.209 & 0.025  & 0.013 & 0.341             & 0.068 & 0.993 & 0.006 & 0.277 & 0.025 & 3.175 & 0.964 \\ \hline
ngc5139       & 13.096    & 0.628 & -1.470     & 0.078 & 5.420          & 0.046 & 0.144  & 0.017 & 0.207             & 0.089 & 0.996 & 0.004 & 0.353 & 0.018 & 3.200 & 0.509 \\ \hline
ngc5272       & 11.073    & 0.538 & -1.443     & 0.047 & 10.139         & 0.073 & 0.010  & 0.007 & 0.346             & 0.064 & 0.995 & 0.005 & 0.279 & 0.016 & 2.939 & 0.946 \\ \hline
ngc5286       & 13.073    & 0.608 & -1.669     & 0.075 & 11.042         & 0.132 & 0.272  & 0.018 & 0.366             & 0.059 & 0.995 & 0.005 & 0.288 & 0.020 & 2.869 & 0.277 \\ \hline
ngc5466       & 10.794    & 0.758 & -1.972     & 0.099 & 16.089         & 0.148 & 0.031  & 0.015 & 0.236             & 0.084 & 0.994 & 0.006 & 0.261 & 0.023 & 3.103 & 0.941 \\ \hline
ngc5904       & 11.416    & 0.452 & -1.236     & 0.068 & 7.455          & 0.054 & 0.029  & 0.009 & 0.294             & 0.077 & 0.995 & 0.005 & 0.266 & 0.015 & 2.766 & 0.866 \\ \hline
ngc5927       & 12.734    & 0.443 & -0.548     & 0.060 & 8.137          & 0.095 & 0.419  & 0.016 & 0.248             & 0.046 & 0.995 & 0.005 & 0.273 & 0.015 & 2.926 & 0.161 \\ \hline
ngc5986       & 13.395    & 0.716 & -1.511     & 0.077 & 10.494         & 0.121 & 0.301  & 0.019 & 0.223             & 0.088 & 0.995 & 0.005 & 0.289 & 0.021 & 2.985 & 0.254 \\ \hline
ngc6093       & 13.668    & 0.602 & -1.639     & 0.085 & 10.299         & 0.111 & 0.219  & 0.017 & 0.229             & 0.081 & 0.995 & 0.005 & 0.290 & 0.021 & 2.985 & 0.318 \\ \hline
ngc6101       & 12.729    & 0.908 & -1.926     & 0.096 & 14.396         & 0.167 & 0.120  & 0.017 & 0.223             & 0.084 & 0.994 & 0.005 & 0.247 & 0.022 & 2.981 & 0.712 \\ \hline
ngc6121       & 13.236    & 0.777 & -1.153     & 0.048 & 1.849          & 0.018 & 0.422  & 0.018 & 0.371             & 0.052 & 0.988 & 0.008 & 0.260 & 0.017 & 3.049 & 0.188 \\ \hline
ngc6144       & 14.046    & 0.603 & -1.785     & 0.086 & 8.057          & 0.112 & 0.460  & 0.018 & 0.214             & 0.081 & 0.993 & 0.006 & 0.254 & 0.023 & 2.934 & 0.166 \\ \hline
ngc6171       & 12.728    & 0.728 & -1.062     & 0.063 & 5.604          & 0.064 & 0.424  & 0.015 & 0.360             & 0.059 & 0.992 & 0.007 & 0.222 & 0.014 & 2.901 & 0.179 \\ \hline
ngc6205       & 11.853    & 0.539 & -1.443     & 0.055 & 7.416          & 0.074 & 0.024  & 0.009 & 0.320             & 0.074 & 0.995 & 0.005 & 0.323 & 0.018 & 3.126 & 0.926 \\ \hline
ngc6218       & 12.155    & 0.864 & -1.303     & 0.048 & 5.097          & 0.045 & 0.198  & 0.018 & 0.349             & 0.063 & 0.994 & 0.006 & 0.264 & 0.018 & 2.992 & 0.422 \\ \hline
ngc6254       & 11.121    & 0.583 & -1.577     & 0.080 & 5.065          & 0.056 & 0.261  & 0.019 & 0.306             & 0.080 & 0.995 & 0.005 & 0.311 & 0.021 & 3.436 & 0.326 \\ \hline
ngc6304       & 11.082    & 0.656 & -0.458     & 0.057 & 6.018          & 0.116 & 0.484  & 0.018 & 0.281             & 0.046 & 0.995 & 0.005 & 0.342 & 0.019 & 3.320 & 0.176 \\ \hline
ngc6341       & 13.070    & 0.562 & -2.211     & 0.091 & 8.489          & 0.067 & 0.030  & 0.010 & 0.221             & 0.082 & 0.994 & 0.005 & 0.301 & 0.022 & 2.930 & 0.942 \\ \hline
ngc6352       & 12.326    & 0.689 & -0.589     & 0.090 & 5.465          & 0.063 & 0.245  & 0.016 & 0.258             & 0.066 & 0.994 & 0.006 & 0.282 & 0.017 & 3.096 & 0.302 \\ \hline
ngc6362       & 12.618    & 0.832 & -1.022     & 0.056 & 7.636          & 0.063 & 0.070  & 0.015 & 0.344             & 0.057 & 0.993 & 0.006 & 0.251 & 0.017 & 2.610 & 0.754 \\ \hline
ngc6366       & 10.348    & 1.927 & -0.668     & 0.079 & 3.370          & 0.035 & 0.744  & 0.016 & 0.293             & 0.064 & 0.990 & 0.007 & 0.281 & 0.018 & 3.076 & 0.218 \\ \hline
ngc6388       & 12.246    & 0.548 & -0.598     & 0.075 & 11.082         & 0.130 & 0.380  & 0.017 & 0.065             & 0.040 & 0.995 & 0.004 & 0.344 & 0.017 & 3.117 & 0.188 \\ \hline
ngc6397       & 13.077    & 1.207 & -2.038     & 0.109 & 2.473          & 0.018 & 0.195  & 0.018 & 0.299             & 0.080 & 0.993 & 0.006 & 0.305 & 0.022 & 3.277 & 0.575 \\ \hline
ngc6426       & 13.428    & 0.872 & -2.345     & 0.086 & 20.449         & 0.305 & 0.393  & 0.017 & 0.280             & 0.080 & 0.993 & 0.007 & 0.251 & 0.025 & 3.122 & 0.241 \\ \hline
ngc6441       & 13.632    & 0.727 & -0.579     & 0.068 & 12.604         & 0.144 & 0.490  & 0.017 & 0.097             & 0.038 & 0.995 & 0.005 & 0.317 & 0.019 & 3.066 & 0.140 \\ \hline
ngc6496       & 12.210    & 0.924 & -0.504     & 0.087 & 9.526          & 0.124 & 0.227  & 0.018 & 0.248             & 0.064 & 0.994 & 0.006 & 0.286 & 0.020 & 3.054 & 0.368 \\ \hline
ngc6535       & 13.634    & 0.881 & -1.913     & 0.092 & 6.275          & 0.103 & 0.419  & 0.018 & 0.304             & 0.074 & 0.975 & 0.012 & 0.241 & 0.022 & 3.482 & 0.269 \\ \hline
ngc6541       & 13.548    & 0.699 & -1.698     & 0.077 & 7.579          & 0.089 & 0.124  & 0.018 & 0.361             & 0.061 & 0.994 & 0.005 & 0.291 & 0.021 & 2.899 & 0.596 \\ \hline
ngc6584       & 12.055    & 0.725 & -1.410     & 0.063 & 13.580         & 0.147 & 0.083  & 0.015 & 0.213             & 0.086 & 0.994 & 0.005 & 0.268 & 0.020 & 3.227 & 0.766 \\ \hline
ngc6624       & 12.828    & 0.693 & -0.502     & 0.077 & 7.942          & 0.096 & 0.268  & 0.017 & 0.090             & 0.050 & 0.995 & 0.005 & 0.286 & 0.019 & 3.028 & 0.286 \\ \hline
ngc6637       & 12.967    & 0.687 & -0.622     & 0.088 & 8.833          & 0.089 & 0.179  & 0.017 & 0.227             & 0.067 & 0.995 & 0.005 & 0.299 & 0.017 & 2.457 & 0.381 \\ \hline
ngc6652       & 13.081    & 0.828 & -0.763     & 0.098 & 9.346          & 0.122 & 0.128  & 0.017 & 0.233             & 0.079 & 0.994 & 0.006 & 0.270 & 0.019 & 2.717 & 0.595 \\ \hline
ngc6656       & 12.295    & 0.762 & -1.723     & 0.088 & 3.279          & 0.036 & 0.351  & 0.018 & 0.315             & 0.077 & 0.994 & 0.005 & 0.297 & 0.022 & 3.365 & 0.249 \\ \hline
ngc6681       & 13.373    & 0.757 & -1.490     & 0.052 & 9.316          & 0.099 & 0.105  & 0.017 & 0.371             & 0.057 & 0.994 & 0.005 & 0.276 & 0.020 & 2.915 & 0.693 \\ \hline
ngc6715       & 10.814    & 0.690 & -1.532     & 0.097 & 26.148         & 0.306 & 0.159  & 0.018 & 0.276             & 0.085 & 0.994 & 0.005 & 0.320 & 0.022 & 3.519 & 0.515 \\ \hline
ngc6717       & 12.872    & 1.048 & -1.190     & 0.086 & 7.418          & 0.109 & 0.217  & 0.017 & 0.194             & 0.085 & 0.989 & 0.008 & 0.254 & 0.018 & 3.204 & 0.442 \\ \hline
ngc6723       & 13.131    & 0.609 & -1.051     & 0.074 & 8.208          & 0.085 & 0.077  & 0.014 & 0.289             & 0.069 & 0.994 & 0.005 & 0.248 & 0.018 & 2.395 & 0.659 \\ \hline
ngc6752       & 10.941    & 0.664 & -1.479     & 0.065 & 4.116          & 0.036 & 0.062  & 0.013 & 0.326             & 0.076 & 0.994 & 0.005 & 0.308 & 0.018 & 3.265 & 0.831 \\ \hline
ngc6779       & 13.692    & 0.659 & -1.945     & 0.090 & 10.363         & 0.127 & 0.259  & 0.017 & 0.211             & 0.079 & 0.994 & 0.005 & 0.291 & 0.024 & 2.867 & 0.293 \\ \hline
ngc6809       & 12.820    & 0.721 & -1.766     & 0.081 & 5.334          & 0.042 & 0.117  & 0.016 & 0.335             & 0.067 & 0.994 & 0.005 & 0.257 & 0.020 & 2.761 & 0.560 \\ \hline
ngc6838       & 10.920    & 0.955 & -0.735     & 0.066 & 3.975          & 0.046 & 0.232  & 0.018 & 0.374             & 0.057 & 0.994 & 0.006 & 0.289 & 0.017 & 3.210 & 0.429 \\ \hline
ngc6934       & 12.070    & 0.816 & -1.429     & 0.066 & 15.672         & 0.152 & 0.110  & 0.016 & 0.206             & 0.089 & 0.994 & 0.005 & 0.271 & 0.019 & 2.691 & 0.624 \\ \hline
ngc6981       & 11.600    & 0.617 & -1.392     & 0.068 & 16.561         & 0.156 & 0.055  & 0.014 & 0.192             & 0.086 & 0.994 & 0.005 & 0.277 & 0.019 & 2.726 & 0.853 \\ \hline
ngc7006       & 12.244    & 0.925 & -1.555     & 0.125 & 39.121         & 0.496 & 0.087  & 0.016 & 0.324             & 0.073 & 0.993 & 0.006 & 0.260 & 0.024 & 3.021 & 0.750 \\ \hline
ngc7078       & 13.743    & 0.551 & -2.329     & 0.079 & 10.689         & 0.091 & 0.089  & 0.016 & 0.288             & 0.077 & 0.995 & 0.005 & 0.301 & 0.022 & 2.899 & 0.732 \\ \hline
ngc7089       & 12.546    & 0.627 & -1.455     & 0.067 & 11.658         & 0.103 & 0.051  & 0.012 & 0.209             & 0.088 & 0.996 & 0.004 & 0.286 & 0.018 & 2.862 & 0.851 \\ \hline
ngc7099       & 13.054    & 0.645 & -2.170     & 0.081 & 8.444          & 0.082 & 0.045  & 0.013 & 0.326             & 0.068 & 0.995 & 0.005 & 0.283 & 0.023 & 2.741 & 0.865 \\ \hline
palomar1      & 7.221     & 3.533 & -0.712     & 0.073 & 11.095         & 0.259 & 0.162  & 0.016 & 0.085             & 0.048 & 0.944 & 0.026 & 0.234 & 0.022 & 4.023 & 0.697 \\ \hline
palomar12     & 8.114     & 0.766 & -0.912     & 0.055 & 18.491         & 0.266 & 0.044  & 0.011 & 0.061             & 0.042 & 0.930 & 0.019 & 0.231 & 0.015 & 3.482 & 0.890 \\ \hline
palomar15     & 12.947    & 1.199 & -2.053     & 0.180 & 43.577         & 0.986 & 0.426  & 0.018 & 0.205             & 0.085 & 0.981 & 0.013 & 0.293 & 0.036 & 3.235 & 0.265 \\ \hline
pyxis         & 9.368     & 1.588 & -1.216     & 0.150 & 35.990         & 0.567 & 0.294  & 0.017 & 0.197             & 0.083 & 0.991 & 0.008 & 0.270 & 0.029 & 3.425 & 0.428 \\ \hline
ruprecht106   & 11.325    & 1.044 & -1.663     & 0.099 & 20.714         & 0.325 & 0.197  & 0.017 & 0.207             & 0.084 & 0.993 & 0.007 & 0.247 & 0.022 & 3.185 & 0.572 \\ \hline
terzan7       & 6.794     & 0.790 & -0.385     & 0.078 & 23.935         & 0.476 & 0.088  & 0.014 & 0.065             & 0.049 & 0.991 & 0.007 & 0.323 & 0.024 & 3.412 & 0.865 \\ \hline
terzan8       & 13.282    & 1.029 & -2.124     & 0.115 & 27.471         & 0.363 & 0.136  & 0.017 & 0.243             & 0.082 & 0.992 & 0.007 & 0.294 & 0.027 & 3.230 & 0.659 \\ \hline
\end{tabular}
}
\caption{Parameter constraints (mean and standard deviation) for the 69 GC of the sample.}
\label{table:param_constraints}
\end{table}

\end{document}